\newcommand{\ifprep}{\iftrue}   
\newcommand{\ifnotprep}{\iffalse}   
\documentclass[aps,prc,twoside,superscriptaddress,showpacs,floatfix]{revtex4}
\usepackage{placeins}
\usepackage{xspace}
\usepackage{longtable}
\usepackage{graphicx}
\usepackage[figuresright]{rotating}
\usepackage{afterpage}

\newcommand{\GeVc}{\ensuremath{\mbox{GeV}/c}\xspace}
\newcommand{\MeVc}{\ensuremath{\mbox{MeV}/c}\xspace}
\newcommand{\T}{\ensuremath{\mbox{T}}\xspace}

\newcommand{\mm}{\ensuremath{\mbox{mm}}\xspace}
\newcommand{\cmcub}{\ensuremath{\mbox{cm}^3}\xspace}

\newcommand{\mrad}{\ensuremath{\mbox{mrad}}\xspace}
\newcommand{\rad}{\ensuremath{\mbox{rad}}\xspace}

\newcommand{\ms}{\ensuremath{\mbox{ms}}\xspace}

\newcommand{\barn}{\ensuremath{\mbox{barn}}\xspace}

\newcommand{\dedx}{\ensuremath{\mbox{d}E/\mbox{d}x}\xspace}
\newcommand{\pip}{\ensuremath{\pi^+}\xspace}
\newcommand{\pim}{\ensuremath{\pi^-}\xspace}
\newcommand{\pipm}{\ensuremath{\pi^{\pm}}\xspace}
\newcommand{\piz}{\ensuremath{\pi^0}\xspace}
\newcommand{\bfpip}{\ensuremath{\mathbf {\pi^+}}\xspace}
\newcommand{\bfpim}{\ensuremath{\mathbf {\pi^-}}\xspace}

\newcommand{\dzeroprime}{\ensuremath{d'_0}\xspace}

\newcommand{\evtspill}{\ensuremath{N_{\mathrm{evt}}}\xspace}
\newcommand{\pt}{\ensuremath{p_{\mathrm{T}}}\xspace}

\newcommand{\intlen}{\ensuremath{\lambda_{\mathrm{I}}}\xspace}

\def\be{\begin{equation}}
\def\ee{\end{equation}}
\def\bea{\begin{eqnarray}}
\def\eea{\end{eqnarray}}

\begin{document}
\title{{\Large EUROPEAN ORGANIZATION FOR NUCLEAR RESEARCH} \\
\ifprep
\vskip 1cm
\begin{flushright}
{\rm CERN-PH-EP/2009-022} \\
{\rm 4 September 2009}
\end{flushright}
\fi
\vskip 2cm

{\LARGE \bf Comparison of large-angle production of charged pions 
with incident protons on cylindrical long and short targets
} \\

\vskip 2cm
{\bf HARP Collaboration}
\vskip 2cm
\begin{flushleft}
{\rm  

 The HARP collaboration has presented 
 measurements of the  double-differential \pipm production
 cross-section in the range of momentum $100~\MeVc \leq p \le 800~\MeVc$ 
 and angle $0.35~\rad \leq \theta  \le 2.15~\rad$
 with proton beams hitting thin nuclear targets.
 In many applications the extrapolation to long targets is necessary. 
 In this paper the analysis of data taken with long (one interaction
 length) solid cylindrical targets made of carbon, tantalum and lead is
 presented. 
 The data were taken  with the large acceptance HARP detector in the T9
 beam line of the CERN Proton Synchrotron.
 The secondary pions were produced by beams 
 of protons with momenta 5~\GeVc, 8~\GeVc and 12~\GeVc.
 The tracking and identification of the
 produced particles were performed using a small-radius
 cylindrical time projection chamber (TPC) placed inside a solenoidal
 magnet. 
 Incident protons were identified by an elaborate system of beam
 detectors.
 Results are obtained for the double-differential yields per target nucleon
  $
  {{\mathrm{d}^2 \sigma}}/{{\mathrm{d}p\mathrm{d}\theta }}
  $.
 The measurements are compared with predictions of the MARS and GEANT4 Monte Carlo
 simulations. 
}
\end{flushleft}

\ifprep
\vskip 5cm
\centerline{\em{(to be published in Physical Review C)}}
\fi
\clearpage
}

\author{M.~Apollonio} 
\altaffiliation{Now at Imperial College, University of London, UK.}
\affiliation{Universit\`{a} degli Studi e Sezione INFN, Trieste, Italy}
\author{A.~Artamonov}   
\altaffiliation{ITEP, Moscow, Russian Federation.}
\affiliation{ CERN, Geneva, Switzerland}
\author{A. Bagulya} 
\affiliation{P. N. Lebedev Institute of Physics (FIAN), Russian Academy of
Sciences, Moscow, Russia}
\author{G.~Barr}
\affiliation{Nuclear and Astrophysics Laboratory, University of Oxford, UK} 
\author{A.~Blondel}
\affiliation{Section de Physique, Universit\'{e} de Gen\`{e}ve, Switzerland} 
\author{F.~Bobisut} 
\affiliation{Sezione INFN$^{(a)}$ and Universit\'a degli Studi$^{(b)}$, 
Padova, Italy}
\author{M.~Bogomilov}
\affiliation{ Faculty of Physics, St. Kliment Ohridski University, Sofia,
  Bulgaria}
\author{M.~Bonesini}
\affiliation{Sezione INFN Milano Bicocca, Milano, Italy} 
\author{C.~Booth} 
\affiliation{ Dept. of Physics, University of Sheffield, UK}
\author{S.~Borghi}  
\altaffiliation{Now at the University of Glasgow, UK}
\affiliation{Section de Physique, Universit\'{e} de Gen\`{e}ve, Switzerland}
\author{S.~Bunyatov}
\affiliation{Joint Institute for Nuclear Research, JINR Dubna, Russia} 
\author{J.~Burguet--Castell}
\affiliation{Instituto de F\'{i}sica Corpuscular, IFIC, CSIC and Universidad de Valencia, Spain}
\author{M.G.~Catanesi}
\affiliation{Sezione INFN, Bari, Italy} 
\author{A.~Cervera--Villanueva}
\affiliation{Instituto de F\'{i}sica Corpuscular, IFIC, CSIC and Universidad de Valencia, Spain}
\author{P.~Chimenti}  
\affiliation{Universit\`{a} degli Studi e Sezione INFN, Trieste, Italy}
\author{L.~Coney} 
\affiliation{Columbia University, New York, USA}
\author{E.~Di~Capua}
\affiliation{Universit\`{a} degli Studi e Sezione INFN, Ferrara, Italy} 
\author{U.~Dore}
\affiliation{ Universit\`{a} ``La Sapienza'' e Sezione INFN Roma I, Roma,
  Italy}
\author{J.~Dumarchez}
\affiliation{ LPNHE, Universit\'{e}s de Paris VI et VII, Paris, France}
\author{R.~Edgecock}
\affiliation{Rutherford Appleton Laboratory, Chilton, Didcot, UK} 
\author{M.~Ellis}     
\altaffiliation{Now at FNAL, Batavia, Illinois, USA.} 
\affiliation{Rutherford Appleton Laboratory, Chilton, Didcot, UK}
\author{F.~Ferri} 
\affiliation{Sezione INFN Milano Bicocca, Milano, Italy}
\author{U.~Gastaldi}
\affiliation{Laboratori Nazionali di Legnaro dell' INFN, Legnaro, Italy}
\author{S.~Giani} 
\affiliation{ CERN, Geneva, Switzerland}
\author{G.~Giannini} 
\affiliation{Universit\`{a} degli Studi e Sezione INFN, Trieste, Italy}
\author{D.~Gibin}
\affiliation{Sezione INFN$^{(a)}$ and Universit\'a degli Studi$^{(b)}$, 
Padova, Italy}
\author{S.~Gilardoni}       
\affiliation{ CERN, Geneva, Switzerland} 
\author{P.~Gorbunov}  
\altaffiliation{ITEP, Moscow, Russian Federation.}
\affiliation{ CERN, Geneva, Switzerland}
\author{C.~G\"{o}\ss ling}
\affiliation{ Institut f\"{u}r Physik, Universit\"{a}t Dortmund, Germany}
\author{J.J.~G\'{o}mez--Cadenas} 
\affiliation{Instituto de F\'{i}sica Corpuscular, IFIC, CSIC and Universidad de Valencia, Spain}
\author{A.~Grant}  
\affiliation{ CERN, Geneva, Switzerland}
\author{J.S.~Graulich}
\altaffiliation{Now at Section de Physique, Universit\'{e} de Gen\`{e}ve, Switzerland.}
\affiliation{Institut de Physique Nucl\'{e}aire, UCL, Louvain-la-Neuve,
  Belgium} 
\author{G.~Gr\'{e}goire}
\affiliation{Institut de Physique Nucl\'{e}aire, UCL, Louvain-la-Neuve,
  Belgium} 
\author{V.~Grichine}  
\affiliation{P. N. Lebedev Institute of Physics (FIAN), Russian Academy of
Sciences, Moscow, Russia}
\author{A.~Grossheim} 
\altaffiliation{Now at TRIUMF, Vancouver, Canada.}
\affiliation{ CERN, Geneva, Switzerland} 
\author{A.~Guglielmi$^{(a)}$}
\affiliation{Sezione INFN$^{(a)}$ and Universit\'a degli Studi$^{(b)}$, 
Padova, Italy}
\author{L.~Howlett}
\affiliation{ Dept. of Physics, University of Sheffield, UK}
\author{A.~Ivanchenko}
\altaffiliation{ On leave from Novosibirsk University,  Russia.}
\affiliation{ CERN, Geneva, Switzerland}
\author{V.~Ivanchenko} 
\altaffiliation{On leave  from Ecoanalitica, Moscow State University,
Moscow, Russia}
\affiliation{ CERN, Geneva, Switzerland}
\author{A.~Kayis-Topaksu}
\altaffiliation{Now at \c{C}ukurova University, Adana, Turkey.}
\affiliation{ CERN, Geneva, Switzerland}
\author{M.~Kirsanov}
\affiliation{Institute for Nuclear Research, Moscow, Russia}
\author{D.~Kolev} 
\affiliation{ Faculty of Physics, St. Kliment Ohridski University, Sofia,
  Bulgaria}
\author{A.~Krasnoperov} 
\affiliation{Joint Institute for Nuclear Research, JINR Dubna, Russia}
\author{J. Mart\'{i}n--Albo}
\affiliation{Instituto de F\'{i}sica Corpuscular, IFIC, CSIC and Universidad de Valencia, Spain}
\author{C.~Meurer}
\affiliation{Institut f\"{u}r Physik, Forschungszentrum Karlsruhe, Germany}
\noaffiliation{}
\author{M.~Mezzetto$^{(a)}$}
\affiliation{Sezione INFN$^{(a)}$ and Universit\'a degli Studi$^{(b)}$, 
Padova, Italy}
\author{G.~B.~Mills}
\affiliation{Los Alamos National Laboratory, Los Alamos, USA}
\author{M.C.~Morone}
\altaffiliation{Now at University of Rome Tor Vergata, Italy.}   
\affiliation{Section de Physique, Universit\'{e} de Gen\`{e}ve, Switzerland}
\author{P.~Novella} 
\affiliation{Instituto de F\'{i}sica Corpuscular, IFIC, CSIC and Universidad de Valencia, Spain}
\author{D.~Orestano}
\affiliation{Sezione INFN$^{(c)}$ and Universit\'a$^{(d)}$  Roma Tre, 
Roma, Italy}
\author{V.~Palladino}
\affiliation{Universit\`{a} ``Federico II'' e Sezione INFN, Napoli, Italy}
\author{J.~Panman}
\thanks{Corresponding author (J.~Panman).~E-mail: 
jaap.panman@cern.ch}
\affiliation{ CERN, Geneva, Switzerland}
 \author{I.~Papadopoulos}  
\affiliation{ CERN, Geneva, Switzerland}
\author{F.~Pastore} 
\affiliation{Sezione INFN$^{(c)}$ and Universit\'a$^{(d)}$  Roma Tre, 
Roma, Italy}
\author{S.~Piperov}
\affiliation{ Institute for Nuclear Research and Nuclear Energy,
Academy of Sciences, Sofia, Bulgaria}
\author{N.~Polukhina}
\affiliation{P. N. Lebedev Institute of Physics (FIAN), Russian Academy of
Sciences, Moscow, Russia}
\author{B.~Popov} 
\altaffiliation{Also supported by LPNHE, Paris, France.}
\affiliation{Joint Institute for Nuclear Research, JINR Dubna, Russia}
\author{G.~Prior}  
\altaffiliation{Now at CERN}
\affiliation{Section de Physique, Universit\'{e} de Gen\`{e}ve, Switzerland}
\author{E.~Radicioni}
\affiliation{Sezione INFN, Bari, Italy}
\author{D.~Schmitz}
\affiliation{Columbia University, New York, USA}
\author{R.~Schroeter}
\affiliation{Section de Physique, Universit\'{e} de Gen\`{e}ve, Switzerland}
\author{G.~Skoro}
\affiliation{ Dept. of Physics, University of Sheffield, UK}
\author{M.~Sorel}
\affiliation{Instituto de F\'{i}sica Corpuscular, IFIC, CSIC and Universidad de Valencia, Spain}
\author{E.~Tcherniaev}
\affiliation{ CERN, Geneva, Switzerland}
 \author{P.~Temnikov}
\affiliation{ Institute for Nuclear Research and Nuclear Energy,
Academy of Sciences, Sofia, Bulgaria}
\author{V.~Tereschenko}  
\affiliation{Joint Institute for Nuclear Research, JINR Dubna, Russia}
\author{A.~Tonazzo}
\affiliation{Sezione INFN$^{(c)}$ and Universit\'a$^{(d)}$  Roma Tre, 
Roma, Italy}
\author{L.~Tortora$^{(c)}$}
\affiliation{Sezione INFN$^{(c)}$ and Universit\'a$^{(d)}$  Roma Tre, 
Roma, Italy}
\author{R.~Tsenov}
\affiliation{ Faculty of Physics, St. Kliment Ohridski University, Sofia,
  Bulgaria}
\author{I.~Tsukerman}  
\altaffiliation{ITEP, Moscow, Russian Federation.}
\affiliation{ CERN, Geneva, Switzerland}
\author{G.~Vidal--Sitjes}  
\altaffiliation{Now at Imperial College, University of London, UK.}
\affiliation{Universit\`{a} degli Studi e Sezione INFN, Ferrara, Italy}
\author{C.~Wiebusch}   
\altaffiliation{Now at III Phys. Inst. B, RWTH Aachen, Germany.}
\affiliation{ CERN, Geneva, Switzerland}
\author{P.~Zucchelli}
\altaffiliation{Now at SpinX Technologies, Geneva, Switzerland; On leave
from INFN, Sezione di Ferrara, Italy.} 
\affiliation{ CERN, Geneva, Switzerland}

 \collaboration{\bf HARP Collaboration}
 \noaffiliation

 \maketitle

\clearpage

\section{Introduction}

The main objectives of the HARP experiment~\cite{harp-prop} are to
measure charged pion production yields to help 
design the proton driver of a future neutrino factory~\cite{ref:nufact}, 
to provide measurements to improve calculations of the atmospheric
neutrino flux~\cite{Battistoni,Stanev,Gaisser,Engel} 
and to measure particle yields as input for the flux
calculation of accelerator neutrino experiments~\cite{ref:physrep}, 
such as K2K~\cite{ref:k2k,ref:k2kfinal},
MiniBooNE~\cite{ref:miniboone} and SciBooNE~\cite{ref:sciboone}. 
In addition to these specific aims, the data provided by HARP are
valuable for validating hadron production models used in simulation
programs. 
The HARP experiment has taken data with beams of pions and protons
with momenta from 1.5~\GeVc to 15~\GeVc hitting targets made of a large
range of materials.
To provide a large angular and momentum coverage of the produced charged
particles the experiment comprises two spectrometers, a forward
spectrometer built around a dipole magnet and a large-angle spectrometer
constructed in a solenoidal magnet.

A large amount of data collected by the HARP experiment with thin 
(5\% of nuclear interaction length) and cryogenic targets have already 
been analyzed and published~\cite{ref:harp:alPaper,ref:harp:bePaper,ref:harp:carbonfw,ref:harp:o2n2,ref:harp:forward_pi,ref:harp:la,ref:harp:la:pions,ref:harp:tantalum,ref:harp:cacotin,ref:harp:bealpb},
covering all the physics subjects discussed above.
Cross-sections for some of these data-sets based on the same raw data
have been published by a different group~\cite{ref:cdp:la}. 
We disagree with the analysis of Ref.~\cite{ref:cdp:la} (see
Ref.~\cite{ref:tpcmom} for differences in detector calibration).

In this paper {\em effective} measurements of the double-differential
production cross-section,   
$
{{\mathrm{d}^2 \sigma^{\pi}}}/{{\mathrm{d}p\mathrm{d}\theta }}
$
for \pipm production valid for solid cylindrical long targets (one
interaction length) made of carbon, tantalum and lead are presented. 
The secondary pions were produced by beams of protons in a momentum
range from 5~\GeVc to  12~\GeVc impinging perpendicularly on one of the
flat surfaces of the target rods.
In earlier papers the measurements of the double-differential
production cross-sections were presented for data taken using protons
hitting thin beryllium, carbon, aluminium, copper, tin, tantalum and
lead thin targets of 5\% nuclear interaction length.  
Final results can be found in Ref.~\cite{ref:harp:la} for the proton
beam data.
The results presented in this paper provide a means to check the
ability of hadron production models to simulate pion production with
realistic extended targets by comparing short\footnote{In previous
papers we used the term ``thin'' instead.} and long target data taken
with the same experiment.
The choice of target materials covers an often used low $A$ material,
carbon, and examples of large $A$ targets like tantalum and lead.
We have limited ourselves to the momentum range of the incoming beam
where the most statistics is available (5~\GeVc--12~\GeVc).

Effective measurements of particle production using extended targets are
not unambiguously defined.
In this case the results have to be understood as follows.
The absorption and re-interactions of the beam proton in the target are
not corrected for, thus the measurements encompass  the full
behaviour of the beam particle in the one interaction length
(\intlen) of the target. 
This holds e.g. when the proton is (elastically) scattered or when
protons or pions are produced in a relatively small forward cone (up to
$\approx$200~\mrad).
As long as particles are produced within this forward cone they are
regarded as part of the beam.
Particles emitted outside this cone are regarded as ``products''.
The effect of absorption and interactions of these secondary pions and protons
emitted outside this cone are corrected for as losses of the produced pions or
as background in the case of production of tertiaries.
Thus, the analysis procedure aims to correct for these effects such that the
effective target is transparent for the secondary ``product'' pions and
one \intlen long for the ``beam particles''.
Some ambiguities still persist, and for the precise, quantitative
interpretation of the results it is important to take into account the
consequences of this procedure.  
One should note that the diameter of the targets is 30.3~mm
both for the long and short targets.
The relevant target parameters are listed in Table~\ref{tab:targets}.
Nevertheless, we believe that this definition is the most useful one to
provide input for the simulation of the production targets of
super-beams and neutrino factories.
Another procedure could be envisaged if one would know the target
characteristics for a given installation exactly.
In that case measurements on a replica target could be made and the
particles emerging from the target could be registered without bothering
what happened inside the target.
Since these measurements are aiming at design studies for future
installations, the latter approach cannot be followed.

\begin{table}[bh] 
\caption{Parameters of the long targets. All targets have a diameter of 30.3~mm.} 
\label{tab:targets}
\begin{center}
\begin{tabular}{lrrr} \hline
 \bf{Target material} & C &  Ta & Pb \\ 
\hline 
 \bf{Length (cm)} & 38.01 & 11.14 &  17.05 \\
 \bf{Density (g/\cmcub)} & 1.72 & 16.67 &  11.34 \\
\end{tabular}
\end{center}
\end{table}

Data were taken in the T9 beam of the CERN PS. 
Contrary to the short-target data, no interaction trigger was
applied while taking data with long targets. 
All good beam triggers were recorded, regardless of the activity in the
downstream trigger detectors.
The collected statistics for the different nuclear targets are reported in
Table~\ref{tab:events-p}.  
The analysis proceeds by selecting tracks in the Time Projection
Chamber (TPC) in events with an incident proton.
Momentum and polar angle measurements and particle identification are
based on the measurements of track position and energy deposition in
the TPC.
An unfolding method is used to correct for experimental resolution,
efficiency and acceptance and to obtain the double-differential pion
production yields.
Otherwise, the analysis follows the same methods as used for the
determination of \pipm production by protons on short targets.  
These analysis methods are documented in Ref.~\cite{ref:harp:tantalum} with
improvements described in Ref.~\cite{ref:harp:la} and will be only
briefly outlined here. 

The long-target data were taken with relatively high beam
intensities (typically 5000 particles per spill). 
Under these circumstances a space charge due to ions built up
inside the drift volume of the TPC.
These charges were responsible for a time-dependent (``dynamic'')
distortion of the track images measured by the TPC.
These effects are more severe for the long-target data than for the
short-target data due to the high interaction rate.
A first set of results on pion production at large angles by protons 
using short-target data had been published
\cite{ref:harp:tantalum,ref:harp:cacotin, ref:harp:bealpb} based on the
analysis of the data limited to the beginning of each accelerator spill
for this reason. 
It was no longer necessary to limit the analysis of the short-target
data set to the events taken in the beginning of each spill after  
corrections had been developed~\cite{ref:dyndist}.  
These corrections were fully applied in the analysis of
Ref.~\cite{ref:harp:la}.
However, in the case of the long-target data the distortion effects are
larger, and the corrections could only be reliably applied to a part of
the 400~\ms spill, so that typically 30\%--50\% of the statistics are
available for the analysis reported in this paper.
More details will be given in Section~\ref{sec:distortions}.

\section{Experimental apparatus and data selection}
\label{sec:apparatus}
 
\begin{figure}[tb]
 \begin{center}
  \hspace{0mm} 
  \includegraphics[width=12cm,angle=0]{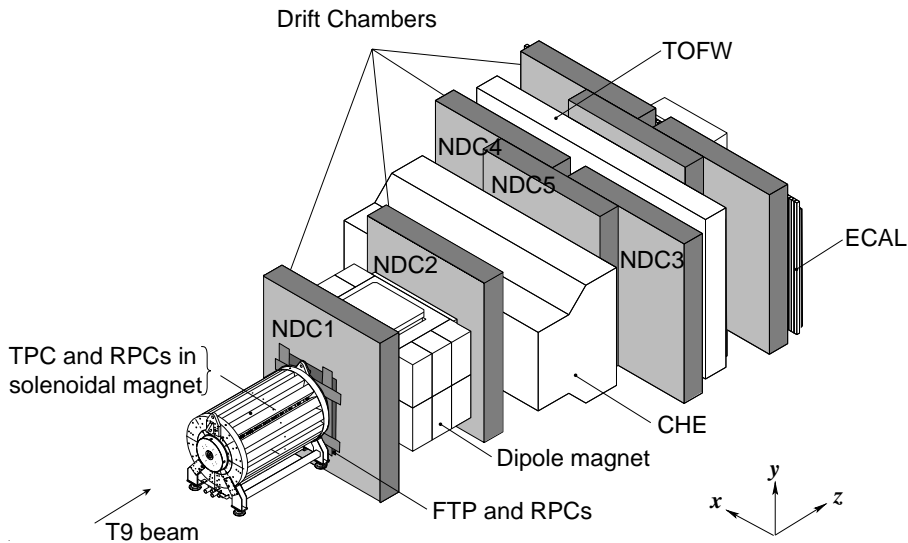}
 \end{center}
\caption{Schematic layout of the HARP detector. 
The convention for the coordinate system is shown in the lower-right
corner. 
The three most downstream (unlabelled) drift chamber modules are only partly
equipped with electronics and are not used for tracking. 
The detector covers a total length of 13.5 m along the beam axis 
and has a maximum width of 6.5 m perpendicular to the beam.
The beam muon identifier is visible as the most downstream detector
 (white box).
}
\label{fig:harp}
\end{figure}

The HARP detector is shown in Fig.~\ref{fig:harp} and is 
described in detail in Ref.~\cite{ref:harpTech}.
The forward spectrometer, mainly used in the analysis for the conventional
neutrino beams and atmospheric neutrino flux, comprises a dipole magnet,
large planar drift chambers 
(NDC)~\cite{NOMAD_NIM_DC}, a time-of-flight wall (TOFW) \cite{ref:tofPaper}, 
a threshold Cherenkov counter
(CHE) and an electromagnetic calorimeter (ECAL).
In the large-angle region a cylindrical TPC with a radius of 408~\mm
is  positioned inside a solenoidal magnet with a field of 0.7~\T. 
The TPC detector was designed to measure and identify tracks in the
angular region from 0.25~\rad to 2.5~\rad with respect to the beam axis.
The target is placed inside the inner field cage (IFC) of the TPC such that,
in addition to particles produced in the forward direction, 
backward-going tracks can be measured.
The TPC is used
for tracking, momentum determination and measurement of the
energy deposition \dedx for particle identification~\cite{ref:tpc:ieee}.
A set of resistive plate chambers (RPC) forms a barrel inside the solenoid 
around the TPC to measure the arrival time of the secondary
particles~\cite{ref:rpc}. 
Charged particle identification (PID) can be achieved by measuring the 
ionization per unit length in the gas (\dedx) as a function of the total
momentum of the particle. 
Additional PID can be performed through a time-of-flight measurement
with the RPCs; this method is used to provide an independent calibration
of the PID based on \dedx. 

In addition to the data taken with the solid targets of
5\% and 100\% \intlen
runs were also taken with an empty target holder to check backgrounds. 
Data taken with a liquid hydrogen target at 3~\GeVc, 5~\GeVc and
8~\GeVc incident-beam momentum together with cosmic-ray data were used 
to provide an absolute calibration of the efficiency, momentum scale and
resolution of the detector. 

\begin{figure}[tbp!]
\begin{center}
 \includegraphics[width=0.30\textwidth,angle=0]{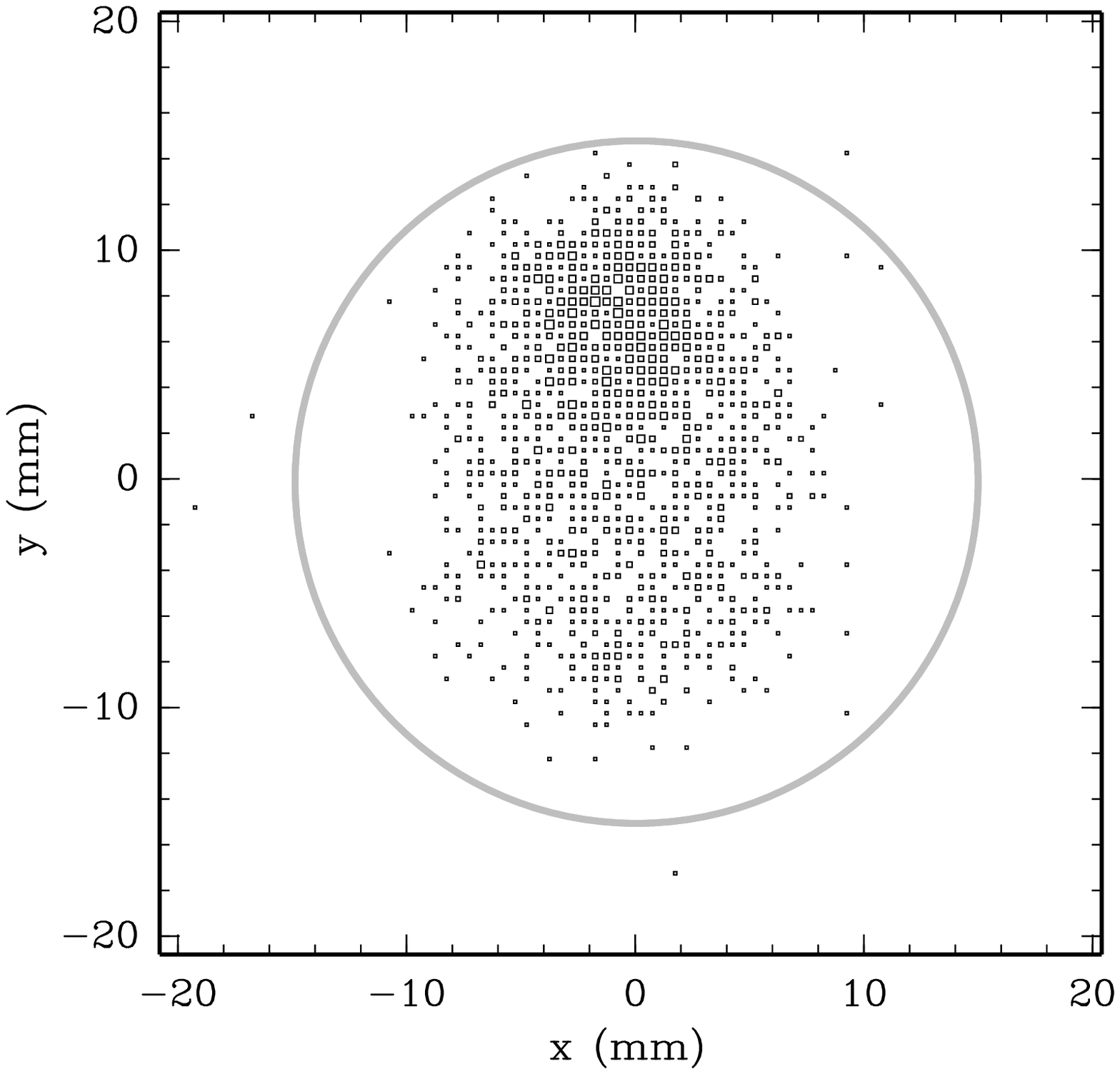}
~
 \includegraphics[width=0.30\textwidth,angle=0]{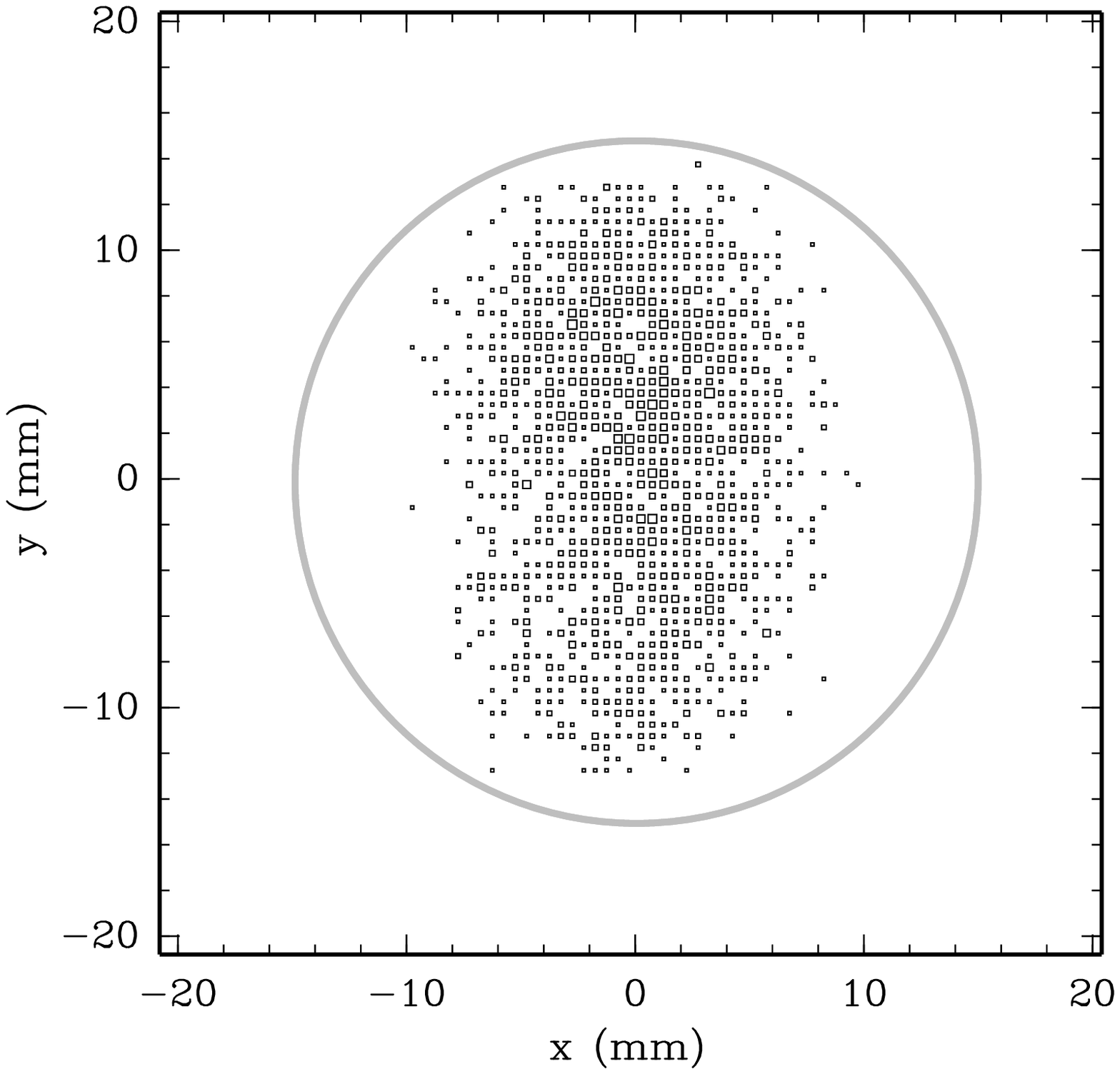}
~
 \includegraphics[width=0.30\textwidth,angle=0]{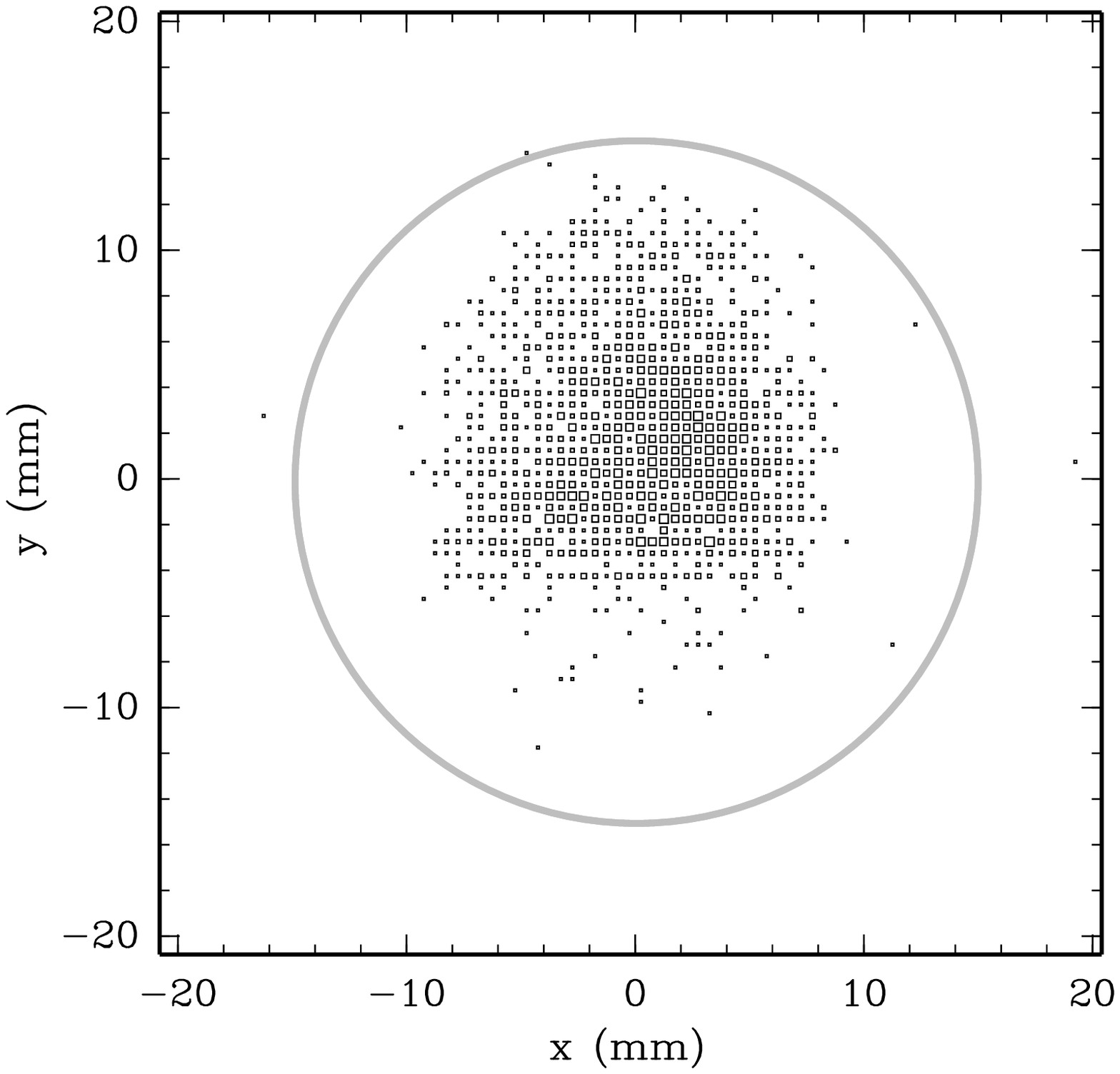}
\end{center}
\caption{
The position distributions of the incident-beam particles
 in the $x$--$y$ plane in the 5~\GeVc beam (left panel), 8~\GeVc beam
 (middle panel) and 12~\GeVc beam  (right panel).
The outline of the target is shown as a shaded circle.
}
\label{fig:mwpc}
\end{figure}

The momentum of the T9 beam is known with a precision of
the order of 1\%~\cite{ref:t9}. 
The beam profiles projected to the target position as measured by the
beam MWPCs are shown for the three different beam momenta in
Fig.~\ref{fig:mwpc}. 
The absolute normalization of the number of incident protons is
performed by just counting the incoming beam particles with the same
trigger as used for the analysis of the secondary particles. 
This is possible since no selection on the interaction is performed in
the trigger for the data sets used in the present analysis.
The dimensions and mass of the solid targets were carefully measured.
The purity of the target materials exceeded 99.9\%.
The uncertainties in thickness and density of the targets are well below
1\%. 

\begin{table}[tbp!] 
\caption{Total number of events and tracks used in the various nuclear 
  100\%~$\lambda_{\mathrm{I}}$ target data sets taken with the proton beam and the number of
  protons on target as calculated from the pre-scaled incident-proton triggers.} 
\label{tab:events-p}
\begin{center}
\begin{tabular}{lrrrr} \hline
 \bf{Data set}           &         &\bf{C}&\bf{Ta}&\bf{Pb} \\ \hline
 \bf{Total DAQ events}     
                         & 5~\GeVc & 768202 &  875079 & 1034775    \\
                         & 8~\GeVc & 708144 & 1312215 &  871706    \\
                         &12~\GeVc & 236362 & 1375923 & 1085263    \\
 \bf{Accepted beam protons   }
                         & 5~\GeVc & 253888 &  296512 &  341120    \\
                         & 8~\GeVc & 339968 &  729728 &  490048    \\
                         &12~\GeVc & 175232 & 1046464 &  822016    \\
 \bf{Fraction of triggers used}
                         & 5~\GeVc &   0.32 &    0.31 &    0.33    \\
                         & 8~\GeVc &   0.49 &    0.29 &    0.30    \\
                         &12~\GeVc &   0.55 &    0.27 &    0.28    \\
 \bf{$\bfpim$ selected with PID} 
                         & 5~\GeVc &  10550 &    7020 &    8185    \\
                         & 8~\GeVc &  31110 &   32590 &   20336    \\
                         &12~\GeVc &  22269 &   65346 &   44498    \\
 \bf{$\bfpip$ selected with PID} 
                         & 5~\GeVc &  15496 &    9008 &    9851    \\
                         & 8~\GeVc &  41252 &   37747 &   23097    \\
                         &12~\GeVc &  27362 &   75389 &   49725    \\
\end{tabular}
\end{center}
\end{table}

Beam instrumentation provides identification of the incoming
particle, the determination of the time when it hits the target, 
and the impact point and direction of the beam particle
on the target. 
It is based on a set of four multi-wire proportional chambers (MWPC)
to measure the position and direction of the incoming beam particles 
and time-of-flight (TOF) detectors and two
N$_2$-filled Cherenkov counters to identify incoming particles.  
Several trigger detectors are installed to select events with an
interaction and to define the normalization.
The beam of positive particles used for this measurement consists mainly of
positrons, pions and protons, with small components of kaons and
deuterons and heavier ions.
Its composition depends on the selected beam momentum.
The proton fraction in the incoming positive particle beam varies from
35\% at 3 GeV/c to 92\% at 12 GeV/c.  
At the first stage of the analysis a favoured beam particle type is selected
using the beam time-of-flight system and the two Cherenkov
counters.
A value of the pulse height consistent with the absence of a signal in both beam
Cherenkov detectors distinguishes protons (and kaons) from electrons and pions.
We also ask for time measurements to be present which are needed for calculating 
the arrival time of the beam proton at the target. 
The beam TOF system is used to reject ions, such as deuterons, and to
separate protons from pions at low momenta.
In most beam settings the nitrogen pressure in the beam Cherenkov
counters was too low for kaons to be above the threshold.
Kaons are thus counted in the proton sample.
However, the fraction of kaons has been measured in the 12.9~\GeVc beam
using a dedicated combination of the pressure setting of the two
Cherenkov counters and are found to contribute less than 0.5\%, and
hence are negligible in the proton beam sample.
Electrons radiate in the Cherenkov counters and would be counted as
pions. 
More details on the beam particle selection can be found in 
Refs.~\cite{ref:harpTech} and
\cite{ref:harp:alPaper,ref:harp:bePaper,ref:harp:carbonfw}. 

The length of the accelerator spill is 400~ms with a typical intensity
of 5000 beam particles per spill.
The average number of events recorded by the data acquisition ranges
up to 450 per spill for the data taken with long targets.
The analysis proceeds by first selecting a beam proton  not accompanied
by other beam tracks.  
After the event selection the sample of tracks to be used for analysis
is defined.
Table~\ref{tab:events-p} shows the number of
events and the number of \pipm selected in the analysis.
The large difference between the first and second set of rows (``Total
DAQ events'' and ``Accepted beam protons'') is due to the relatively large
fraction of pions in the beam and to the number of calibration triggers
used. 
The entry ``Fraction of triggers used'' shows the part of the data for
which distortions in the TPC could be calibrated reliably as explained
below.

\subsection{Distortions in the TPC}
\label{sec:distortions}

\begin{figure}[tbp!]
 \begin{center}
  \includegraphics[width=0.43\textwidth,angle=0]{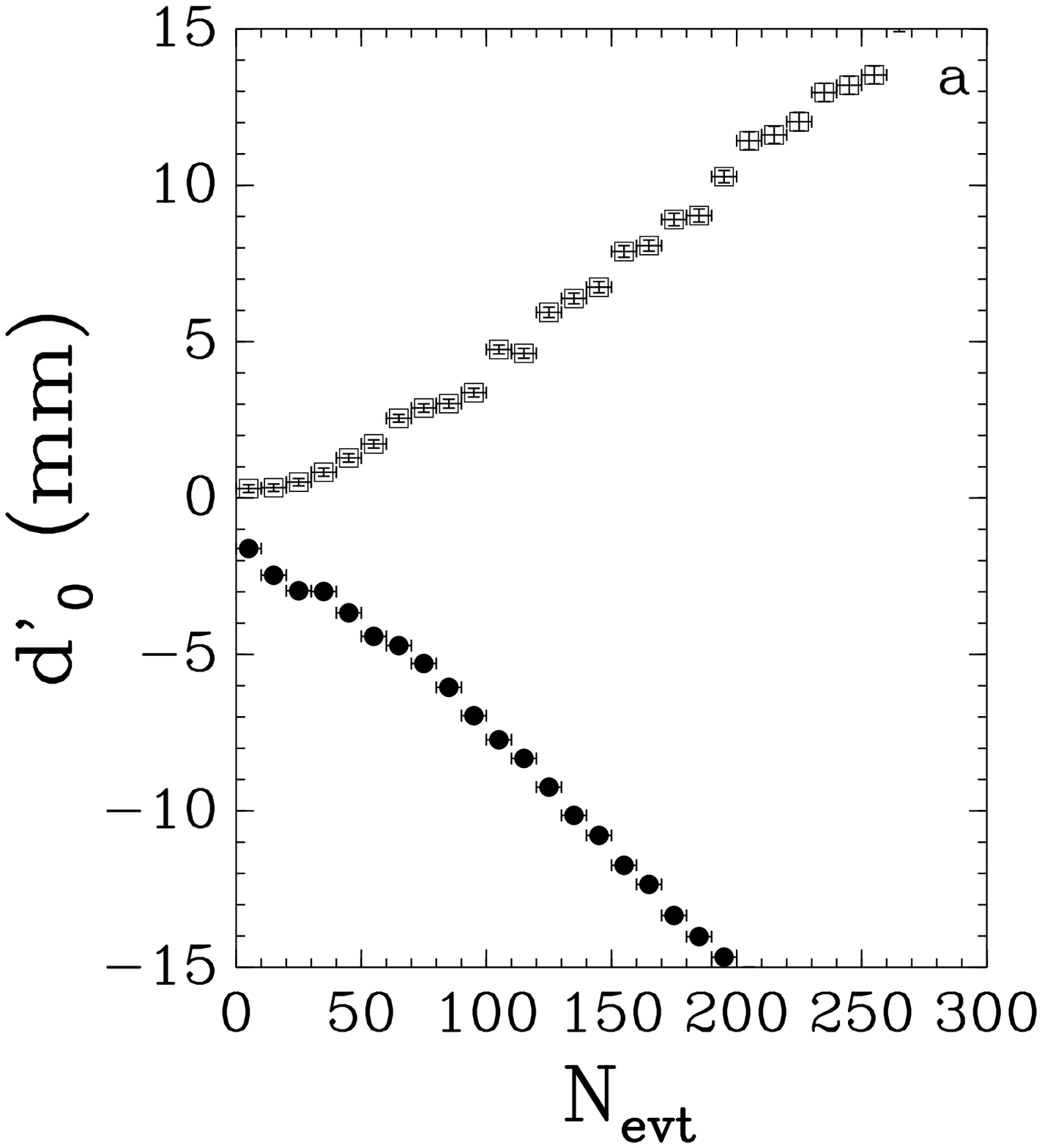}
  ~
  \includegraphics[width=0.43\textwidth,angle=0]{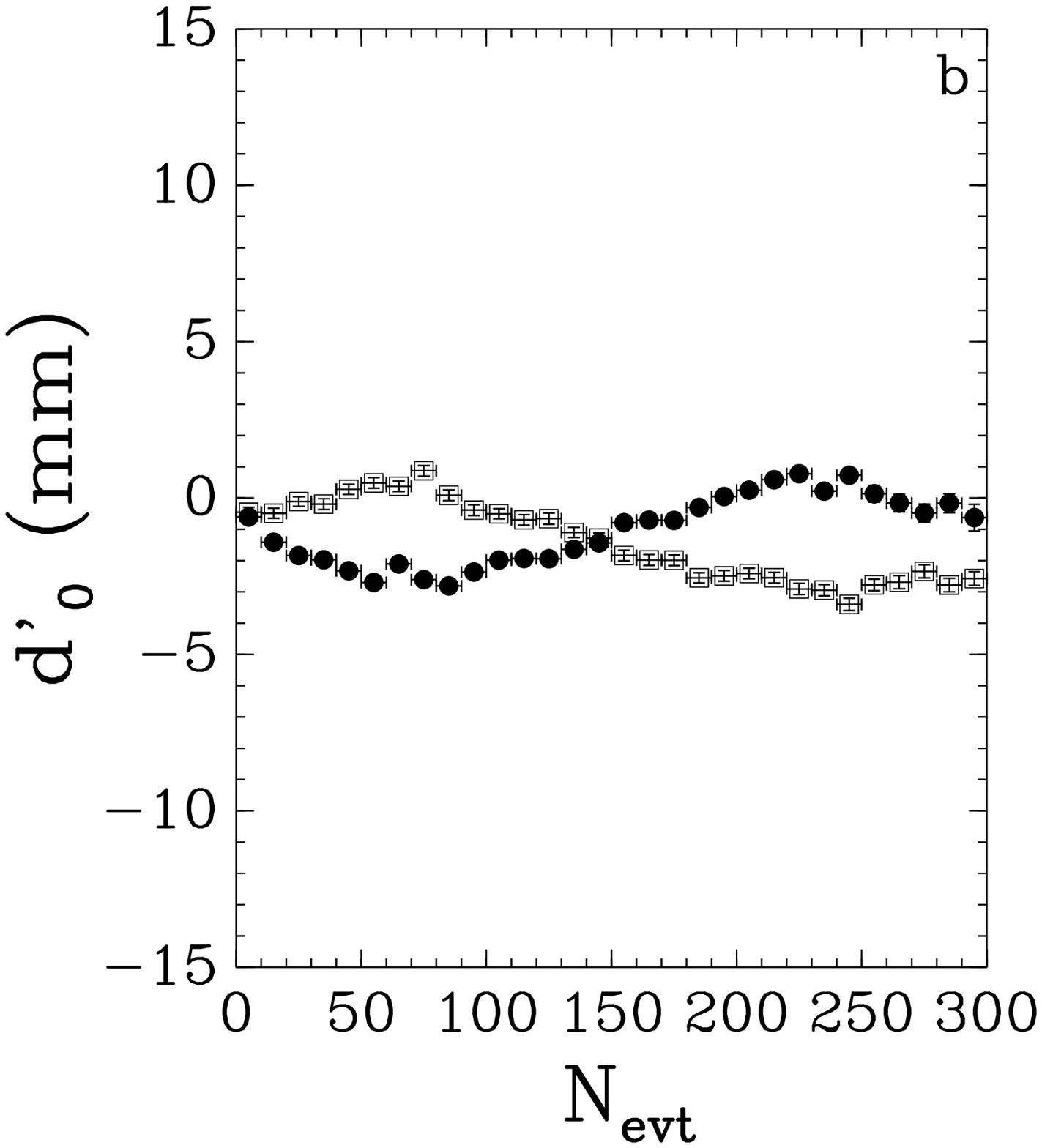}
  \includegraphics[width=0.43\textwidth,angle=0]{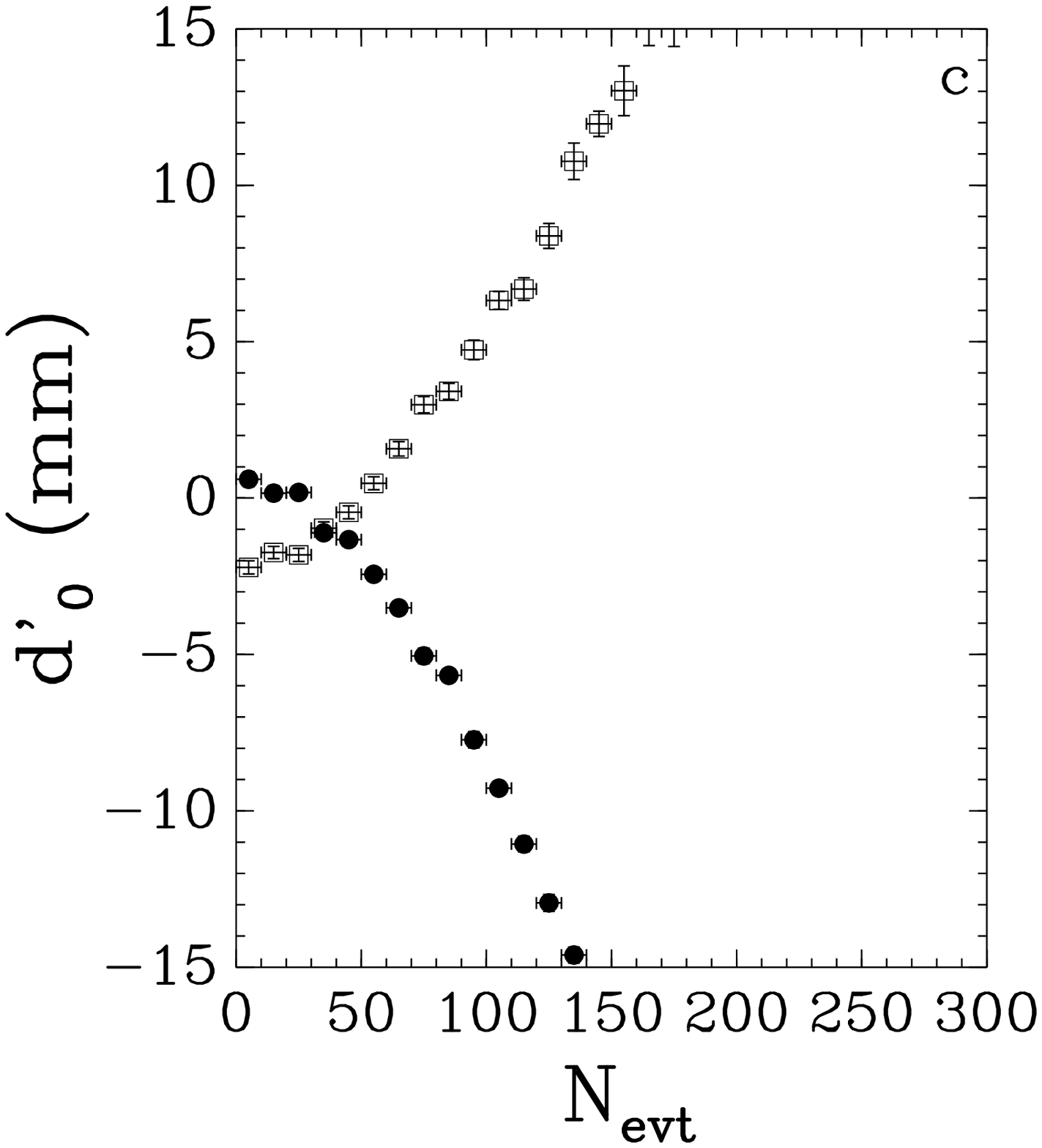}
  ~
  \includegraphics[width=0.43\textwidth,angle=0]{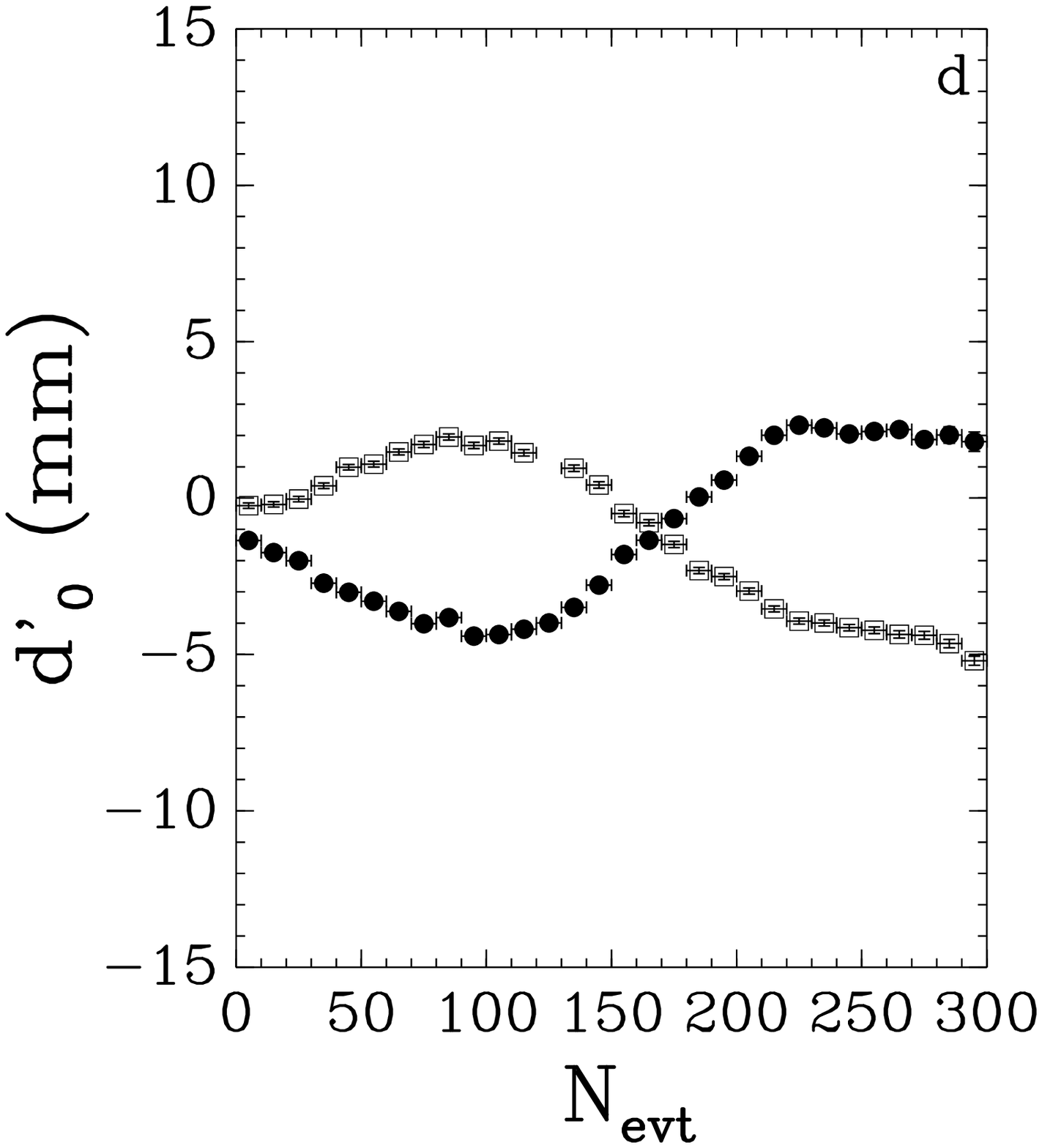}
 \end{center}
\caption{
Average \dzeroprime (filled circles for reconstructed positive tracks,
 open squares for
reconstructed negative tracks) as a function of event number in spill for 12~\GeVc
 C data in the top panels and  for 12~\GeVc  Pb data in the bottom
 panels. 
The left panels (a: p--C; c: p--Pb) show uncorrected data and in the 
right panels (b: p--C; d: p--Pb) dynamic distortion
 corrections have been applied.
After the ``default'' correction for the static distortions (equal for
 each setting) a small residual effect at the beginning of the spill is
 visible at $\evtspill=0$ (left panel).
This is due to the fact that the inner and outer field cages are powered
 with individual HV supplies.
A setting-by-setting correction compatible with the reproducibility of
 the power supplies is applied for the data of the right panel together
 with the dynamic distortion correction.
The value of $\langle \dzeroprime \rangle$ at $N_{evt}=0$ in the right panel has a small negative
value as expected from the fact that the energy-loss is not described in the 
track-model used in the fit. The difference observed in the results for the
two charges shows that the model can correct the
 distortion in the  12~\GeVc C data to within about $\pm2$~mm.
In the bottom left panel (12~\GeVc Pb data) one observes a steep (almost
 linear) behaviour from  $\evtspill \approx 50$~events, while the data
 early in the spill are practically not affected. 
In this case the distortion can be corrected up to about $\pm$3~mm, as
 measured by \dzeroprime.
However, beyond $\evtspill \approx 150$~events other benchmarks show
 that momentum biases are not kept under control.
}
\label{fig:dzeroprime}
\end{figure}
\begin{figure}[tbp!]
 \begin{center}
  \includegraphics[width=0.4\textwidth]{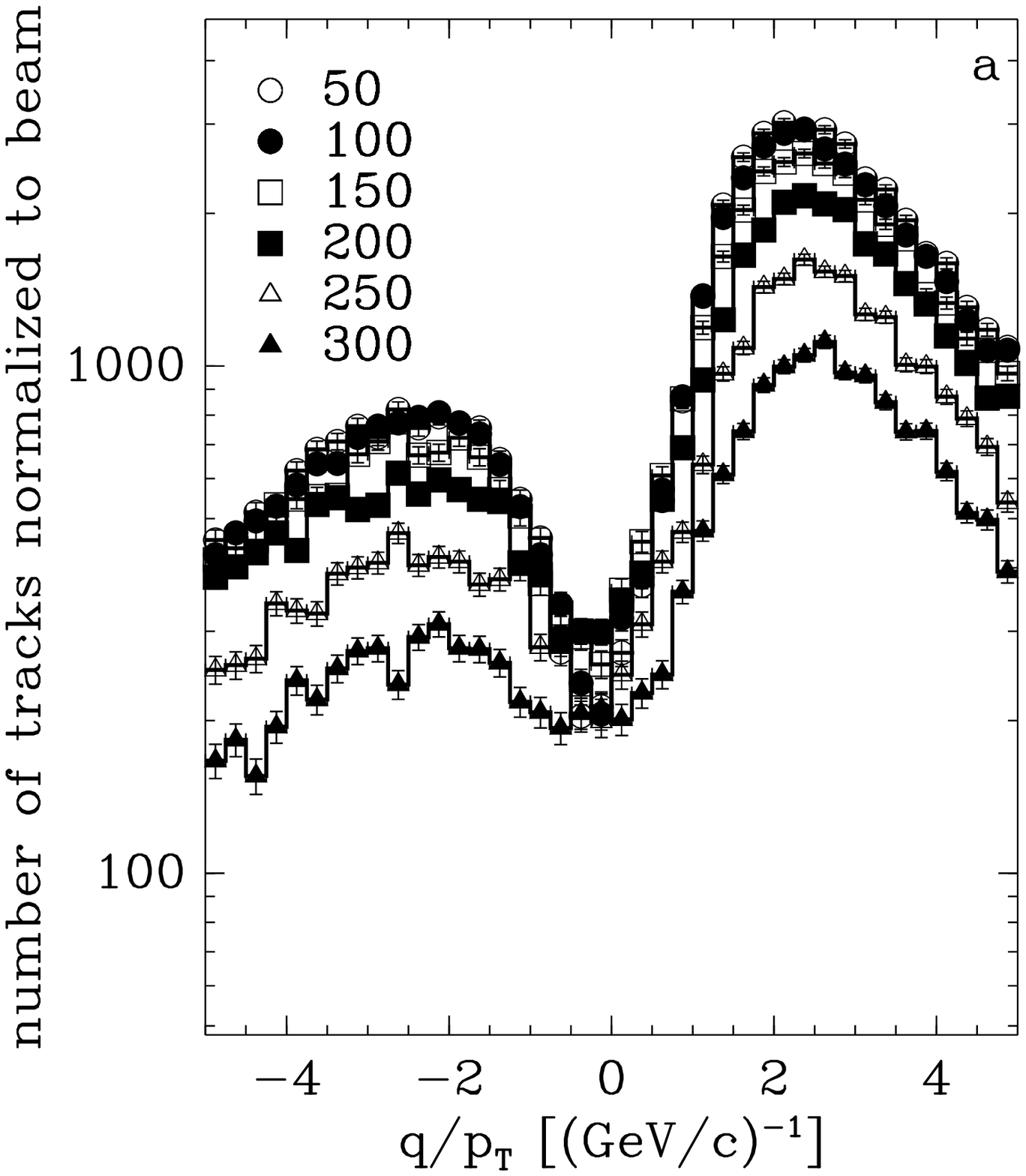}
  ~
  \includegraphics[width=0.4\textwidth]{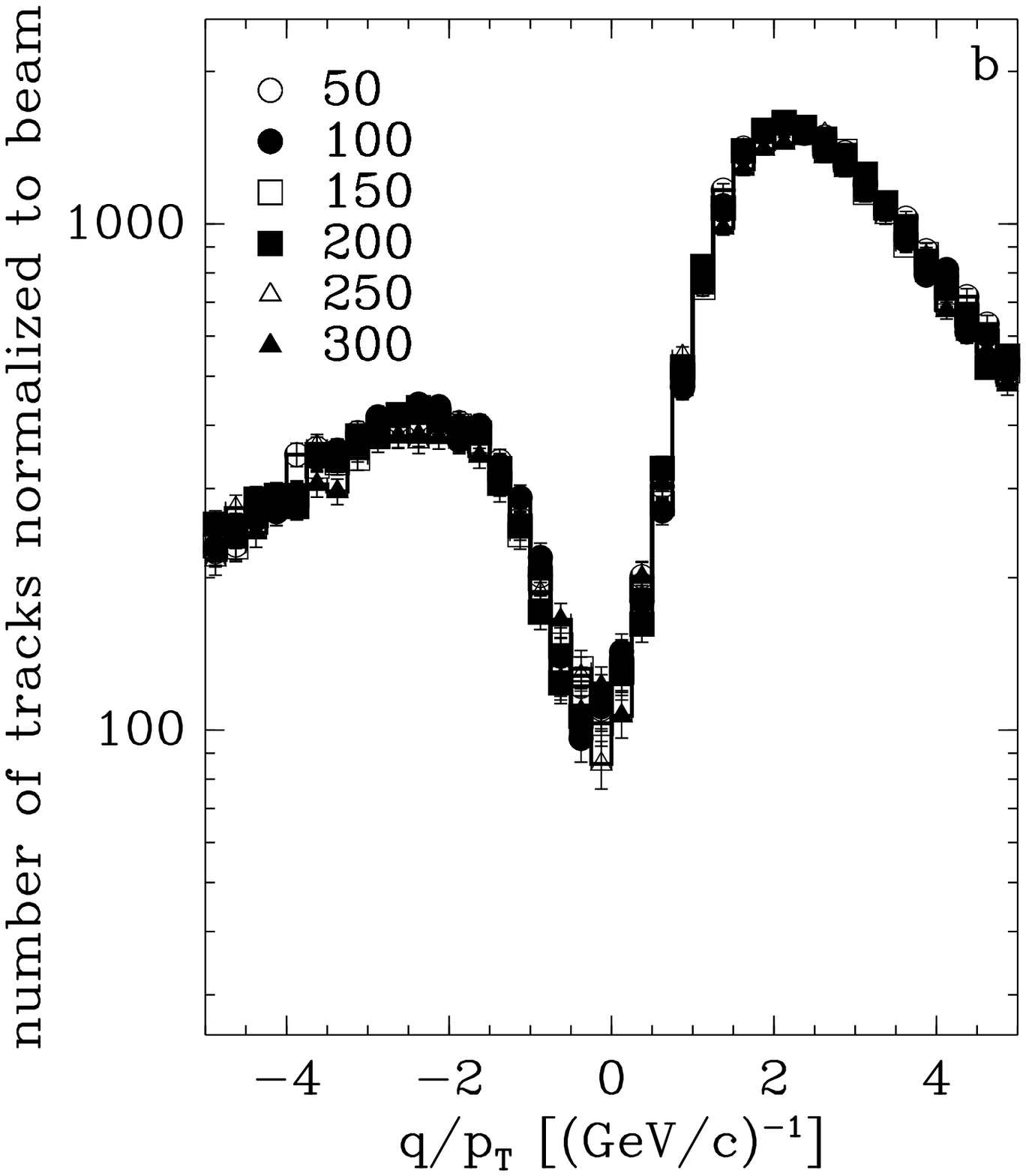}
  \includegraphics[width=0.4\textwidth]{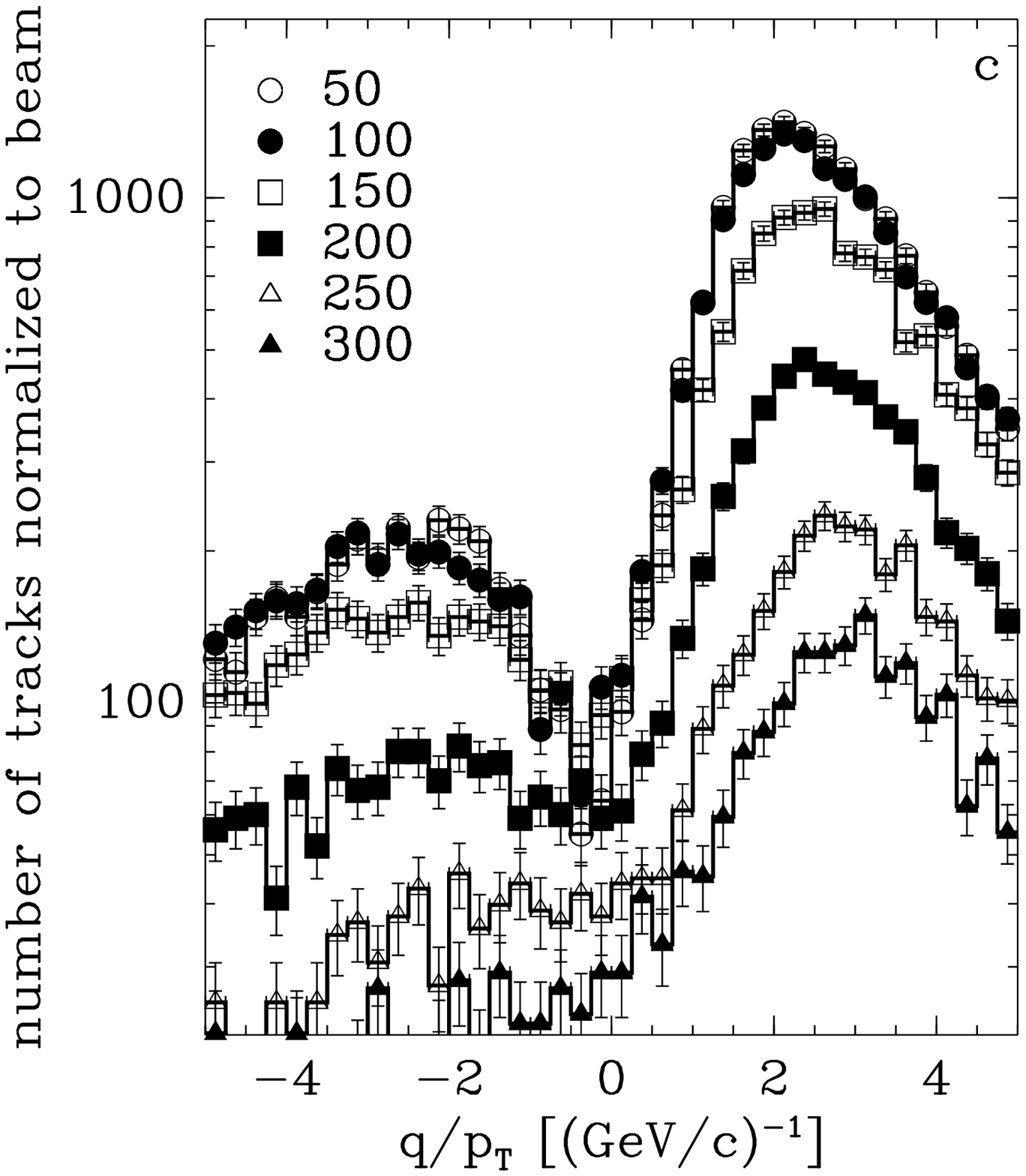}
  ~
  \includegraphics[width=0.4\textwidth]{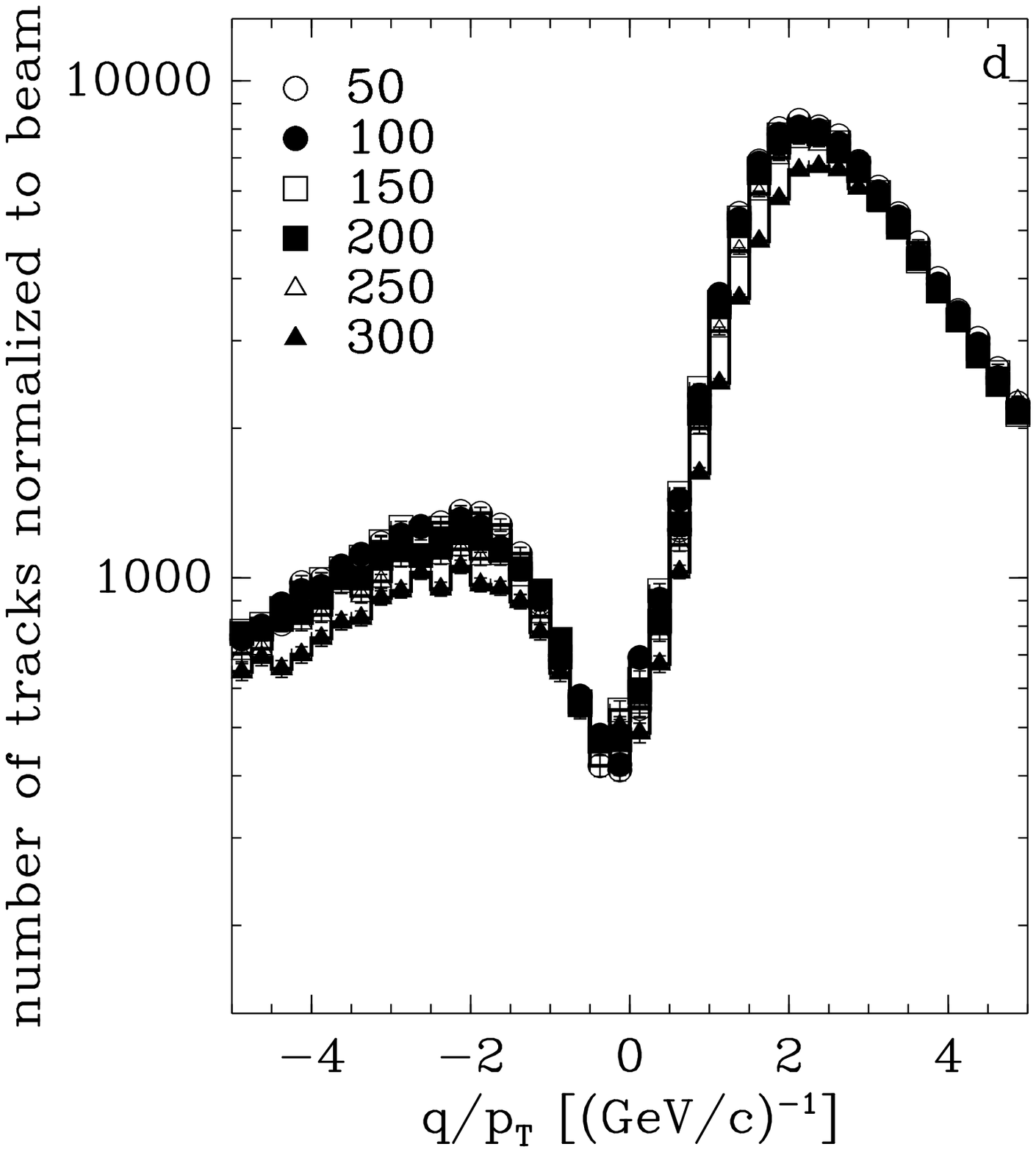}
 \end{center}
\caption{
Distribution in $q/\pt$ for the 12~\GeVc C data (top panels) and  for
 the 12~\GeVc Pb data (bottom panels), where $q$ is the measured charge
 of the particle and \pt its transverse momentum.
The six curves show
 six regions in event number in spill (each in groups of 50 events in
 spill). 
Groups are labelled with the last event number accepted in the group,
 e.g. ``50'' stands for the group with event number from 1 to 50.
The six groups are normalized to the same number of incoming beam
 particles, taking the first group as reference.
 Left panels (a: p--C; c: p--Pb): without dynamic distortion  corrections; 
right panels (b: p--C; d: p--Pb): with  dynamic distortion corrections.  
In the top left panel only the first three groups of 50 events in spill are
 equivalent, while in the top right panel the groups are nearly
 indistinguishable. 
In the bottom left panel (Pb) a very large difference between the groups of 50
 events in spill are observed (only the first two groups are
 equivalent).
The very large loss of tracks at high event numbers is due to the fact
 that particles no longer point back to the incident-beam particle
 and are rejected by this selection criterion.
 In the (bottom) right panel the first five groups are nearly indistinguishable.
}
\label{fig:invpt}
\end{figure}%
\begin{figure}[tbp!]
 \begin{center}
  \includegraphics[width=0.45\textwidth,angle=0]{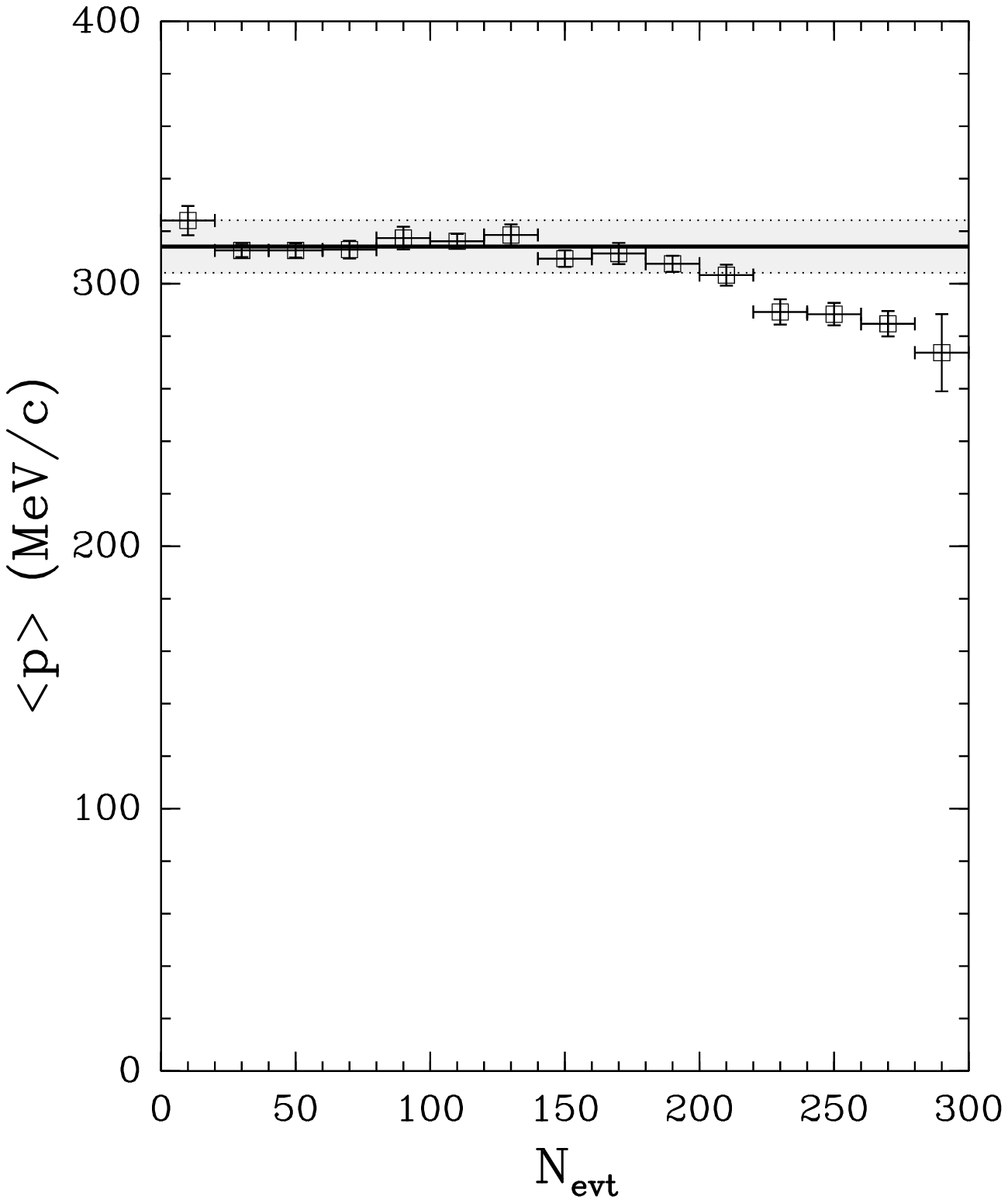}
 \end{center}
\caption{
The average momentum as a function of \evtspill observed for protons selected
 within a high \dedx region for the p--Pb data at 12~\GeVc.
The angle of the particles is restricted in a range with $\sin \theta
 \approx 0.9$. 
The data are corrected for dynamic distortions and stay stable within
 3\% up to about 150 \evtspill.
The corrections become too large to be corrected reliably beyond 150
 events (30\% of the spill).
The shaded band shows a  $\pm3$\% variation.
}
\label{fig:pb:momentum}
\end{figure}

Besides the usual need for calibration of the detector, a number of
hardware shortfalls, discovered mainly after the end of data-taking,
had to be overcome to use the TPC data reliably in the analysis.
The TPC is affected by a relatively large number of dead or noisy 
pads and static and dynamic distortions of the reconstructed trajectories.
The corrections  applied to the measurements are identical to the ones
used for our analysis of short-target data taken with incoming protons and a
description can be found in 
Ref.~\cite{ref:harp:la,ref:dyndist}~\footnote{
The results of Ref.~\cite{ref:cdp:la} have been analysed by another
group with a treatment of the dynamic distortions~\cite{ref:tpc:dydak}
with which we disagree.}.
 
The size of the corrections for dynamic distortions grows as function of
the time within each accelerator spill and for each data set the part of
the spill which can be reliably corrected is checked.
The fraction of events usable for the analysis is typically 30\%--50\%, but
varies for the different data sets (see Table~\ref{tab:events-p}). 
The presence of a possible residual momentum bias in the TPC measurement 
due to the dynamic distortions was investigated using a large set of
calibration methods.  
A dedicated paper~\cite{ref:tpcmom} addresses this point and shows that
our estimation of the knowledge of the absolute momentum scale is better
than $3.5\%$. 
Due to the large event rate in the data taken with the long targets, the
dynamic distortions are more severe than for the short-target data.
It is possible to correct only a relatively small fraction
of the data reliably, i.e. the first $\approx$30\% of the Ta and Pb data
and the first half of the C data.  
The track impact distance\footnote{The \dzeroprime sign indicates
if the helix encircles the beam particle trajectory (positive sign) 
or not (negative sign).}
with respect to the trajectory of the incoming beam particle,  
\dzeroprime,
is a very sensitive probe to measure the
distortion strength~\cite{ref:dyndist}.
As an example of the effect of the distortions and the quality of the
corrections the distributions of \dzeroprime in the 12~\GeVc C and Pb
data are shown in Fig.~\ref{fig:dzeroprime}, as examples of a better and
worse situation, respectively. 

Figure~\ref{fig:invpt} shows as a further check the distribution
in $q/\pt$, where $q$ is the measured charge of the particle and \pt its
transverse momentum.
Tracks have been divided into six groups depending on the number
\evtspill of their event in the spill. 
The six groups correspond to $50 n < \evtspill \le 50 (n+1)$ (for $n$
ranging from zero to five), thus displaying the distribution of tracks
from early events in the spill separately from tracks in events measured
later in the spill.
To make the absolute normalization meaningful,
the distributions have been scaled to an equal number of incident-beam
particles compared to the first group of 50 events. 
In the left panel, no dynamic distortion corrections have been applied
and a clear difference of the distributions is visible especially for Pb. 
One should note that the momentum measurement as well as the efficiency
is modified by the distortions.  
The right panel shows the distributions after the corrections.
The distributions are more equal, although especially for the Pb data
still important differences are observed at the end of the spill.
To understand the asymmetry of positively and negatively charged tracks,
one should keep in mind that no particle identification was performed.
Thus both protons and pions contribute to the positives while the \pim's
are the only component of the negative particles.
Since the statistical errors in this analysis are smaller than the
systematic errors a conservative approach was chosen, and only the first
part of the spill where the dynamic distortion corrections could be
applied was used.
For the case of the 12~\GeVc C and Pb data shown in the examples, only the
first 250 and 150 events of the spill were used, respectively.
For the 12~\GeVc Pb data Figs.~\ref{fig:dzeroprime} and
\ref{fig:invpt} show very large effects in the uncorrected results.  
These are among the data sets where the largest dynamic distortions are
observed.
This is explained by the relatively high interaction rate using the long
target and high multiplicity in p--Pb interactions without sufficient
reduction of the beam intensity during data taking.

It cannot be taken for granted that no residual momentum bias is
incurred when corrections have to be applied corresponding to
$\langle \dzeroprime \rangle$ values larger than 15~\mm.  
Therefore, the results of the corrections have to be checked using a
benchmark which ensures good momentum reconstruction.
A direct test of the effect of the  correction on the measurement
of momentum is shown in Fig.~\ref{fig:pb:momentum} for the worst case
(12~\GeVc Pb).
A sample of relatively high momentum protons was selected setting
a fixed window with relatively high values of \dedx in the TPC.
This high \dedx window ensured that
the particles were correctly identified as protons and simultaneously
selected a momentum band. 
It is visible that the momentum measurement starts to show deviations
beyond event number 150 inside spills.  
Whereas the other estimators do not reveal a problem, the momentum
estimator using \dedx reveals a deviation beyond the accepted systematic
error of $\pm3$\%. 
This explains the relatively small part of the full spill used in the
analysis. 

\section{Data analysis}
\label{sec:analysis}

Only a short outline of the data analysis procedure is presented here,
for further details see Refs. \cite{ref:harp:tantalum,ref:harp:la}. 
The double-differential yieldper target nucleon for the production of a particle of 
type $\alpha$ can be expressed in the laboratory system as:

\begin{equation}
{\frac{{\mathrm{d}^2 \sigma_{\alpha}}}{{\mathrm{d}p_i \mathrm{d}\theta_j }}} =
\frac{1}{{N_{\mathrm{pot}} }}\frac{A}{{N_A \rho t}}
 \sum_{i',j',\alpha'} M_{ij\alpha i'j' \alpha'}^{-1} \cdot
{N_{i'j'}^{\alpha'} } 
\ ,
\label{eq:cross}
\end{equation}

where $\frac{{\mathrm{d}^2 \sigma_{\alpha}}}{{\mathrm{d}p_i \mathrm{d}\theta_j }}$
is expressed in bins of true momentum ($p_i$), angle ($\theta_j$) and
particle type ($\alpha$).

The `raw yield' $N_{i'j'}^{\alpha'}$ 
is the number of particles of observed type $\alpha'$ in bins of reconstructed
momentum ($p_{i'}$) and  angle ($\theta_{j'}$). 
These particles must satisfy the event, track and PID 
selection criteria.
Although, owing to the stringent PID selection,  the background from
misidentified protons in the pion sample is small, the pion and proton
raw yields ($N_{i'j'}^{\alpha'}$, for 
$\alpha'=\pim, \pip, \mathrm{p}$) have been measured simultaneously. 
It is thus possible to correct for the small remaining proton
background in the pion data without prior assumptions concerning the
proton production cross-section.

The matrix $ M_{ij\alpha i'j' \alpha'}^{-1}$ 
corrects for the  efficiency and the resolution of the detector. 
It unfolds the true variables $ij\alpha$ from the reconstructed
variables $i'j'\alpha'$  with a Bayesian technique~\cite{dagostini} 
and corrects  
the observed number of particles to take into account effects such as 
trigger efficiency, reconstruction efficiency, acceptance, absorption,
pion decay, tertiary production, 
PID efficiency, PID misidentification and electron background. 
The method used to correct for the various effects is  described in
more detail in Ref.~\cite{ref:harp:tantalum}.

In order to predict the population of the migration matrix element 
$M_{ij\alpha i'j'\alpha'}$, the resolution, efficiency
and acceptance of the detector are obtained from the Monte Carlo.
This is accurate provided the Monte Carlo
simulation describes these quantities correctly. 
Where some deviations
from the control samples measured from the data are found, 
the data are used to introduce (small) {\em ad hoc} corrections to the
Monte Carlo. 
Using the unfolding approach, possible known biases in the measurements
are taken into account automatically as long as they are described by
the Monte Carlo.
In the experiment simulation, which is based on the GEANT4
toolkit~\cite{ref:geant4}, the materials in the beam-line and the 
detector are accurately described as well as
the relevant features of the detector response and 
the digitization process.
The time-dependent properties of the TPC, such as pulse-height
calibration per channel and the presence of dead channels were
reproduced for each individual data set by running a dedicated
set of high-statistics simulations corresponding to each data set.
In general, the Monte Carlo simulation compares well with the data, as
shown in Ref.~\cite{ref:harp:tantalum}. 
For all important issues physical benchmarks have been used to validate
the analysis.
The absolute efficiency and the measurement of the angle and momentum
was determined with elastic scattering. 
The momentum and angular resolution was determined exploiting the two
halves of cosmic-ray tracks crossing the TPC volume.
The efficiency of the particle identification was checked using two
independent detector systems.
Only the latter needs a small {\em ad hoc} correction compared to the
simulation.  

The factor  $\frac{A}{{N_A \rho t}}$ in Eq.~\ref{eq:cross}
is the inverse of the number of target nuclei per unit area
($A$ is the atomic mass,
$N_A$ is the Avogadro number, $\rho$ and $t$ are the target density
and thickness).
As explained above, we do not make a correction for the attenuation
of the beam in the target, so that the yields are valid for a
$\intlen=100\%$ target. 
The result is normalized to the number of incident protons on the target
$N_{\mathrm{pot}}$. 
The absolute normalization of the result is calculated in the first
instance relative to the number of incident-beam particles accepted by
the selection. 
After unfolding, the factor  $\frac{A}{{N_A \rho t}}$ is applied.
The beam normalization 
has uncertainties smaller than 2\% for all beam momentum settings.

The background due to interactions of the primary
pions outside the target (called `Empty target background') is
measured using data taken without the target mounted in the target
holder.
Owing to the selection criteria which only accept events from the
target region and the good definition of the interaction point this
background is negligible ($< 10^{-5}$).
To subtract backgrounds generated by 
\piz's produced in hadronic interactions of the incident-beam particle,
the assumption is made that the \piz spectrum is similar to the
spectrum of charged pions.
In an iterative procedure the \pim production spectra are used for the
subtraction, while the difference between \pip and \pim production is
used to estimate the systematic error. 
In the region below 125~\MeVc a large fraction of the electrons can be
unambiguously identified.
These tracks are used as a relative normalization between data and
simulation. 
An additional systematic error of 10\% is assigned to the
normalization of the \piz subtraction using the identified electrons
and positrons.
The absorption and decay of particles is simulated by the Monte Carlo.
The generated single particle can re-interact and produce background
particles by hadronic or electromagnetic processes.
These processes are simulated and  additional
particles reconstructed in the TPC in the same event are taken into
account in the unfolding procedure as background.
In the low momentum and large angle region the corrections for tertiary
particles amount to 10\%--15\%.

The effects of the systematic uncertainties on the final results are estimated
by repeating the analysis with the relevant input modified within the
estimated uncertainty intervals.
In many cases this procedure requires the construction of a set of
different migration matrices.
The correlations of the variations between the cross-section bins are
evaluated and expressed in the covariance matrix.
Each systematic error source is represented by its own covariance
matrix.
The sum of these matrices describes the total systematic error.
The magnitude of the overall systematic errors 
will be described in Section \ref{sec:results}.

\section{Experimental results}
\label{sec:results}

\begin{table}[tbp] 
\small{
\begin{center}
\caption{Experimental uncertainties for the analysis of the data taken
 with carbon and tantalum targets in the 
 5~GeV/c, 8~GeV/c and 12~GeV/c beams.  The
 numbers represent the uncertainty in percent of 
 the cross-section integrated over the angle and momentum region
 indicated. 
 The systematic errors for the Pb data are very similar to the Ta data.
} 
\label{tab:errors-p}
\vspace{2mm}
\begin{tabular}{ l l rrr | rrr | rr} \hline
\bf{p (\GeVc) }&\multicolumn{4}{c|}{0.1 -- 0.3}
                            &\multicolumn{3}{c|}{0.3 -- 0.5}
                            &\multicolumn{2}{c}{0.5 -- 0.7} \\
\hline
\bf{Angle (\mrad)}& &350--950&950--1550&1550--2150
            &350--950&950--1550&1550--2150 &350--950&950--1550 \\
\hline
\bf{5 \GeVc }&&&&&&&&\\
\hline
\bf{Total syst.} 
                 & (C)   & 12.4 &  7.0 &  8.8 &  5.2 &  5.7 & 11.4 &  9.0 & 13.9 \\
                 & (Ta)  & 24.4 & 14.6 & 14.8 &  7.6 &  5.2 &  9.1 &  8.9 & 13.4 \\
\bf{Statistics}  
                 & (C)   &  1.4 &  1.3 &  1.7 &  1.0 &  1.7 &  3.4 &  1.3 &  2.7 \\
                 & (Ta)  &  1.9 &  1.5 &  1.9 &  1.1 &  1.5 &  2.7 &  1.3 &  2.3 \\
\hline
\bf{8 \GeVc }&&&&&&&&\\
\hline
\bf{Total syst.} 
                 & (C)   & 13.2 &  7.2 &  9.1 &  5.4 &  5.6 & 10.5 &  8.2 & 13.3 \\
                 & (Ta)  & 24.5 & 13.9 & 14.1 &  7.7 &  4.9 &  8.2 &  8.4 & 12.1 \\
\bf{Statistics}  
                 & (C)   &  1.5 &  1.3 &  1.8 &  1.0 &  1.7 &  3.5 &  1.2 &  2.6 \\
                 & (Ta)  &  1.4 &  1.1 &  1.4 &  0.8 &  1.1 &  2.0 &  0.9 &  1.7 \\
\hline
\bf{12 \GeVc }&&&&&&&&\\
\hline
\bf{Total syst.}   
                 & (C)   & 13.6 &  7.4 &  9.0 &  5.4 &  5.2 & 10.3 &  8.0 & 12.8 \\
                 & (Ta)  & 24.2 & 14.0 & 14.0 &  7.8 &  4.9 &  7.6 &  8.3 & 12.2 \\
\bf{Statistics} 
                 & (C)   &  1.6 &  1.5 &  2.0 &  1.1 &  1.8 &  3.7 &  1.3 &  2.7 \\
                 & (Ta)  &  0.9 &  0.7 &  0.9 &  0.5 &  0.7 &  1.2 &  0.5 &  1.0 \\
\hline
\end{tabular}
\end{center}
}
\end{table}
\begin{table}[tbp] 
\begin{center}
\caption{Contributions to the experimental uncertainties for the carbon
 and tantalum target data. The numbers
 represent the uncertainty in percent of 
 the cross-section integrated over the angle and momentum region
 indicated.
 The overall normalization has an uncertainty
 of 2\%, and is not reported in the table.} 
\label{tab:errors-syst}
\vspace{2mm}
\begin{tabular}{ l rrr | rrr | rr} \hline
\bf{p (\GeVc) }&\multicolumn{3}{c|}{0.1 -- 0.3}
                            &\multicolumn{3}{c|}{0.3 -- 0.5}
                            &\multicolumn{2}{c}{0.5 -- 0.7} \\
\hline
\bf{Angle (\mrad)}&350--950&950--1550&1550--2150
            &350--950&950--1550&1550--2150 &350--950&950--1550 \\
\hline
&\multicolumn{8}{l}{\bf{12 \GeVc p--C}}\\
\hline
Absorption               &    1.1 &  0.1 &  0.6 &  0.6 &  0.4 &  0.7 &  0.1 &  1.1  \\
Tertiaries               &    2.9 &  1.2 &  0.8 &  2.5 &  0.3 &  2.5 &  0.8 &  2.1  \\
Target region cut        &    8.8 &  5.8 &  6.3 &  3.8 &  3.3 &  3.9 &  4.2 &  4.3  \\
Efficiency               &    1.7 &  2.2 &  5.9 &  1.3 &  2.1 &  6.1 &  1.3 &  2.6  \\
Shape of $\pi^0$         &    2.5 &  0.7 &  0.3 &  0.1 &  0.0 &  0.0 &  0.0 &  0.0  \\
Normalization of $\pi^0$ &    1.7 &  0.2 &  0.2 &  0.0 &  0.0 &  0.0 &  0.0 &  0.0  \\
Particle ID              &    0.1 &  0.2 &  0.5 &  1.2 &  0.6 &  0.2 &  5.7 &  4.8  \\
Momentum resolution      &    4.0 &  0.3 &  0.3 &  0.2 &  0.5 &  0.9 &  0.6 &  0.9  \\
Momentum scale           &    8.4 &  3.8 &  2.0 &  2.1 &  3.1 &  6.6 &  3.3 & 10.2  \\
Angle bias               &    0.6 &  0.4 &  1.4 &  0.5 &  1.4 &  1.6 &  0.8 &  2.1  \\
\hline
&\multicolumn{8}{l}{\bf{12 \GeVc p--Ta}}\\
\hline
Absorption               &    2.7 &  1.3 &  0.9 &  2.0 &  0.2 &  0.6 &  1.0 &  0.7  \\
Tertiaries               &    3.2 &  1.2 &  1.2 &  5.2 &  0.4 &  1.9 &  3.1 &  1.7  \\
Target region cut        &   17.5 & 10.0 &  9.8 &  4.7 &  3.8 &  4.8 &  4.2 &  4.8  \\
Efficiency               &    2.2 &  2.5 &  4.1 &  1.3 &  1.8 &  2.4 &  1.2 &  2.7  \\
Shape of $\pi^0$         &    4.9 &  1.9 &  0.8 &  0.0 &  0.0 &  0.0 &  0.0 &  0.0  \\
Normalization of $\pi^0$ &    1.3 &  0.0 &  0.5 &  0.0 &  0.0 &  0.1 &  0.0 &  0.0  \\
Particle ID              &    0.3 &  0.6 &  0.6 &  0.6 &  0.1 &  0.6 &  4.8 &  4.3  \\
Momentum resolution      &    8.9 &  4.2 &  4.1 &  0.9 &  0.4 &  0.4 &  1.5 &  1.7  \\
Momentum scale           &   12.4 &  7.9 &  7.9 &  2.4 &  2.1 &  4.8 &  3.6 &  9.6  \\
Angle bias               &    0.8 &  0.2 &  1.1 &  0.1 &  1.1 &  1.3 &  0.9 &  1.4  \\
\hline
\end{tabular}
\end{center}
\end{table}

\begin{table}[tbp] 
\begin{center}
\caption{Correlation coefficients of the systematic errors 
for momentum bins in an angular range and angular bins 
in a momentum range for the 12~\GeVc p--C data.} 
\label{tab:errors-syst-corr}
\vspace{2mm}
\begin{tabular}{ cc | c | rrrrrrrrrrrrrr} 
\hline
\multicolumn{10}{l}{\bf{momentum bins for $350~\mrad < \theta \le 950~\mrad$}}\\
\hline
 $p_{\mbox{min}}$ (\GeVc)& $p_{\mbox{max}}$ (\GeVc)& fractional error & 
\multicolumn{8}{|l}{correlation coefficients}\\    
\hline
0.20 & 0.25 & 0.094& 1.00&      &      &      &      &      &      &      \\
0.25 & 0.30 & 0.079& 0.96&  1.00&      &      &	     &	    & 	   &      \\
0.30 & 0.35 & 0.070& 0.96&  0.99&  1.00&      &	     &	    &	   &      \\
0.35 & 0.40 & 0.059& 0.93&  0.95&  0.97&  1.00&	     &	    &	   &      \\
0.40 & 0.45 & 0.051& 0.88&  0.92&  0.95&  0.94&  1.00&	    &	   &      \\
0.45 & 0.50 & 0.055& 0.64&  0.64&  0.68&  0.65&  0.85&  1.00&	   &      \\
0.50 & 0.55 & 0.063& 0.49&  0.49&  0.53&  0.49&  0.74&  0.97&  1.00&      \\
0.55 & 0.60 & 0.071& 0.20&  0.20&  0.24&  0.21&  0.51&  0.85&  0.95&  1.00\\
\hline
\multicolumn{11}{l}{\bf{angular bins for $300~\MeVc < p \le 500~\MeVc $}}\\
\hline
 $\theta_{\mbox{min}}$ (\mrad)& $\theta_{\mbox{max}}$ (\mrad)& fractional error & 
\multicolumn{9}{|l}{correlation coefficients}\\    
\hline
0.35& 0.55& 0.068& 1.00&      &      &      &      &      &      &      &       \\     
0.55& 0.75& 0.059& 0.89&  1.00&      &      &	   &	    &	   &    &       \\
0.75& 0.95& 0.055& 0.65&  0.91&  1.00&      &	   &	    &	   &	  &     \\
0.95& 1.15& 0.070& 0.42&  0.78&  0.95&  1.00&	   &	    &	   &	  &     \\
1.15& 1.35& 0.082& 0.26&  0.65&  0.89&  0.97&  1.00&	    &	   &	  &     \\
1.35& 1.55& 0.096& 0.18&  0.57&  0.83&  0.93&  0.99&  1.00&	   &	  &     \\
1.55& 1.75& 0.095& 0.09&  0.47&  0.74&  0.85&  0.94&  0.98&  1.00&	  &     \\
1.75& 1.95& 0.116& 0.35&  0.66&  0.85&  0.90&  0.90&  0.91&  0.89&  1.00&       \\
1.95& 2.15& 0.182& 0.35&  0.56&  0.71&  0.69&  0.68&  0.71&  0.69&  0.91&  1.00 \\
\hline
\end{tabular}
\end{center}
\end{table}

\begin{sidewaysfigure}[tbp!]
 \begin{center}
  \includegraphics[width=0.460\textwidth,angle=0]{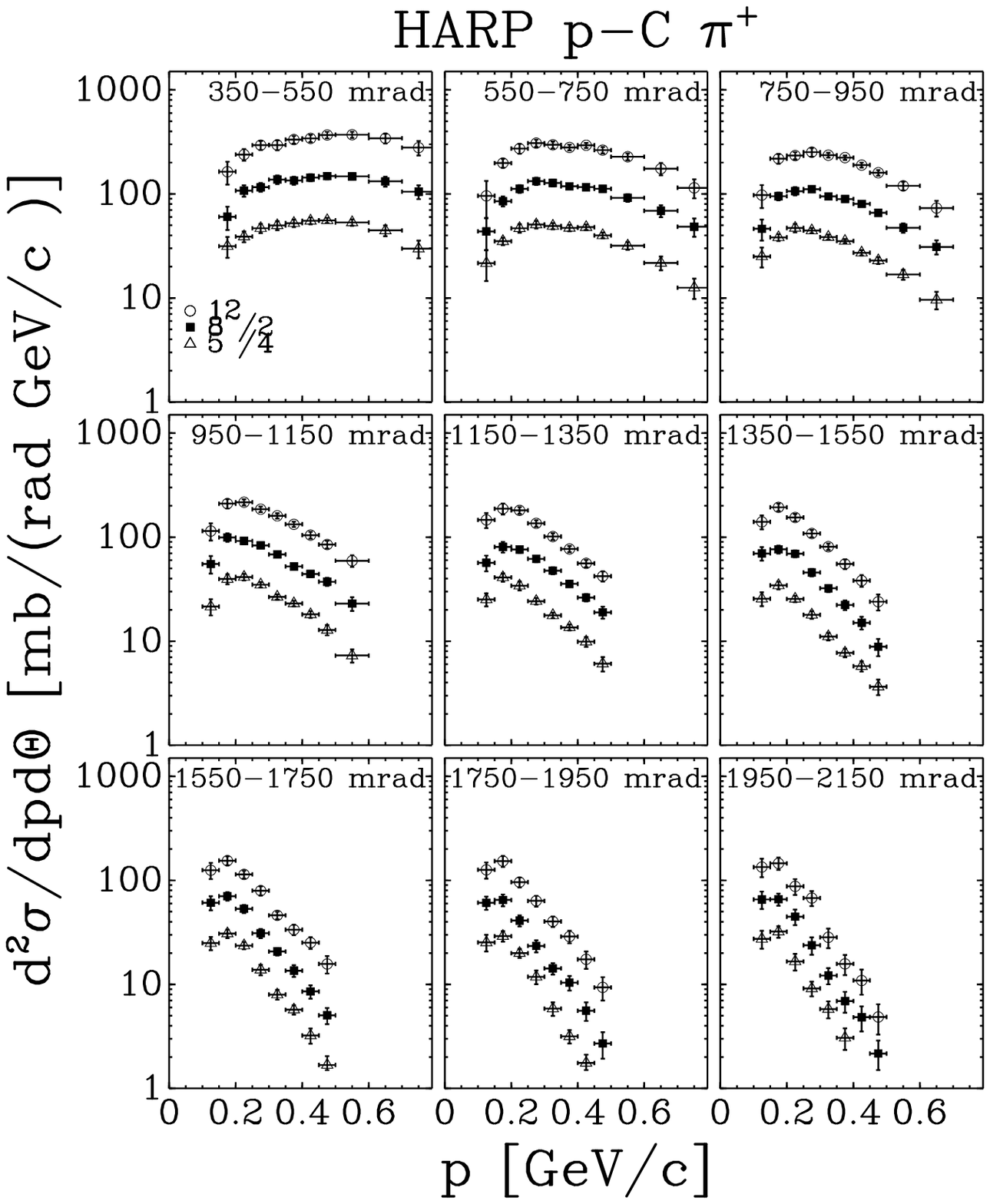}
  \includegraphics[width=0.460\textwidth,angle=0]{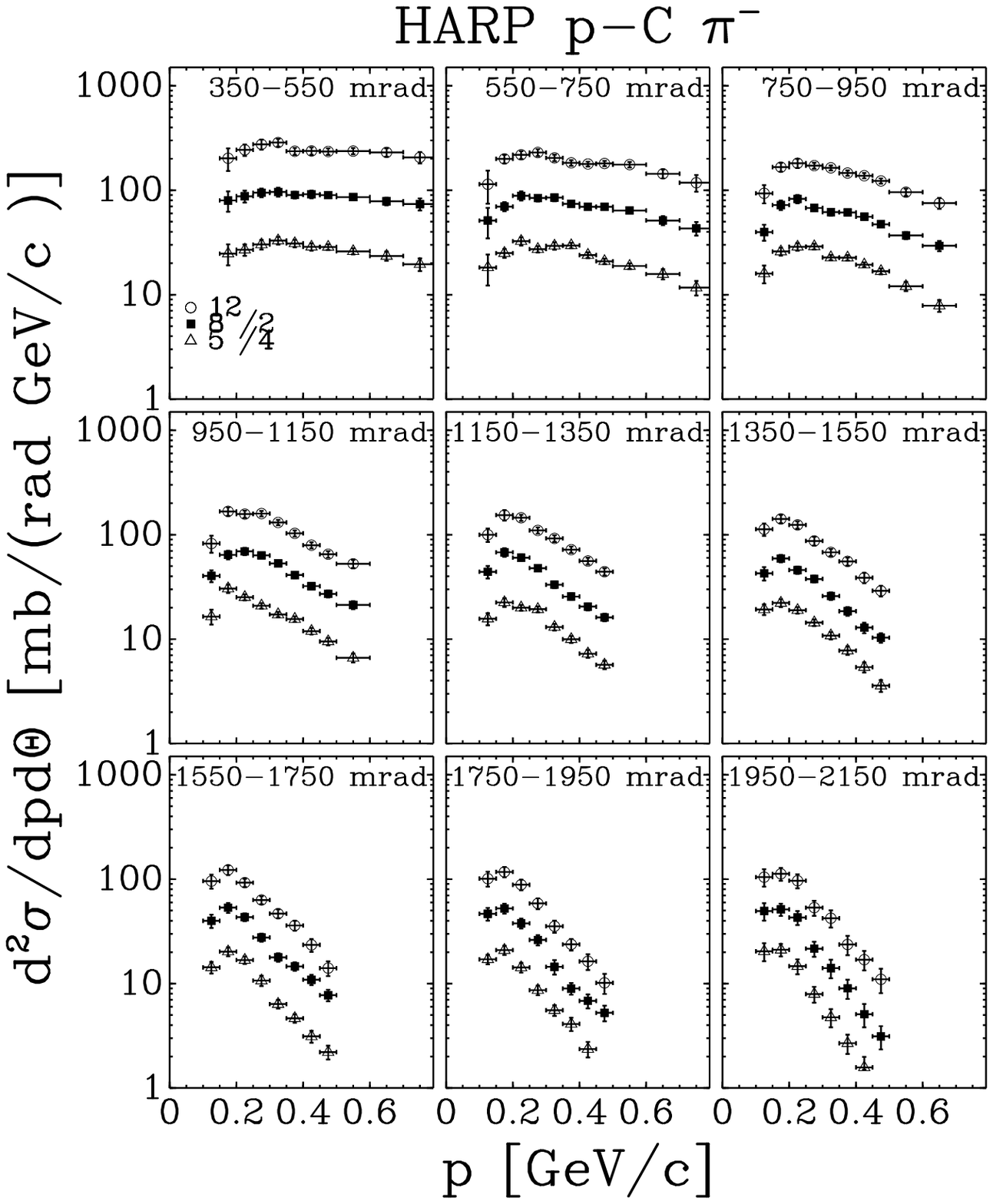}
  \caption{
Double-differential yields per target nucleon for \pip production (left) and  \pim
 production (right) in
p--C interactions as a function of momentum displayed in different
angular bins (shown in \mrad in the panels).
For better visibility the cross-sections have been scaled by a factor
  0.5 (0.25) for the data taken at 8~\GeVc (5~\GeVc). 
The error bars represent the combination of statistical and systematic
 uncertainties. 
}
\label{fig:xs-p-th-pbeam-c}
 \end{center}
\end{sidewaysfigure}

\begin{sidewaysfigure}[tbp!]
 \begin{center}
  \includegraphics[width=0.460\textwidth,angle=0]{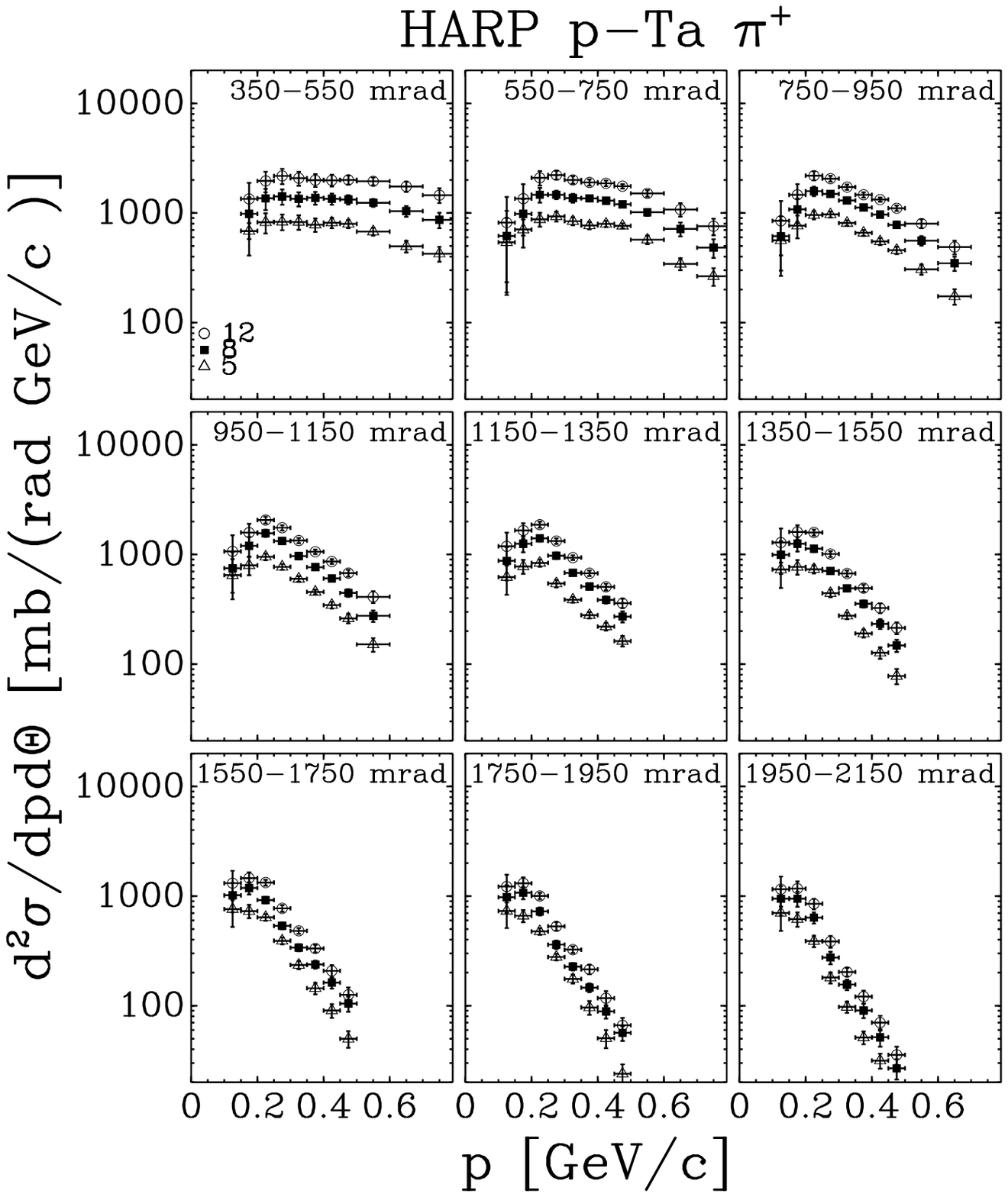}
  \includegraphics[width=0.460\textwidth,angle=0]{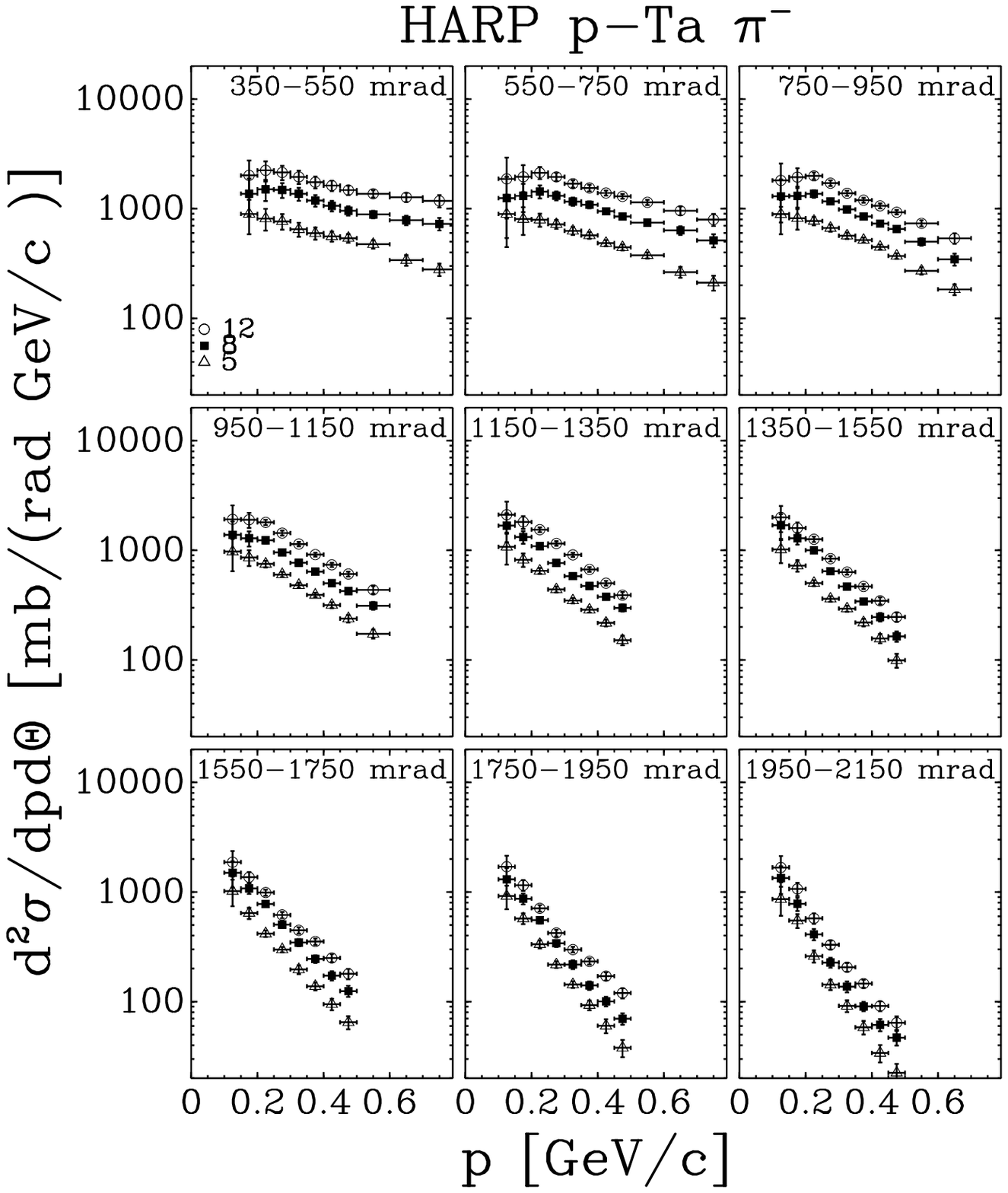}
\caption{
Double-differential yields per target nucleon for \pip production (left) and  \pim
 production (right) in
p--Ta interactions as a function of momentum displayed in different
angular bins (shown in \mrad in the panels).
The error bars represent the combination of statistical and systematic
 uncertainties. 
}
\label{fig:xs-p-th-pbeam-ta}
 \end{center}
\end{sidewaysfigure}

\begin{sidewaysfigure}[tbp!]
 \begin{center}
  \includegraphics[width=0.460\textwidth,angle=0]{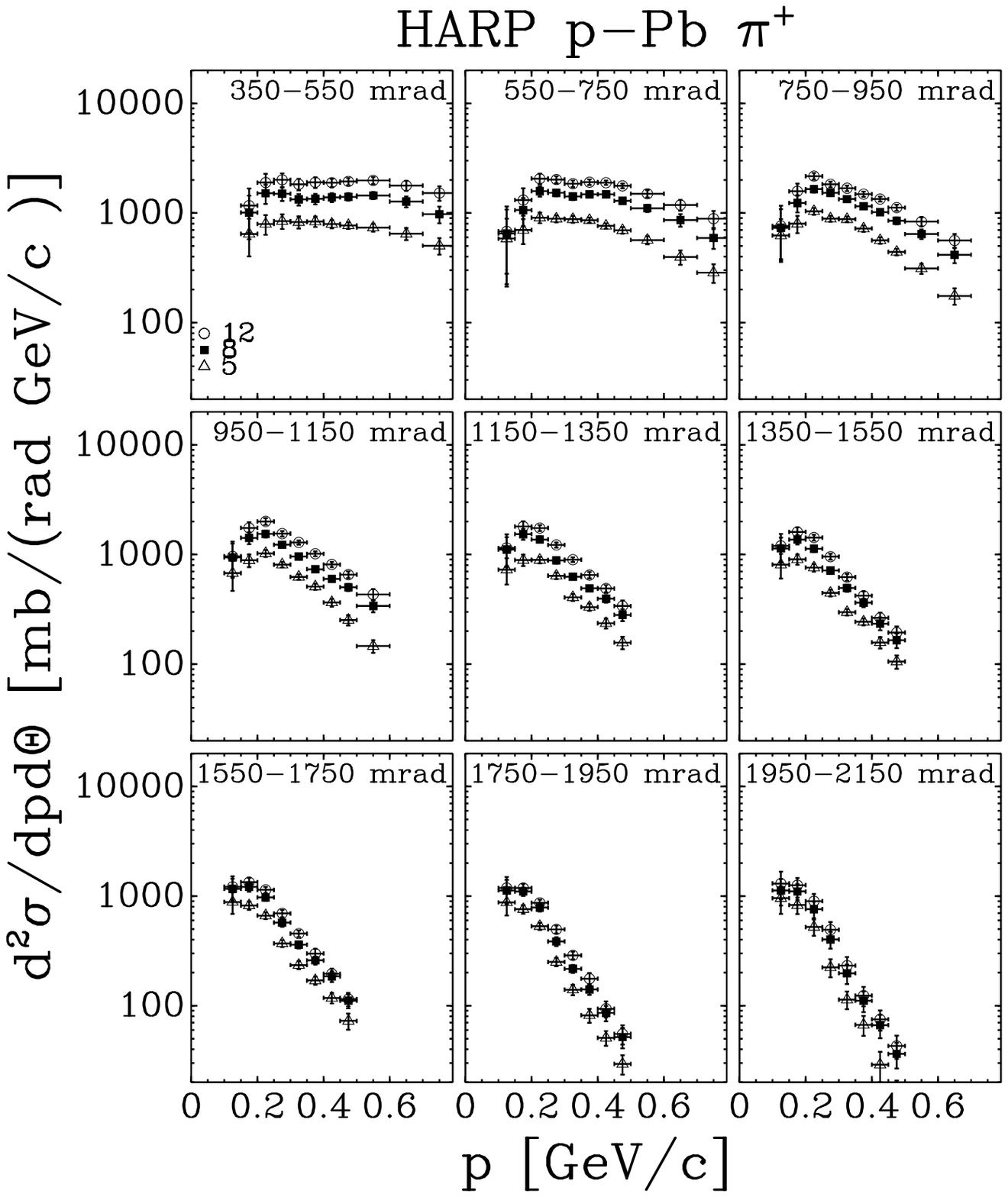}
  \includegraphics[width=0.460\textwidth,angle=0]{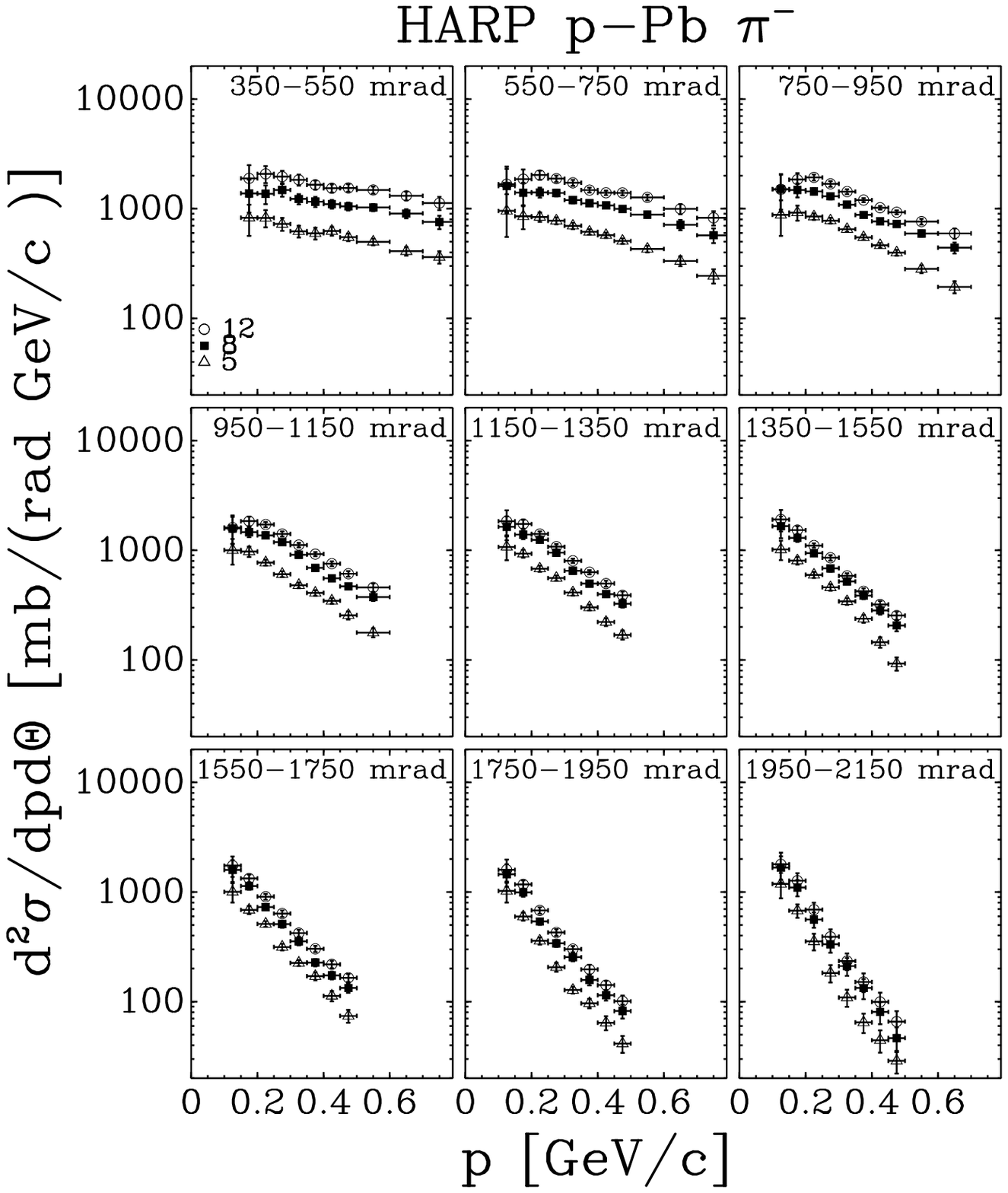}
\caption{
Double-differential yields per target nucleon for \pip production (left) and  \pim
 production (right) in
p--Pb interactions as a function of momentum displayed in different
angular bins (shown in \mrad in the panels).
The error bars represent the combination of statistical and systematic
 uncertainties. 
}
\label{fig:xs-p-th-pbeam-pb}
 \end{center}
\end{sidewaysfigure}

The measured double-differential yields per target nucleon for the 
production of \pip and \pim in the laboratory system as a function of
the momentum and the polar angle for each incident-beam momentum are
shown in Figures \ref{fig:xs-p-th-pbeam-c} to \ref{fig:xs-p-th-pbeam-pb} for 
the three long targets studied here.
The error bars  shown are the
square-roots of the diagonal elements in the covariance matrix,
where statistical and systematic uncertainties are combined
in quadrature.
Correlations cannot be shown in the figures.
The correlation of the statistical errors (introduced by the unfolding
procedure) are typically smaller than 20\% for adjacent momentum bins and
smaller for adjacent angular bins.
The correlations of the systematic errors are larger, typically 90\% for
adjacent bins.
The correlation between systematic errors is shown in Table~\ref{tab:errors-syst-corr}
for selected bins in the carbon data taken with the 12~\GeVc beam.
The overall scale error ($< 2\%$) is not shown.
The double-differential results of this analysis are also tabulated in Appendix A. 

\begin{figure}[tbp!]
 \begin{center}
  \includegraphics[width=0.30\textwidth,angle=0]{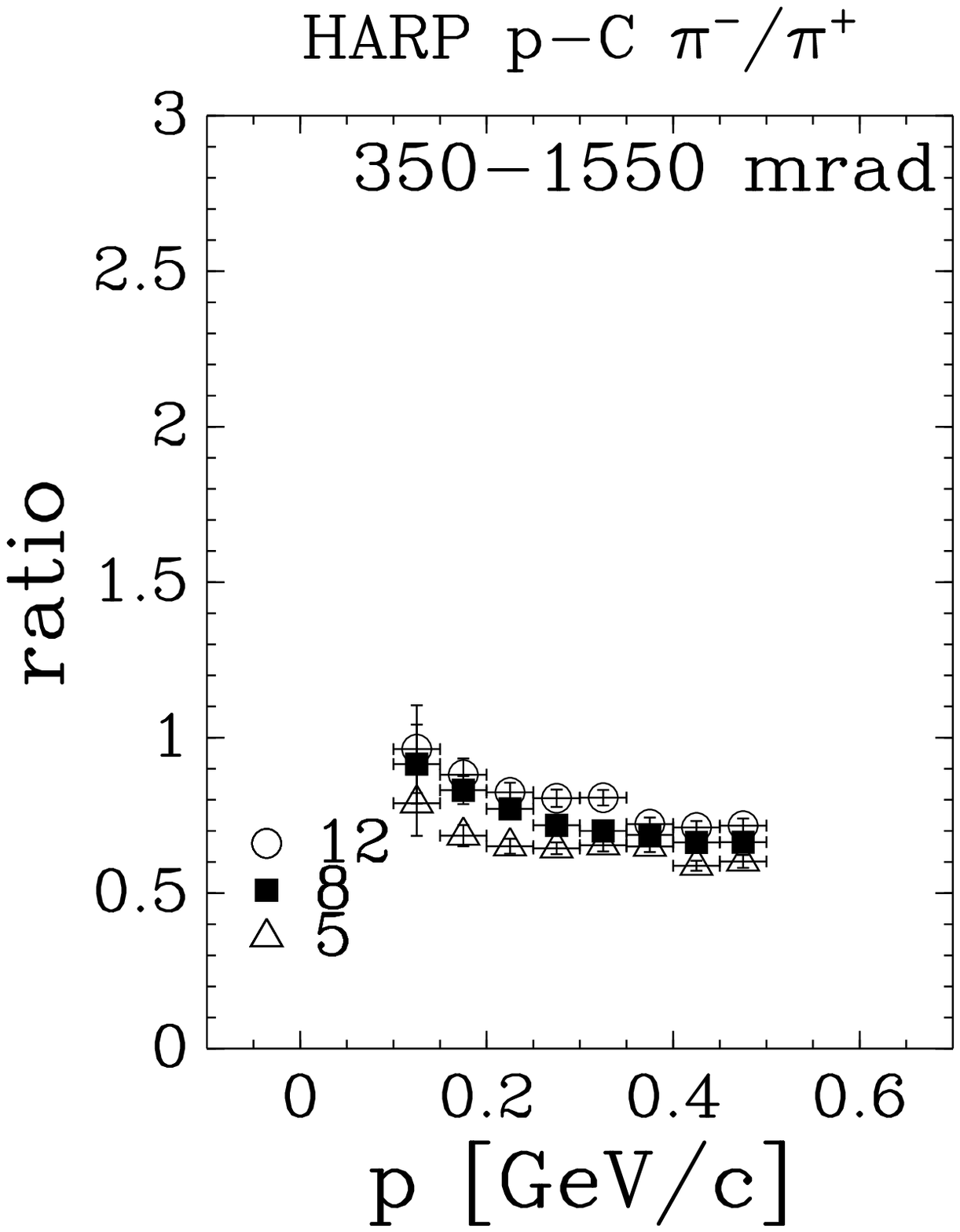}
  ~
  \includegraphics[width=0.30\textwidth,angle=0]{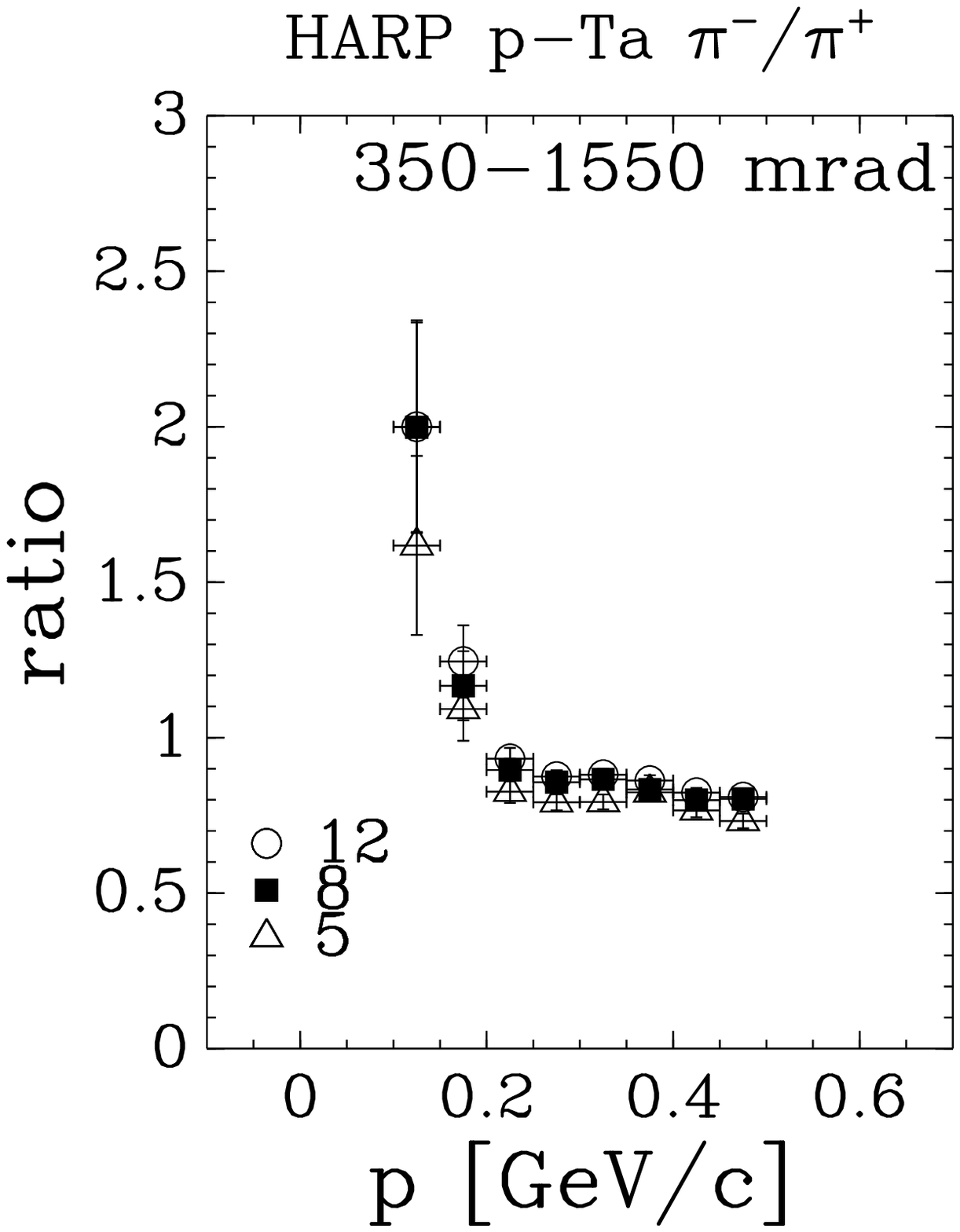}
  ~
  \includegraphics[width=0.30\textwidth,angle=0]{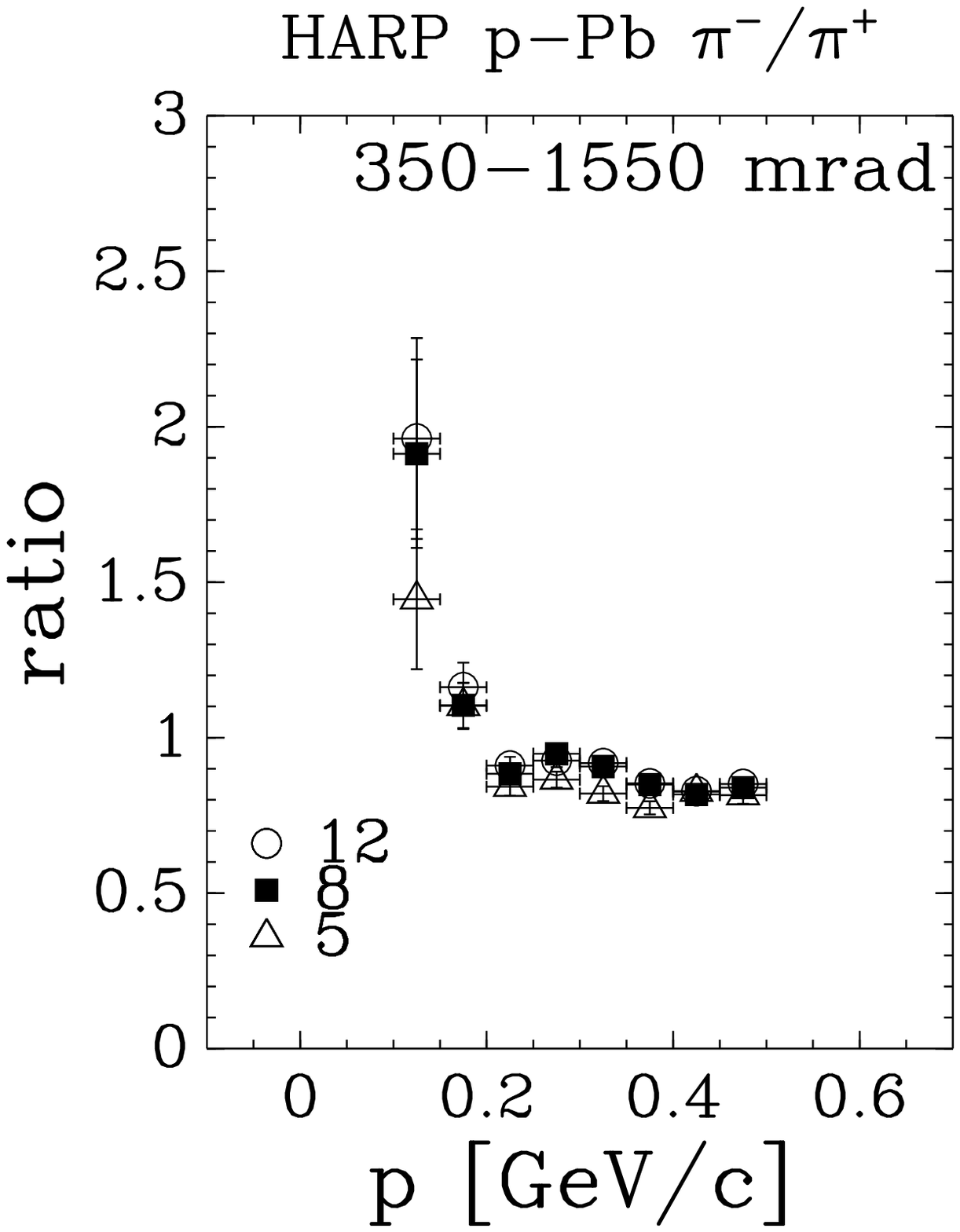}
 \end{center}
\caption{
The ratio of the differential yields for \pim and \pip
 production in
p--C (left panel), p--Ta (middle panel) and p--Pb (right panel) 
interactions as a function of 
the secondary momentum integrated over the
forward angular region.
}
\label{fig:xs-ratio}
\end{figure}

The integrated \pim/\pip ratio in the forward direction is displayed in
Fig.~\ref{fig:xs-ratio} as a function of the secondary momentum. 
The ratios are very similar to the ones observed in the data with the
short targets~\cite{ref:harp:la}.
In the  part of the momentum range shown in most
bins more \pip's are  produced than \pim's.
The \pim/\pip ratio is larger for higher incoming beam momenta than for
lower momenta and drops with increasing secondary momentum.
The large \pim/\pip ratio in the lowest bin of
secondary momentum (100~\MeVc--150~\MeVc) for the heavy nuclear targets
(Pb and Ta) in the beams with 8~\GeVc and 12~\GeVc momentum had already
been observed in the short-target data.
The E910 collaboration had made a similar observation for their lowest
momentum bin (100 \MeVc~ -- 140 \MeVc) in p--Au collisions at 12.3~\GeVc
and 17.5~\GeVc incoming beam momentum~\cite{ref:E910}. 
A plausible explanation has been put forward in
Ref.~\cite{ref:gallmeister}, where it was shown that this effect is due
to an asymmetry in the production of $\Delta$
resonances given the large neutron excess in these heavy nuclei. 
 
The experimental uncertainties are summarized in
Table~\ref{tab:errors-p} for the C and Ta targets. 
The systematic error break-down is shown in Table~\ref{tab:errors-syst}
for the C and Ta data taken with the 12~\GeVc beam.
The numbers for the other momenta are very similar.
The errors for the Pb target are very similar to the ones for the Ta
target.  
The relative sizes of the different systematic error sources are very
similar for \pim and \pip (only \pip is shown) and for the different
beam energies.  
Going from lighter (C) to heavier targets (Ta, Pb)
the corrections for \piz (conversion, concentrated at low secondary
momentum), absorption, tertiaries and target pointing uncertainties are
bigger.    
The discussion and figures shown in
\cite{ref:harp:la} give a reliable indication of the momentum and
angular dependence of the systematic error components.
   
One observes that the statistical
error is small compared to the systematic errors.
The statistical error is calculated by error propagation as part of the
unfolding procedure. 
It takes into account that the unfolding matrix is obtained from the
data themselves~\footnote{The migration matrix is calculated without
prior knowledge of the cross-sections, while the unfolding procedure
determined the unfolding matrix from the migration matrix and the
distributions found in the data.} and hence also contributes to the
statistical error. 
This procedure almost doubles the statistical error, but avoids an important 
systematic error which would otherwise be introduced by assuming a
cross-section model calculating the corrections. 

The largest systematic error corresponds to the uncertainty in the
absolute momentum scale, which was estimated to be around 3\% using elastic
scattering~\cite{ref:harp:tantalum}.
This error dominates at large momenta where the resolution is 
worse and in the bins with the largest angles where the distributions
are steep. 
At low momentum in the forward
direction the uncertainty in the subtraction of the electron and
positron background due to \piz production is large ($\sim 6 - 10 \%$).
This uncertainty is split between the variation in the shape of the
\piz spectrum and the normalization using the recognized electrons. 
The target region definition  and
the uncertainty in the PID efficiency and background from tertiaries
(particles produced in secondary interactions)
are of similar size and are not negligible ($\sim 2-3 \%$).
Relatively small errors are introduced by the uncertainties in
the  absolute knowledge of the angular and the momentum resolution.
The correction for tertiaries is relatively large at low momenta and large 
angles.  
Its uncertainty is responsible for a systematic error of $\sim 3-5 \%$
in these regions. 
Compared to the data taken with the shorter targets the systematic
errors related to the 
efficiency of the cut requiring pions to point to the beam particle
trajectory are much larger, especially for the Ta and Pb targets.
This is not surprising given the large energy loss of these
particles, especially in the forward direction when traversing a large
part of the target.   

As already mentioned above, the overall normalization has an uncertainty
of 2\%, and is not reported in the table.
It is mainly due to the uncertainty in the efficiency that beam
particles 
counted in the normalization actually hit the target, with smaller
components from the target density and beam particle counting procedure.

\begin{figure}[tbp!]
 \begin{center}
  \includegraphics[width=0.80\textwidth,angle=0]{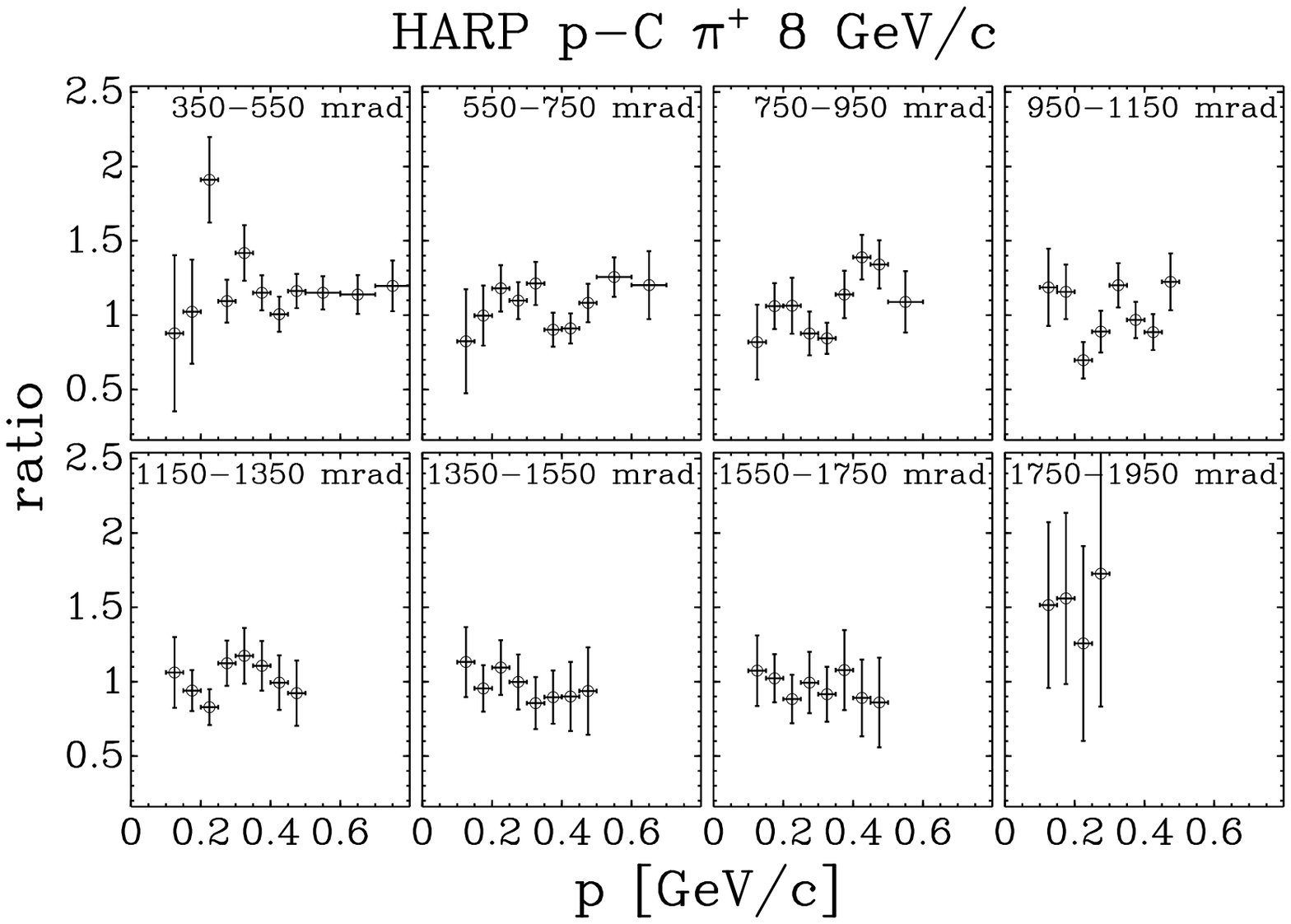}
 \end{center}
\caption{
 Ratio of the double-differential \pip production yields per target nucleon for
 p--C at 8~\GeVc measured with the first 0.1~\intlen of the
 100\%~\intlen target and the 5\%~\intlen target for eight angular bins (shown in mrad).
}
\label{fig:xs-ratio-10-5}
\end{figure}

As a cross-check of the experimental procedures a comparison of the \pip
production yield measured with the first 0.1~\intlen of the long
carbon target and the full 0.05~\intlen of the short carbon target was
made. 
The carbon target is optimal for this cross-check owing to its lower
density which favours the ratio of the resolution of the $z$ measurement
of the origin of the secondary particles and the interaction length.
Nevertheless, an additional systematic error is introduced by 
selecting the outgoing particles from the partial target.
The downstream boundary of the selection has to be made in a region with
high particle population.
The 8~\GeVc carbon data were chosen for their relatively high
statistics. 
The choice to select the first 10\% rather than the first 5\% of the
long target was made as a compromise between the reduced statistics, the
additional systematic error due to the track selection near the boundary
and the aim to remain as close as possible to the 5\%~\intlen target.
Due to the limited statistics once only 10\% of the particles are
selected, the comparison is only meaningful for 
\pip production and for the first eight out of nine angular bins.
The result is shown in Fig.~\ref{fig:xs-ratio-10-5}.
The ratio is compatible with unity within the relatively large
uncertainties. 


%
\afterpage{\clearpage}
\afterpage{\clearpage}
\afterpage{\clearpage}
\afterpage{\clearpage}

\subsection{Comparisons with short target data}
\label{sec:compare}

\begin{sidewaysfigure}[tbp!]
 \begin{center}
  \includegraphics[width=0.460\textwidth,angle=0]{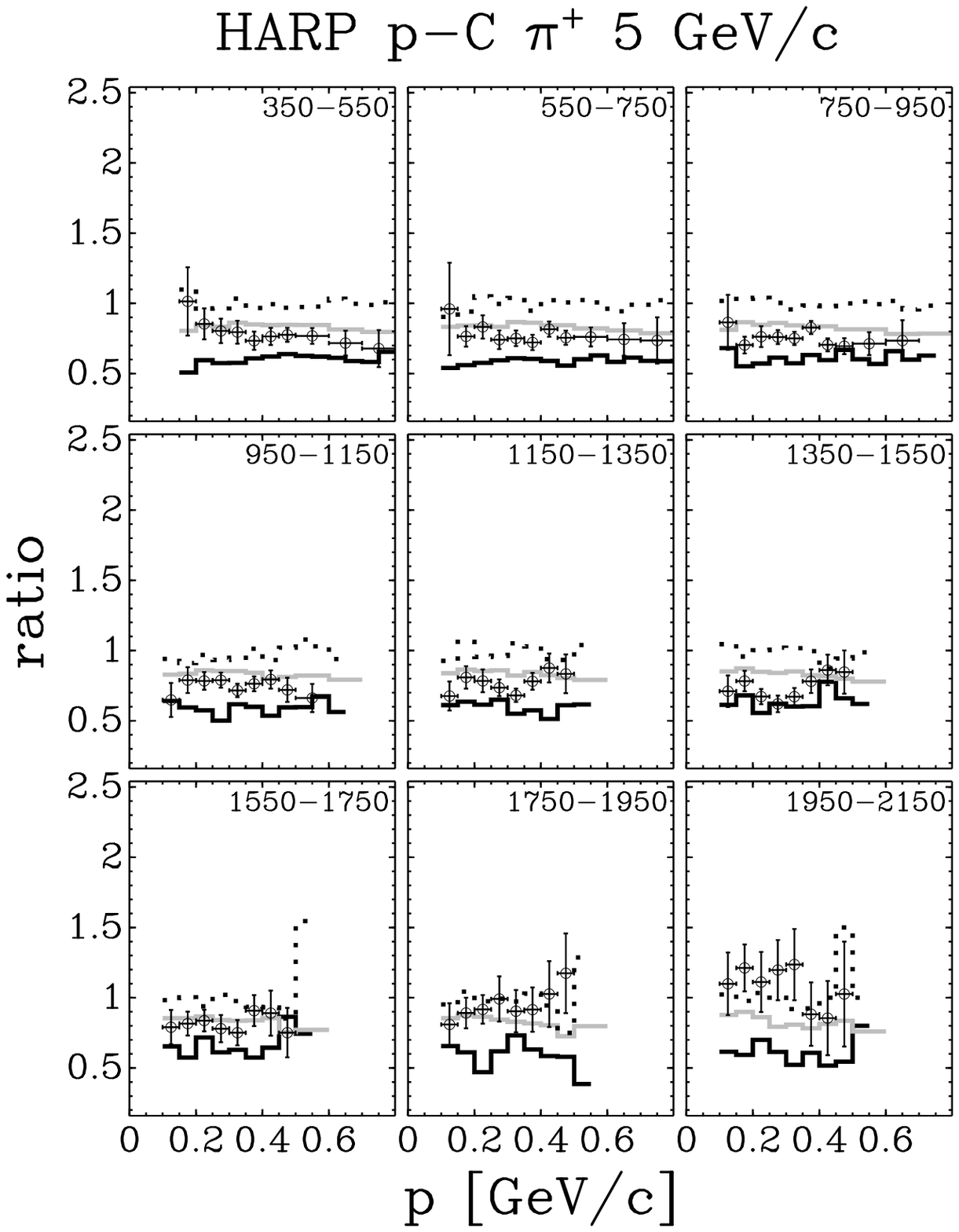}
  \includegraphics[width=0.460\textwidth,angle=0]{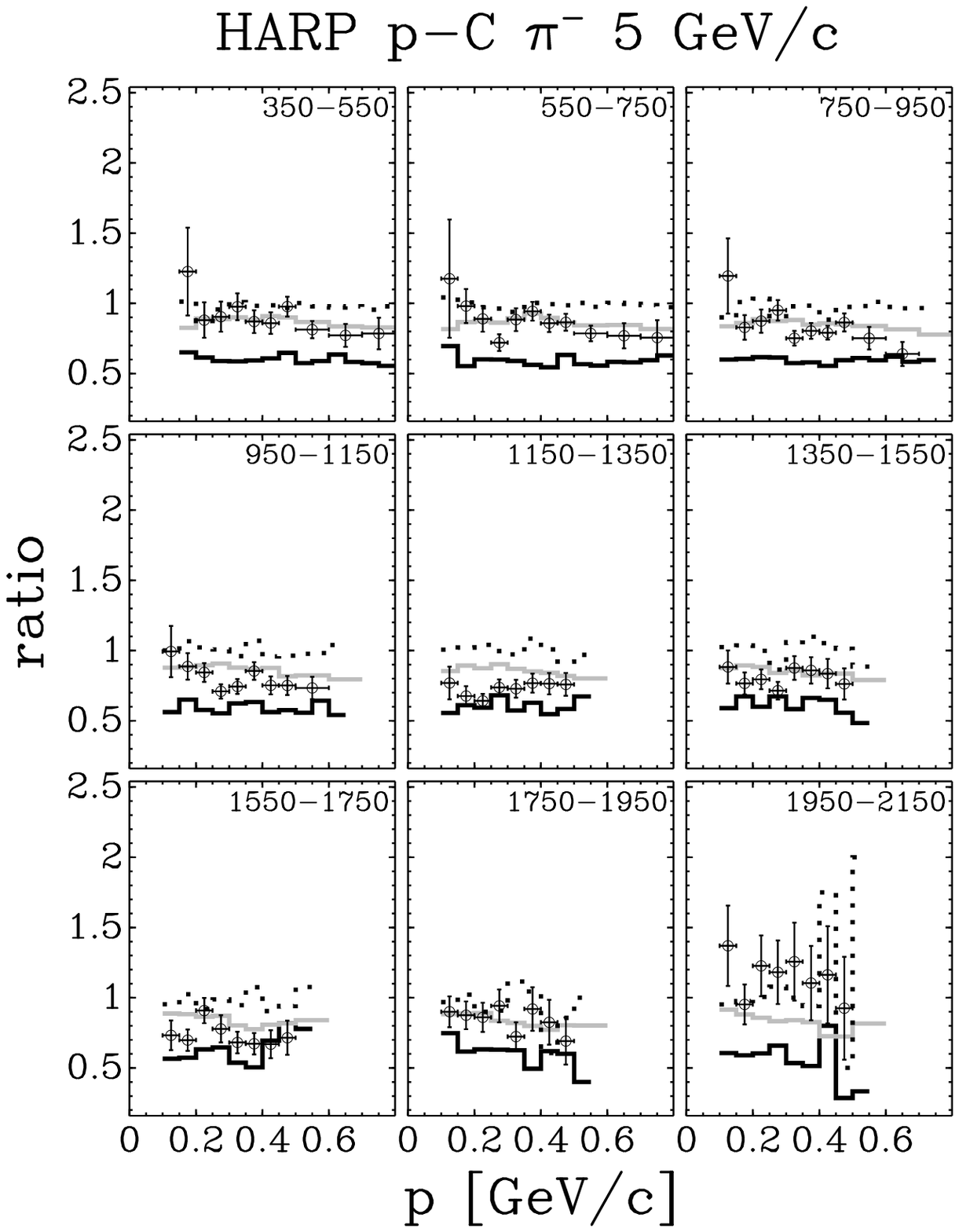}
 \end{center}
\caption{
 Comparison of the ratio of double-differential \pip(left) and \pim (right)
 production yields per target nucleon for p--C at 5~\GeVc taken with the
 100\%~\intlen and 5\%~\intlen target (circles) with MC predictions.
 The black curve represents the MARS prediction and the grey curve the
 GEANT4 simulation, while the dotted line
 shows the ratio of pions produced by a ``first generation'' beam proton
 and all pions as calculated by MARS.
}
\label{fig:modelC5}
\end{sidewaysfigure}

\begin{sidewaysfigure}[tbp!]
 \begin{center}
  \includegraphics[width=0.460\textwidth,angle=0]{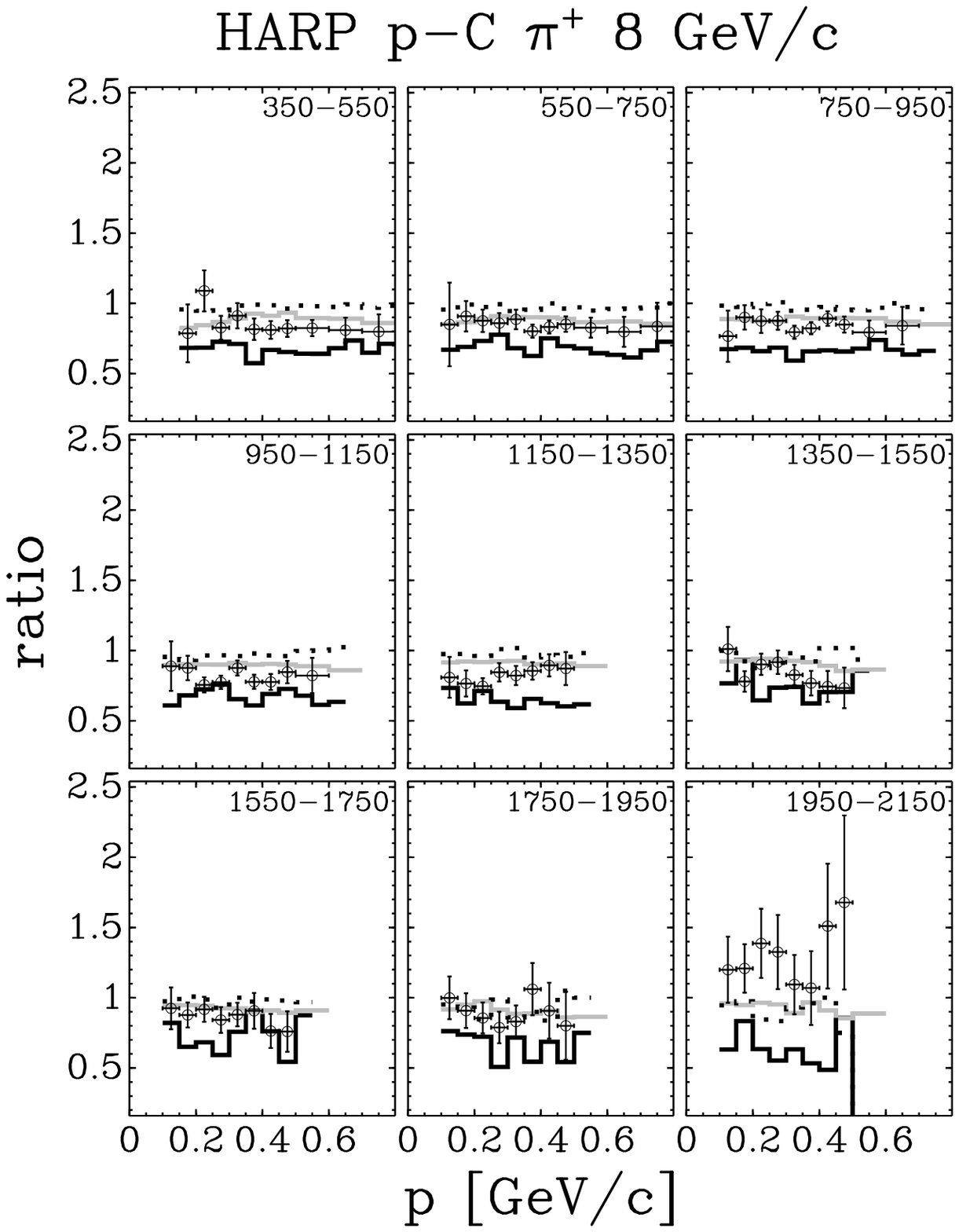}
  \includegraphics[width=0.460\textwidth,angle=0]{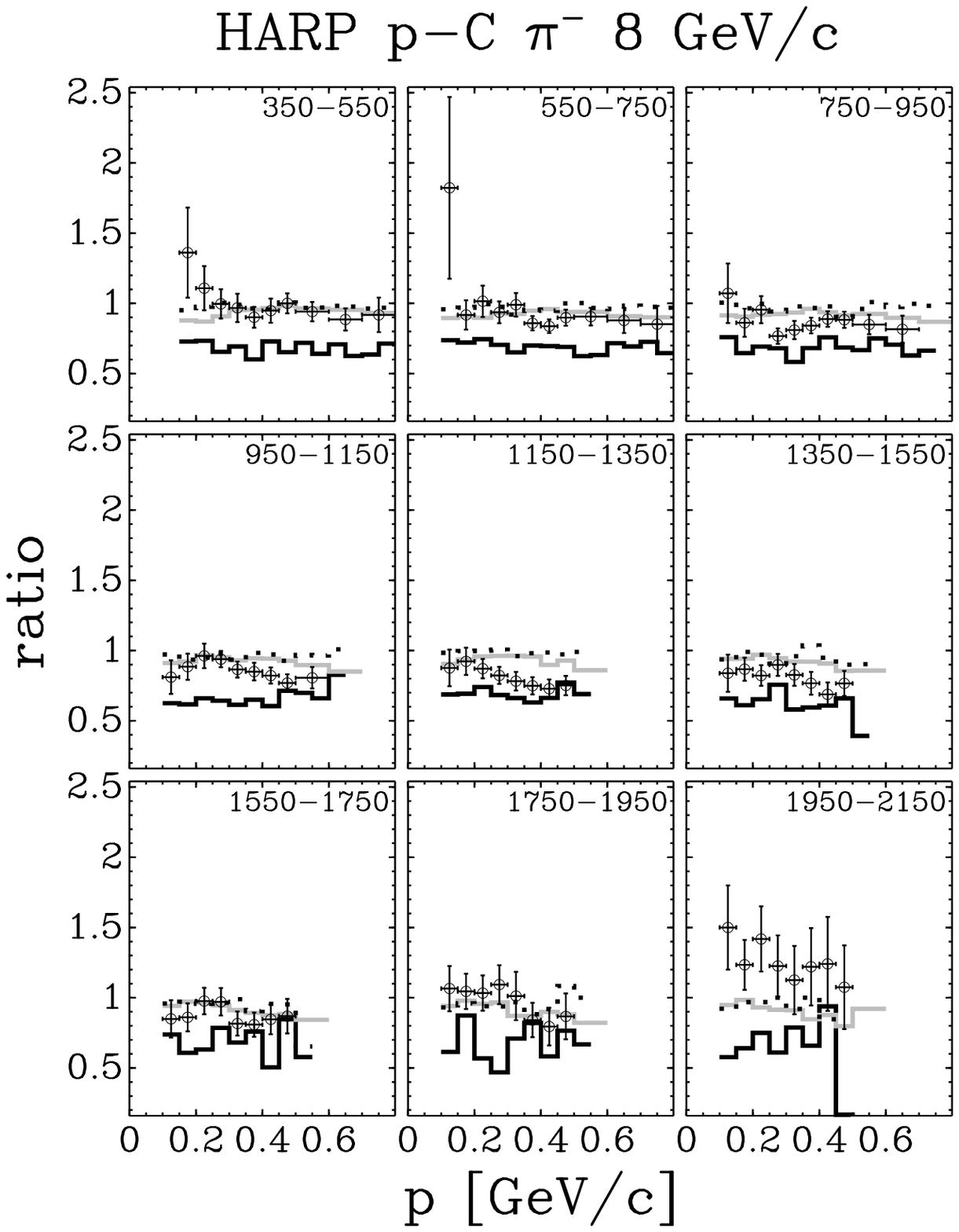}
 \end{center}
\caption{
 Comparison of the ratio of double-differential \pip(left) and \pim (right)
 production yields per target nucleon for p--C at 8~\GeVc taken with the
 100\%~\intlen and 5\%~\intlen target (circles) with MC predictions.
 The black curve represents the MARS prediction and the grey curve the
 GEANT4 simulation, while the dotted line
 shows the ratio of pions produced by a ``first generation'' beam proton
 and all pions as calculated by MARS.
}
\label{fig:modelC8}
\end{sidewaysfigure}

\begin{sidewaysfigure}[tbp!]
 \begin{center}
  \includegraphics[width=0.460\textwidth,angle=0]{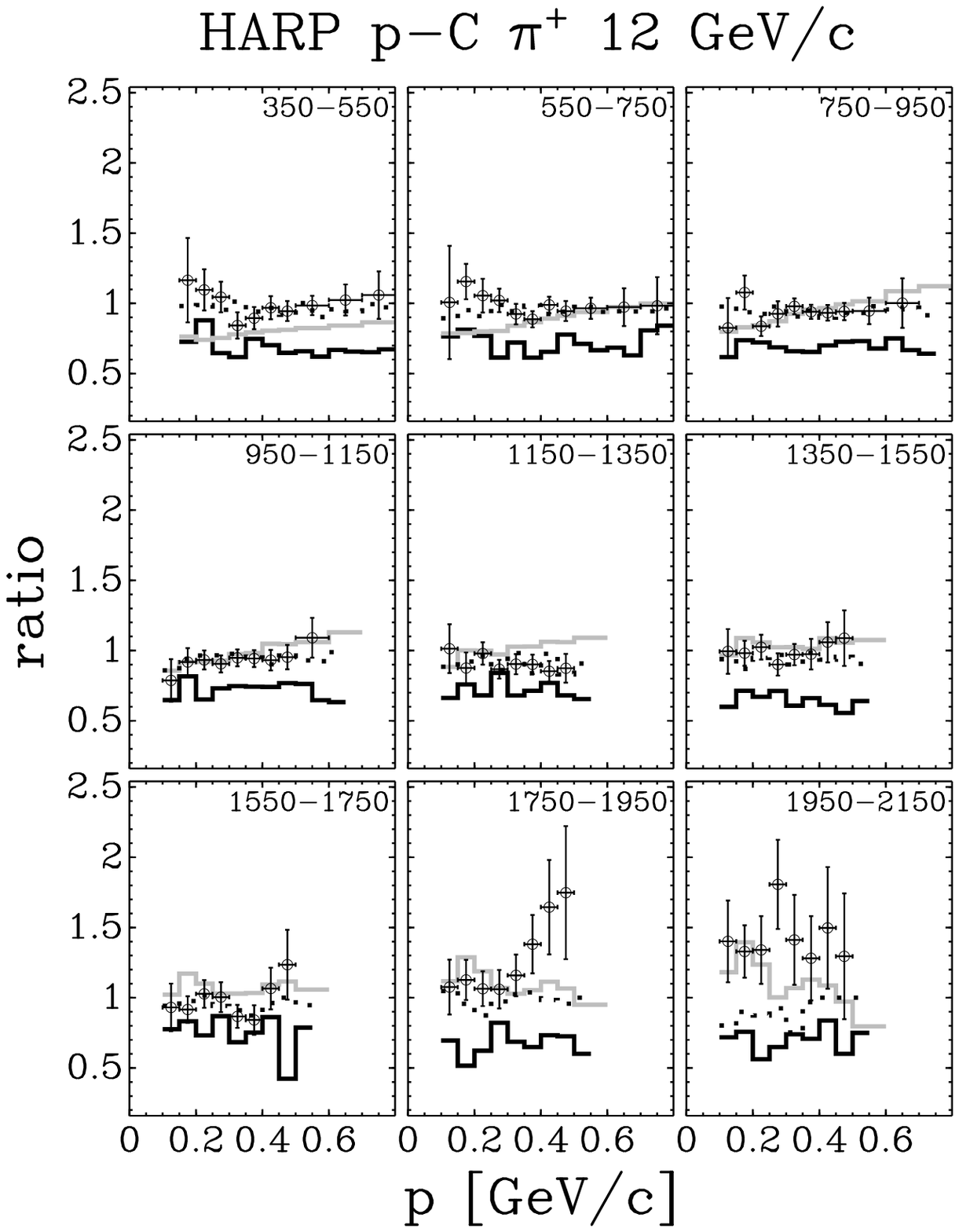}
  \includegraphics[width=0.460\textwidth,angle=0]{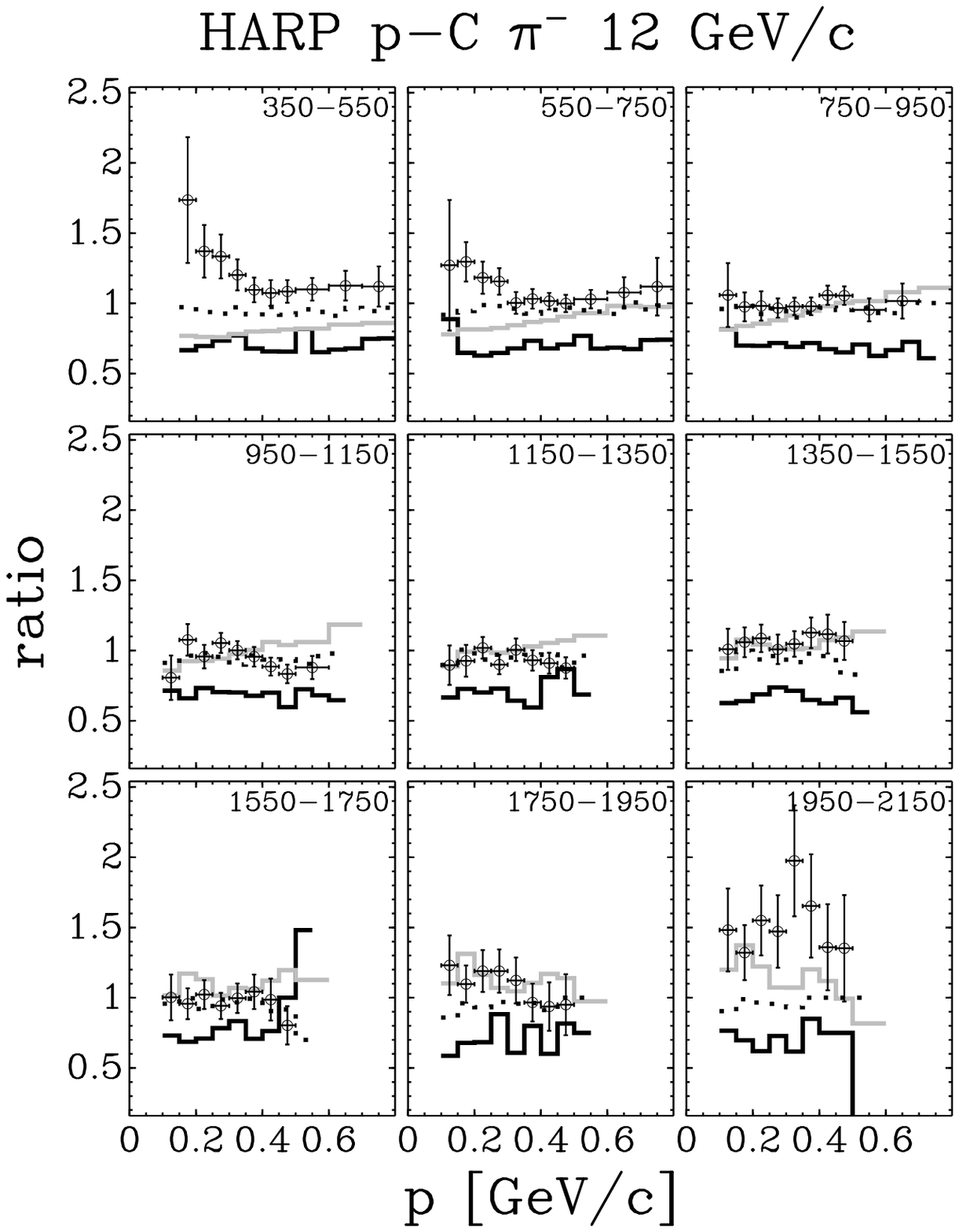}
 \end{center}
\caption{
 Comparison of the ratio of double-differential \pip(left) and \pim (right)
 production yields per target nucleon for p--C at 12~\GeVc taken with the
 100\%~\intlen and 5\%~\intlen target (circles) with MC predictions.
 The black curve represents the MARS prediction and the grey curve the
 GEANT4 simulation, while the dotted line
 shows the ratio of pions produced by a ``first generation'' beam proton
 and all pions as calculated by MARS.
}
\label{fig:modelC12}
\end{sidewaysfigure}

\begin{sidewaysfigure}[tbp!]
 \begin{center}
  \includegraphics[width=0.460\textwidth,angle=0]{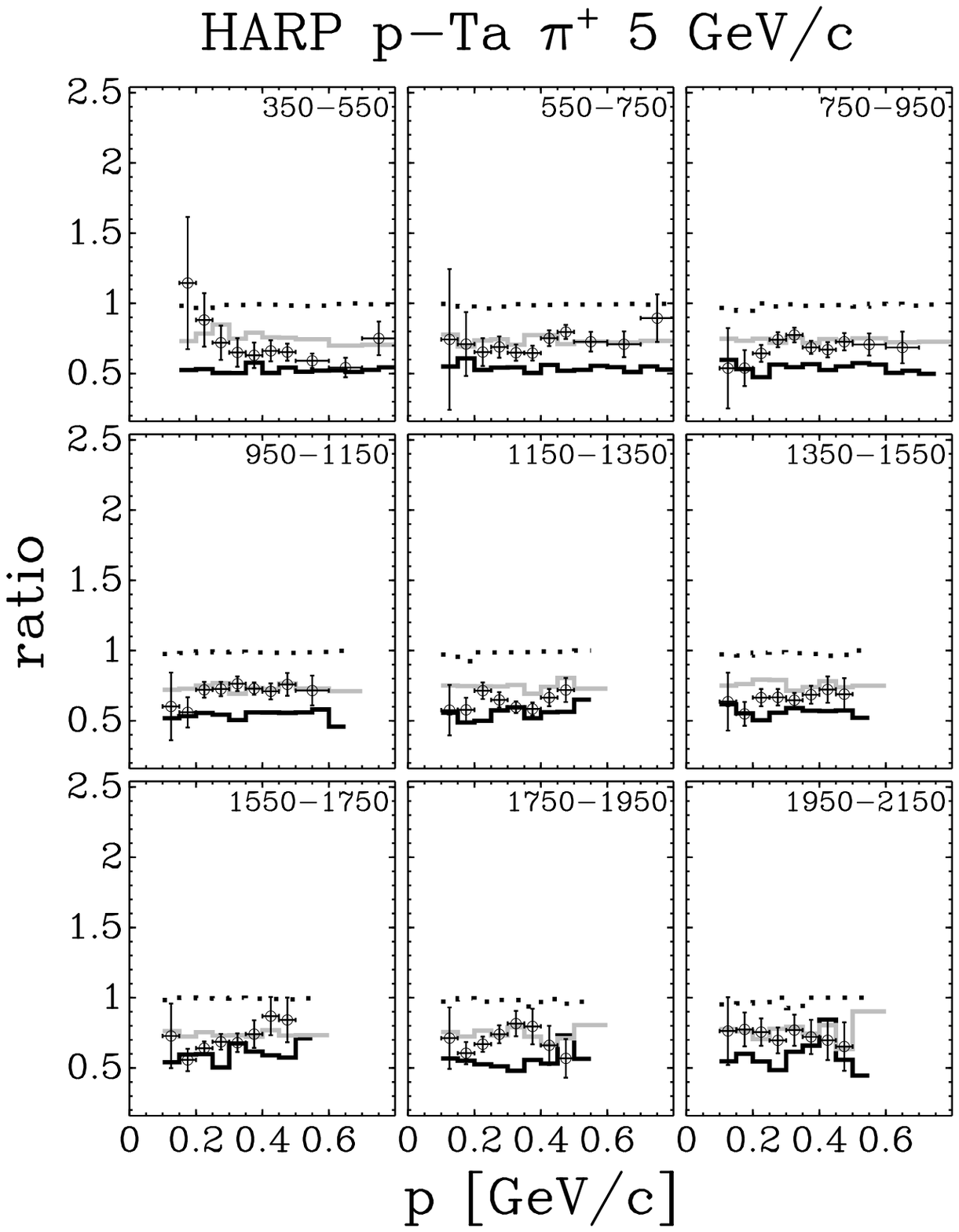}
  \includegraphics[width=0.460\textwidth,angle=0]{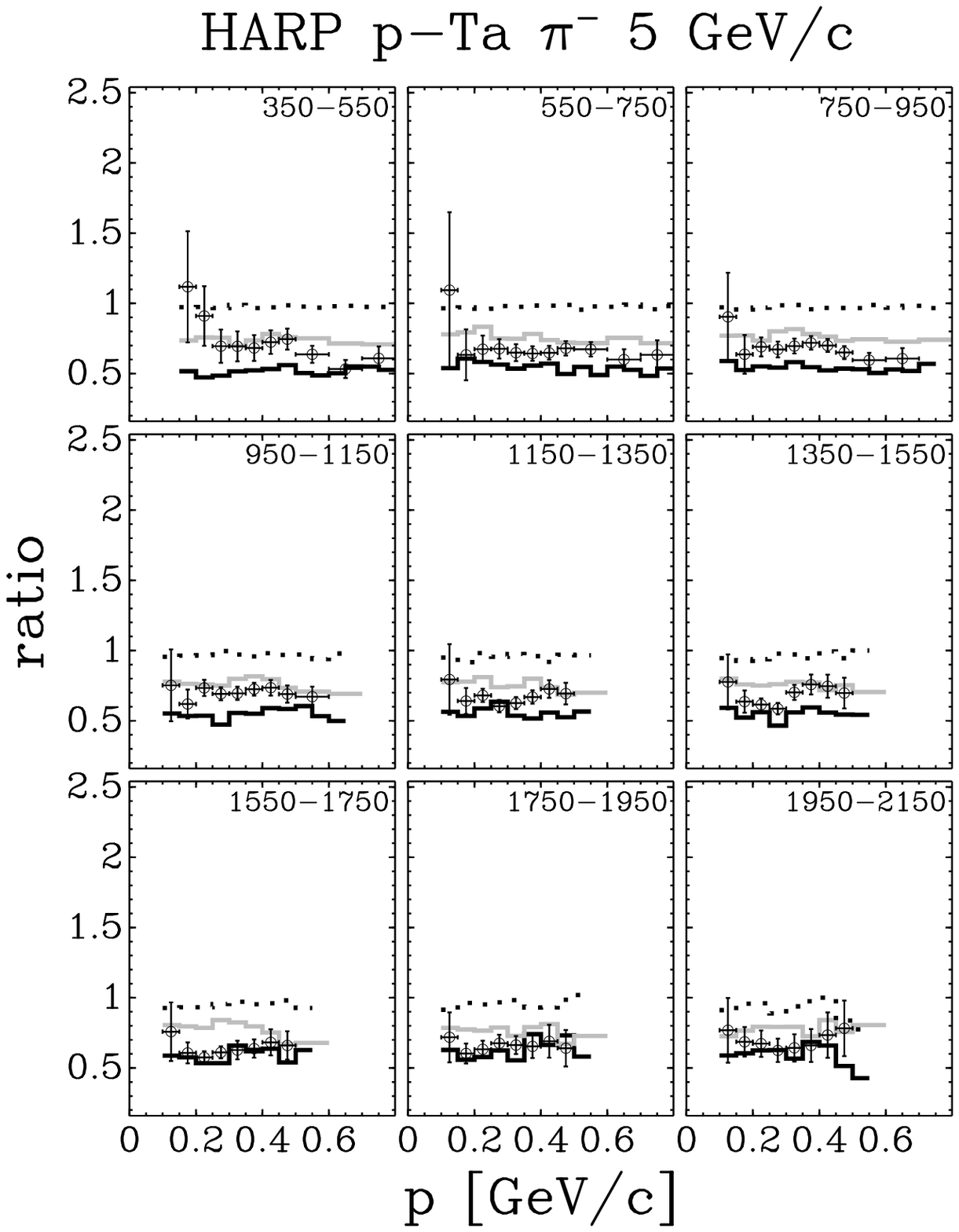}
 \end{center}
\caption{
 Comparison of the ratio of double-differential \pip(left) and \pim (right)
 production yields per target nucleon for p--Ta at 5~\GeVc taken with the
 100\%~\intlen and 5\%~\intlen target (circles) with MC predictions.
 The black curve represents the MARS prediction and the grey curve the
 GEANT4 simulation, while the dotted line
 shows the ratio of pions produced by a ``first generation'' beam proton
 and all pions as calculated by MARS.
}
\label{fig:modelTa5}
\end{sidewaysfigure}

\begin{sidewaysfigure}[tbp!]
 \begin{center}
  \includegraphics[width=0.460\textwidth,angle=0]{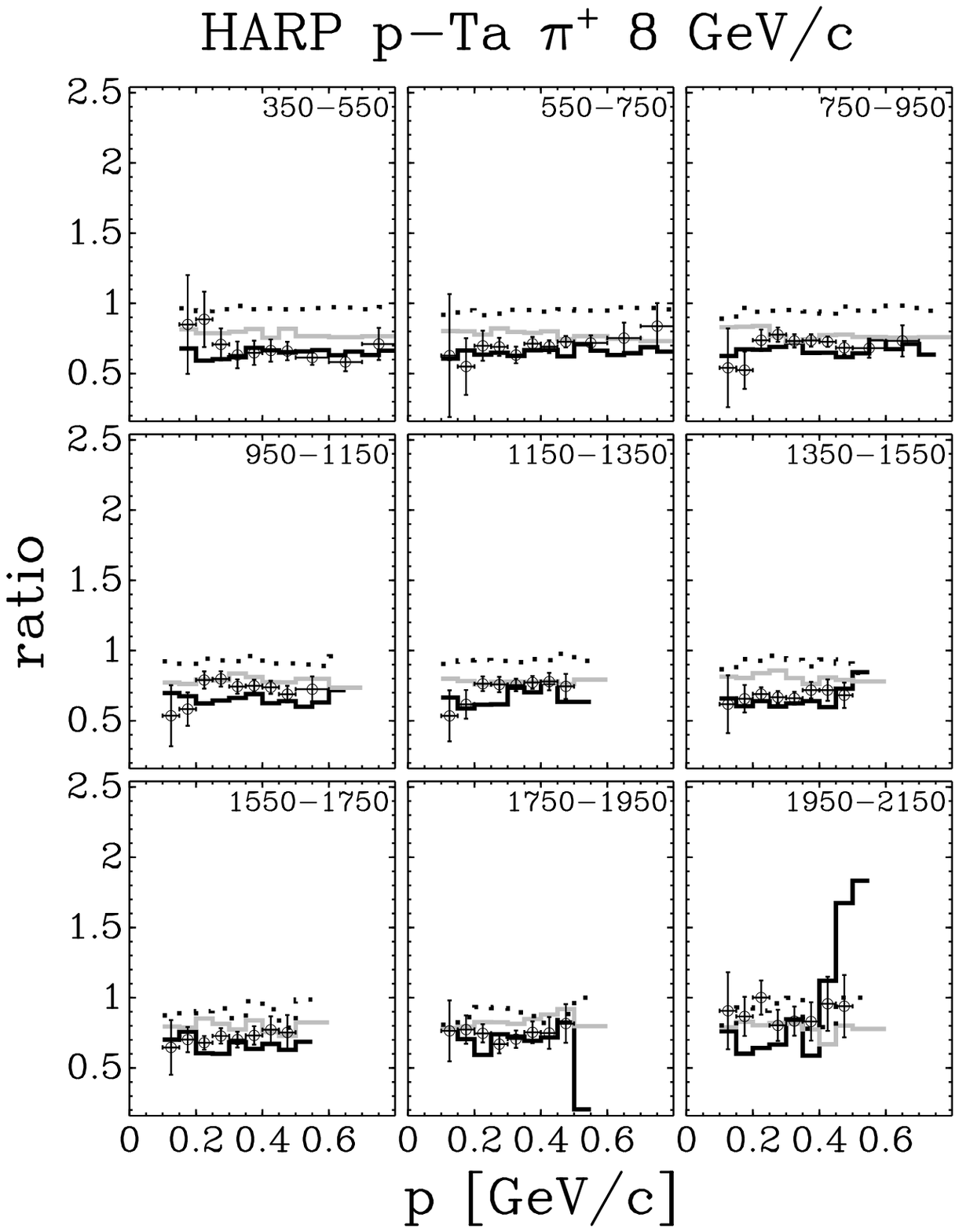}
  \includegraphics[width=0.460\textwidth,angle=0]{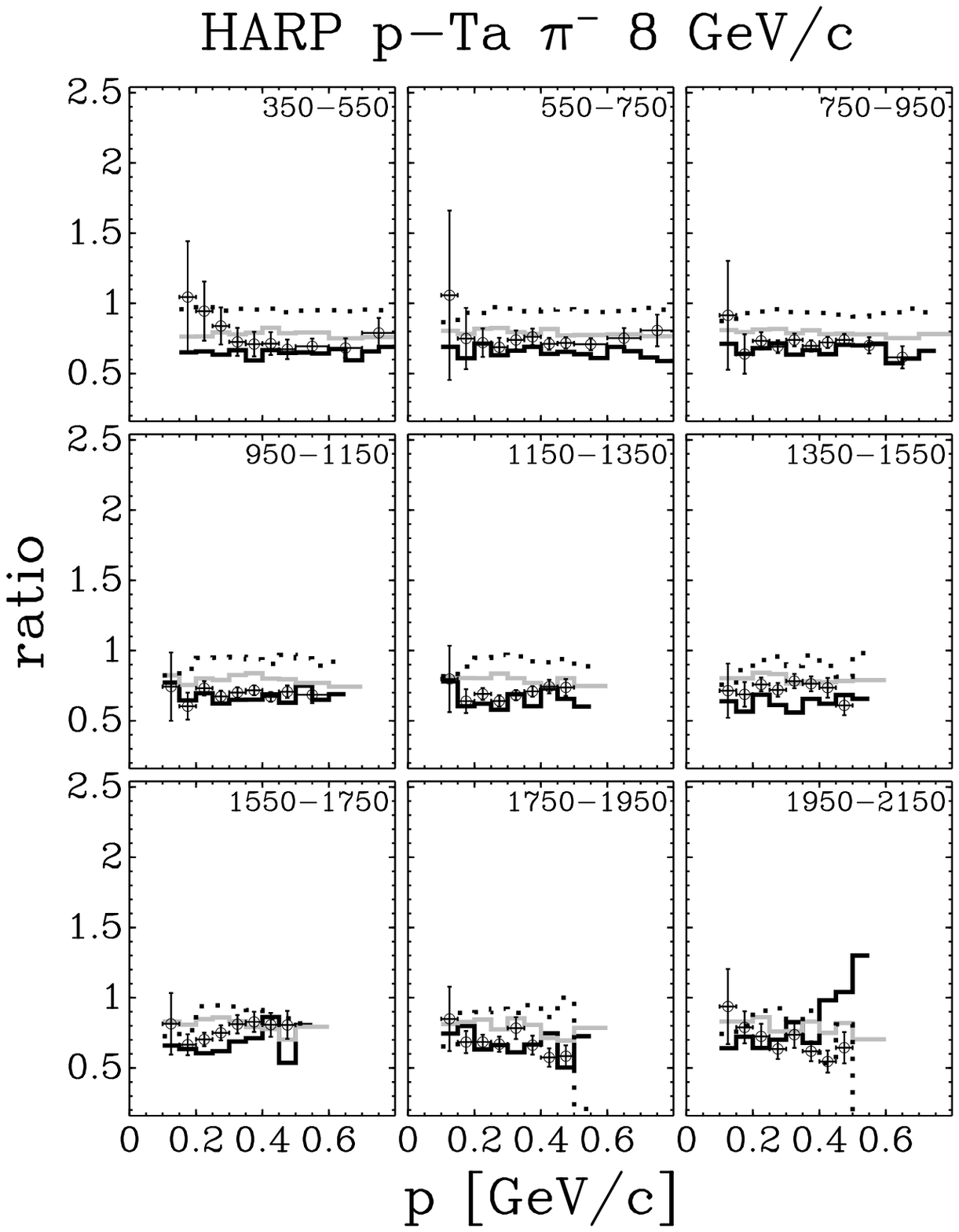}
 \end{center}
\caption{
 Comparison of the ratio of double-differential \pip(left) and \pim (right)
 production yields per target nucleon for p--Ta at 8~\GeVc taken with the
 100\%~\intlen and 5\%~\intlen target (circles) with MC predictions.
 The black curve represents the MARS prediction and the grey curve the
 GEANT4 simulation, while the dotted line
 shows the ratio of pions produced by a ``first generation'' beam proton
 and all pions as calculated by MARS.
}
\label{fig:modelTa8}
\end{sidewaysfigure}

\begin{sidewaysfigure}[tbp!]
 \begin{center}
  \includegraphics[width=0.460\textwidth,angle=0]{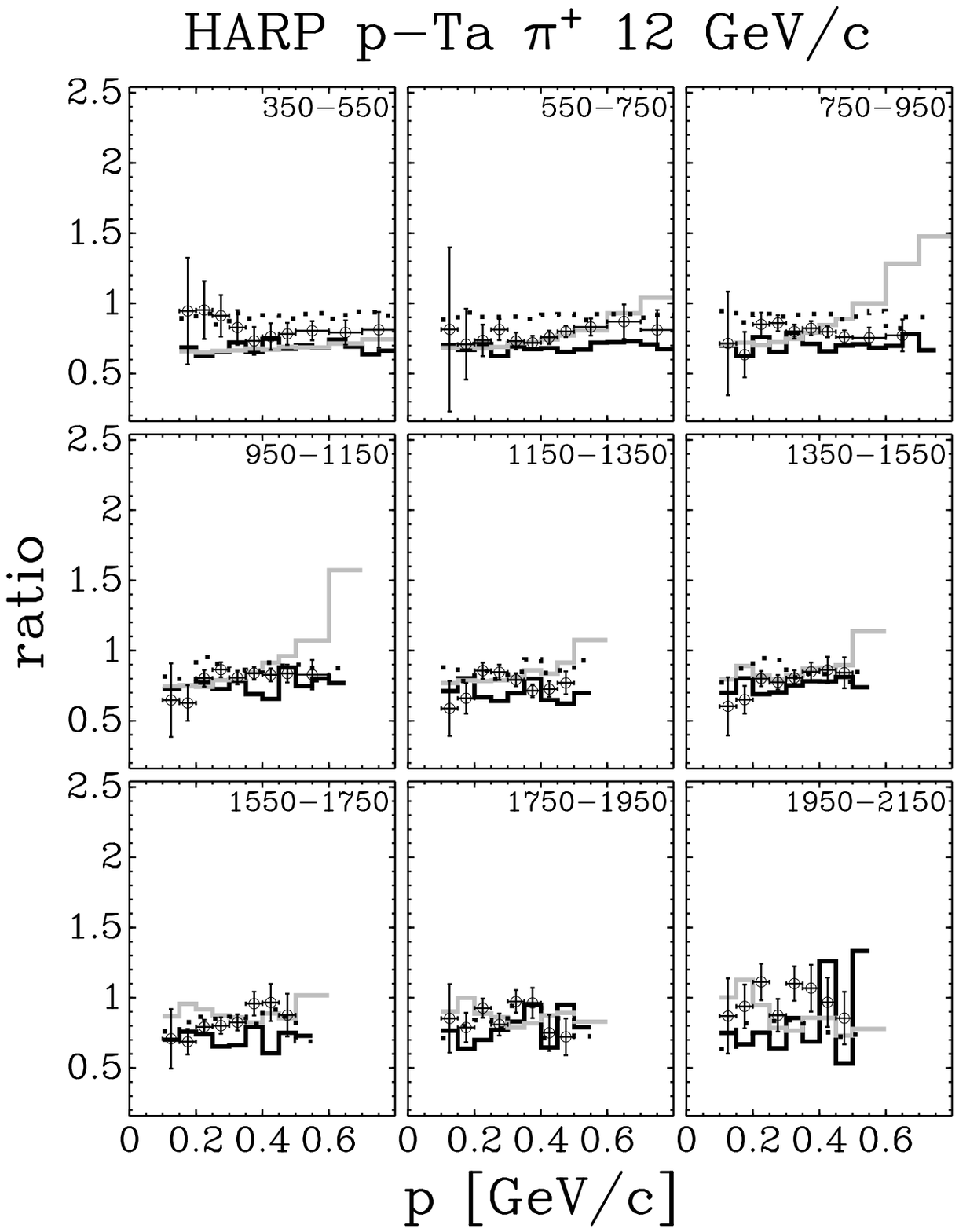}
  \includegraphics[width=0.460\textwidth,angle=0]{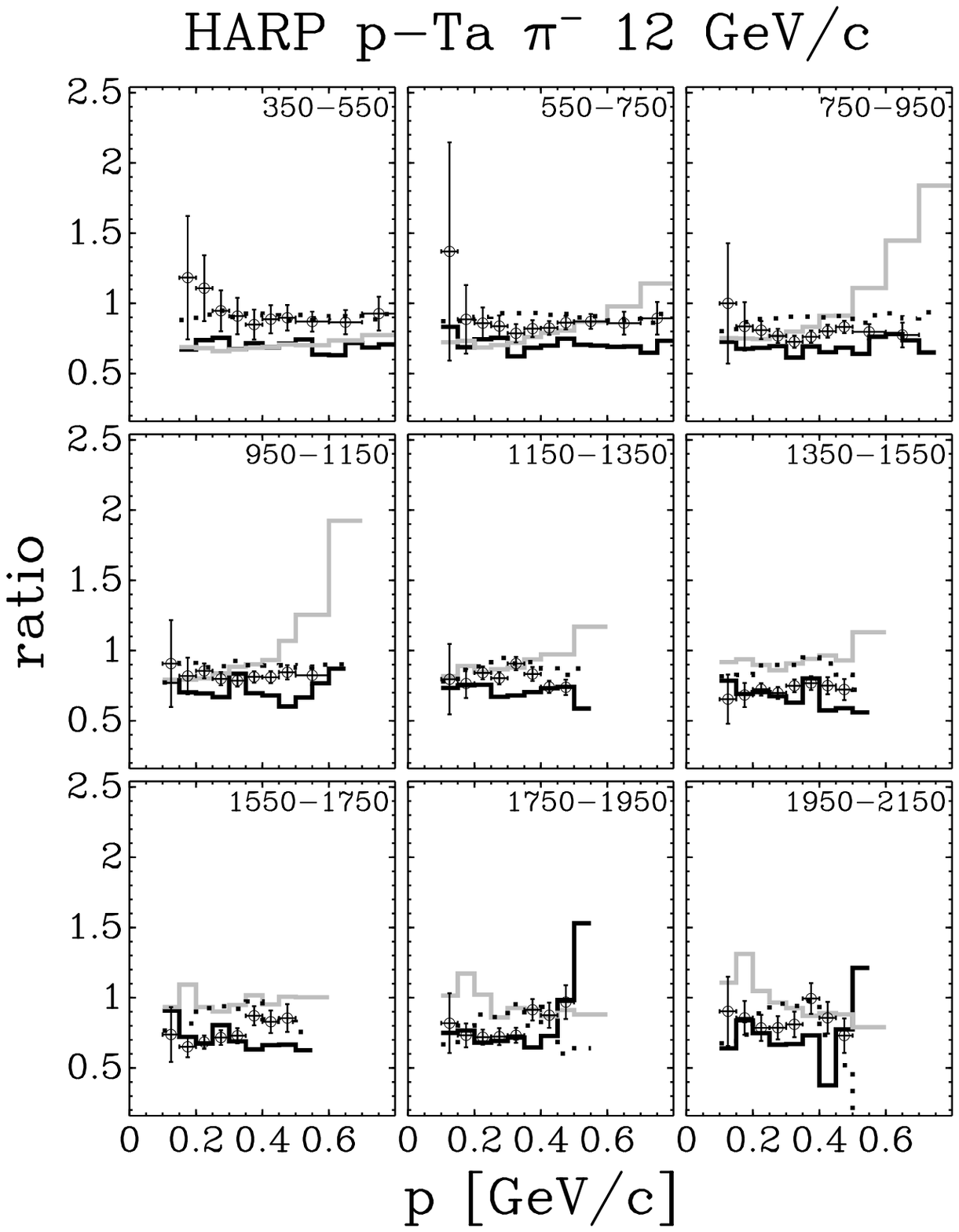}
 \end{center}
\caption{
 Comparison of the ratio of double-differential \pip(left) and \pim (right)
 production yields per target nucleon for p--Ta at 12~\GeVc taken with the
 100\%~\intlen and 5\%~\intlen target (circles) with MC predictions.
 The black curve represents the MARS prediction and the grey curve the
 GEANT4 simulation, while the dotted line
 shows the ratio of pions produced by a ``first generation'' beam proton
 and all pions as calculated by MARS.
}
\label{fig:modelTa12}
\end{sidewaysfigure}

\begin{sidewaysfigure}[tbp!]
 \begin{center}
  \includegraphics[width=0.460\textwidth,angle=0]{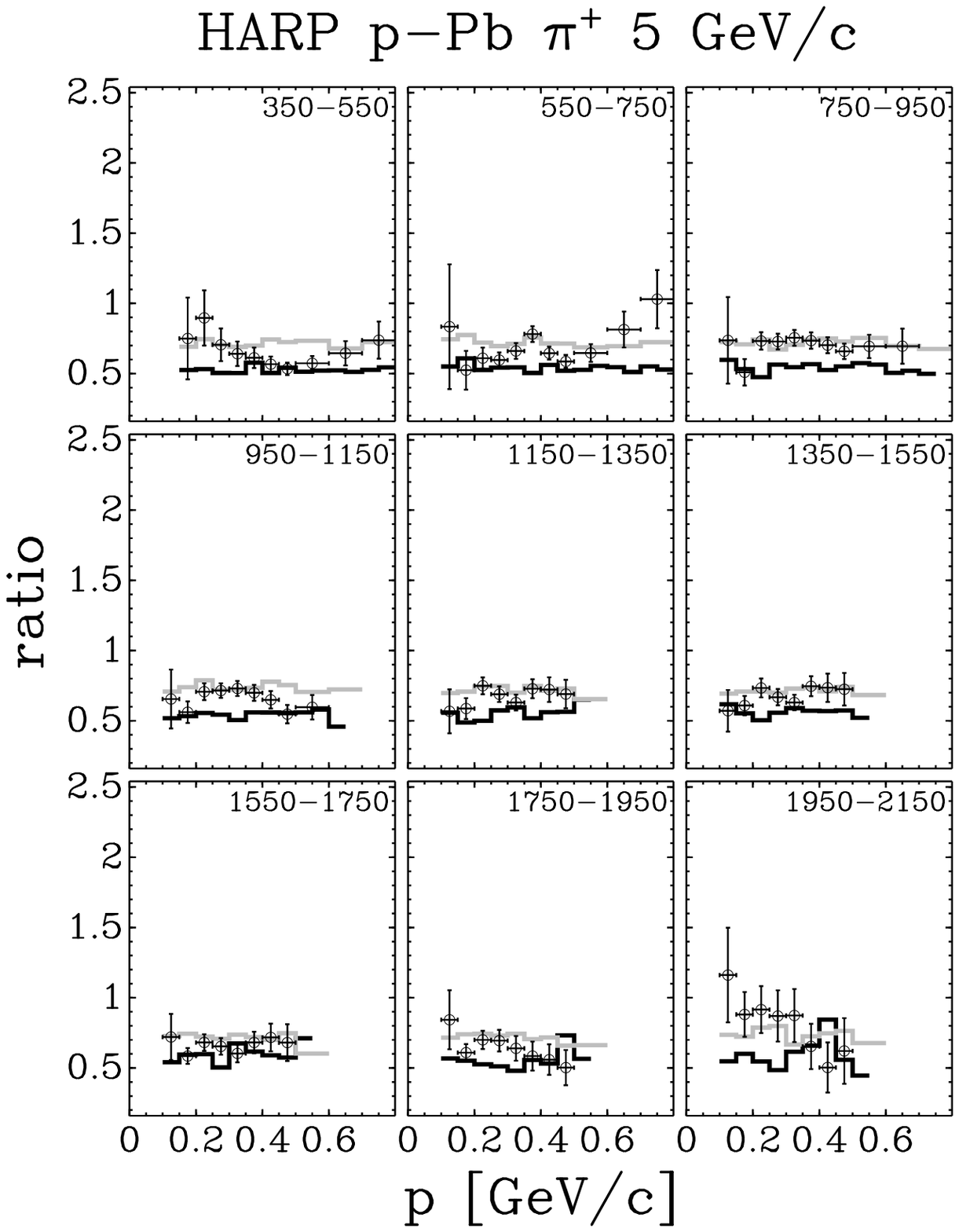}
  \includegraphics[width=0.460\textwidth,angle=0]{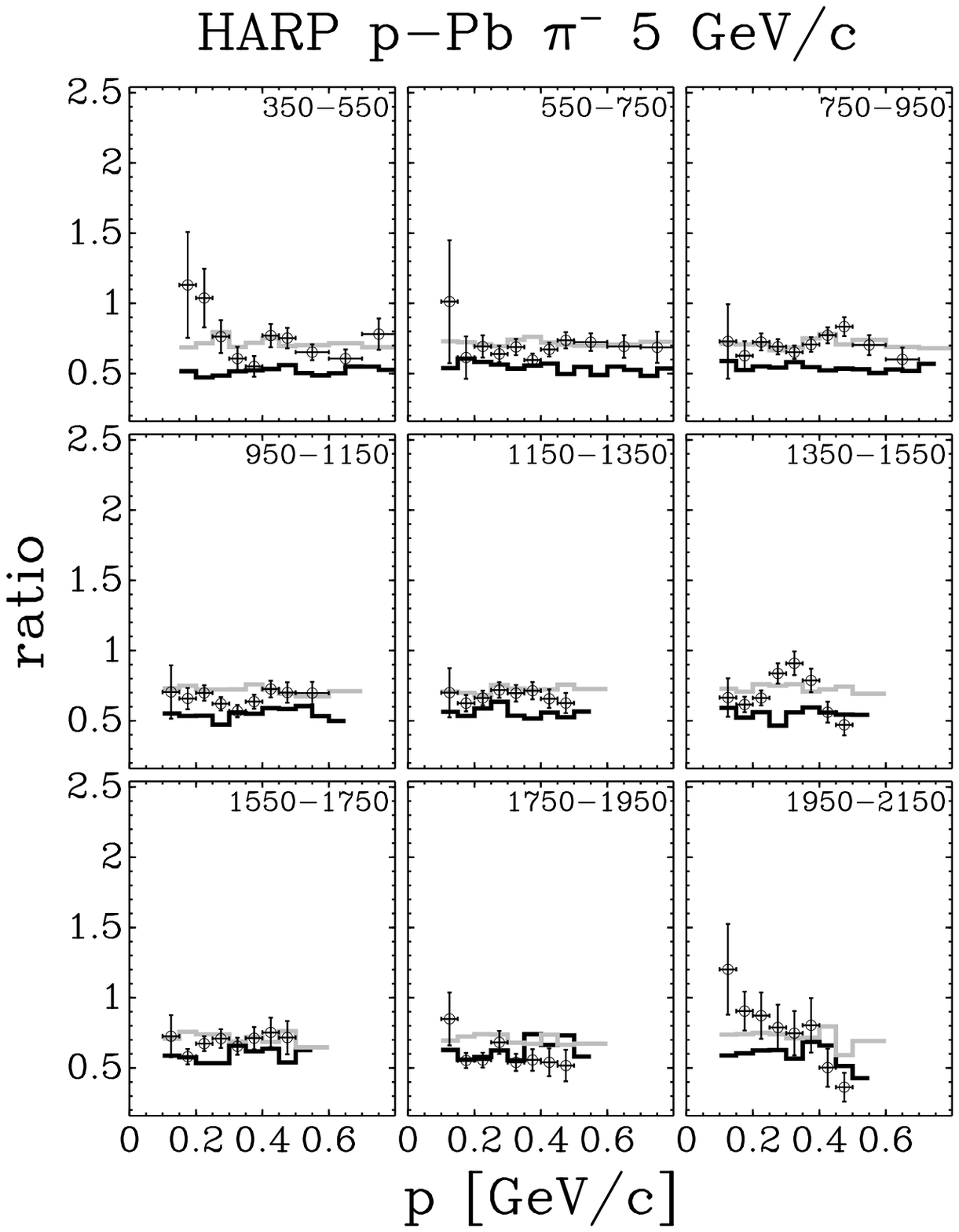}
 \end{center}
\caption{
 Comparison of the ratio of double-differential \pip(left) and \pim (right)
 production yields per target nucleon for p--Pb at 5~\GeVc taken with the
 100\%~\intlen and 5\%~\intlen target (circles) with MC predictions.
 The black curve represents the MARS prediction and the grey curve the
 GEANT4 simulation.
}
\label{fig:modelPb5}
\end{sidewaysfigure}

\begin{sidewaysfigure}[tbp!]
 \begin{center}
  \includegraphics[width=0.460\textwidth,angle=0]{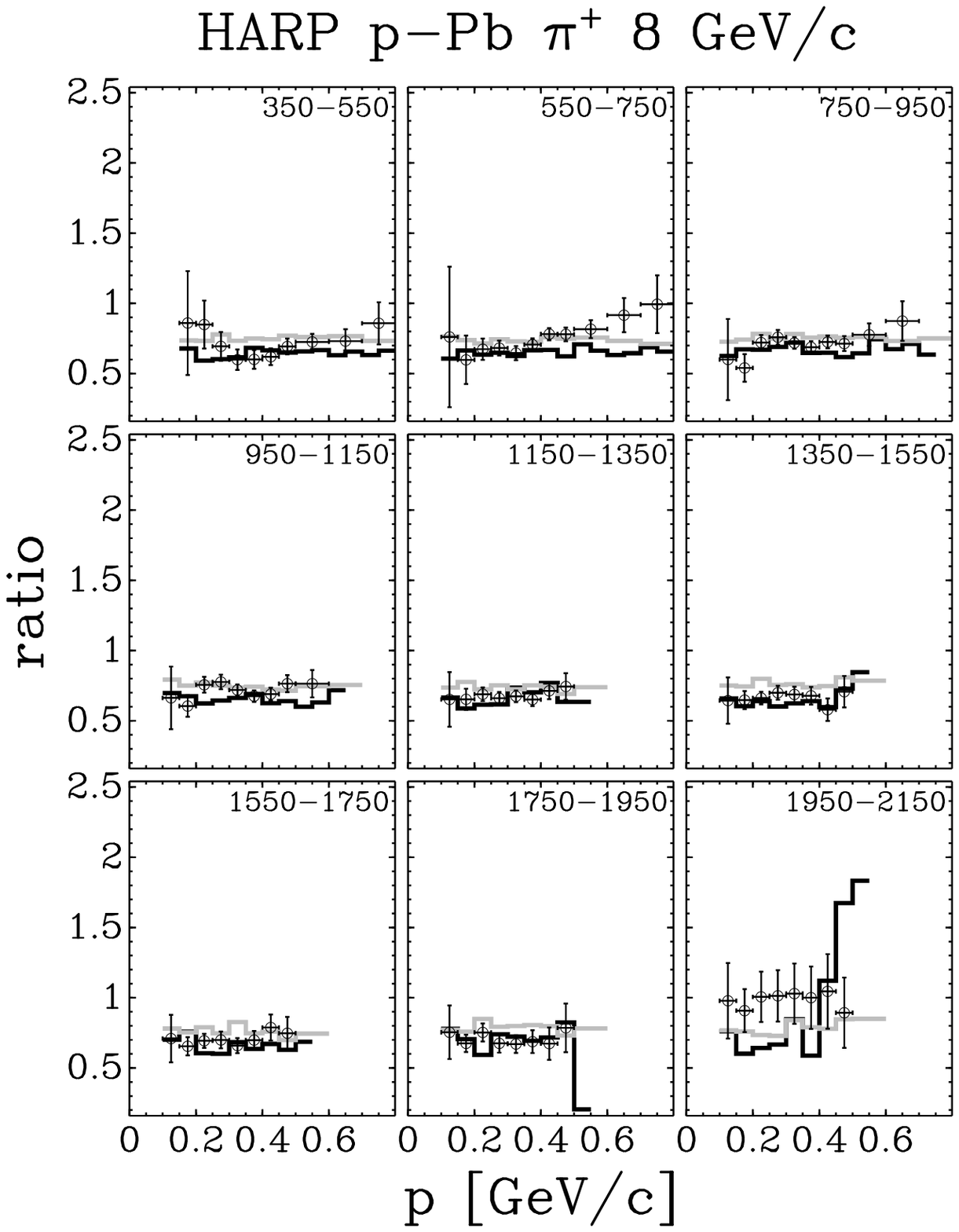}
  \includegraphics[width=0.460\textwidth,angle=0]{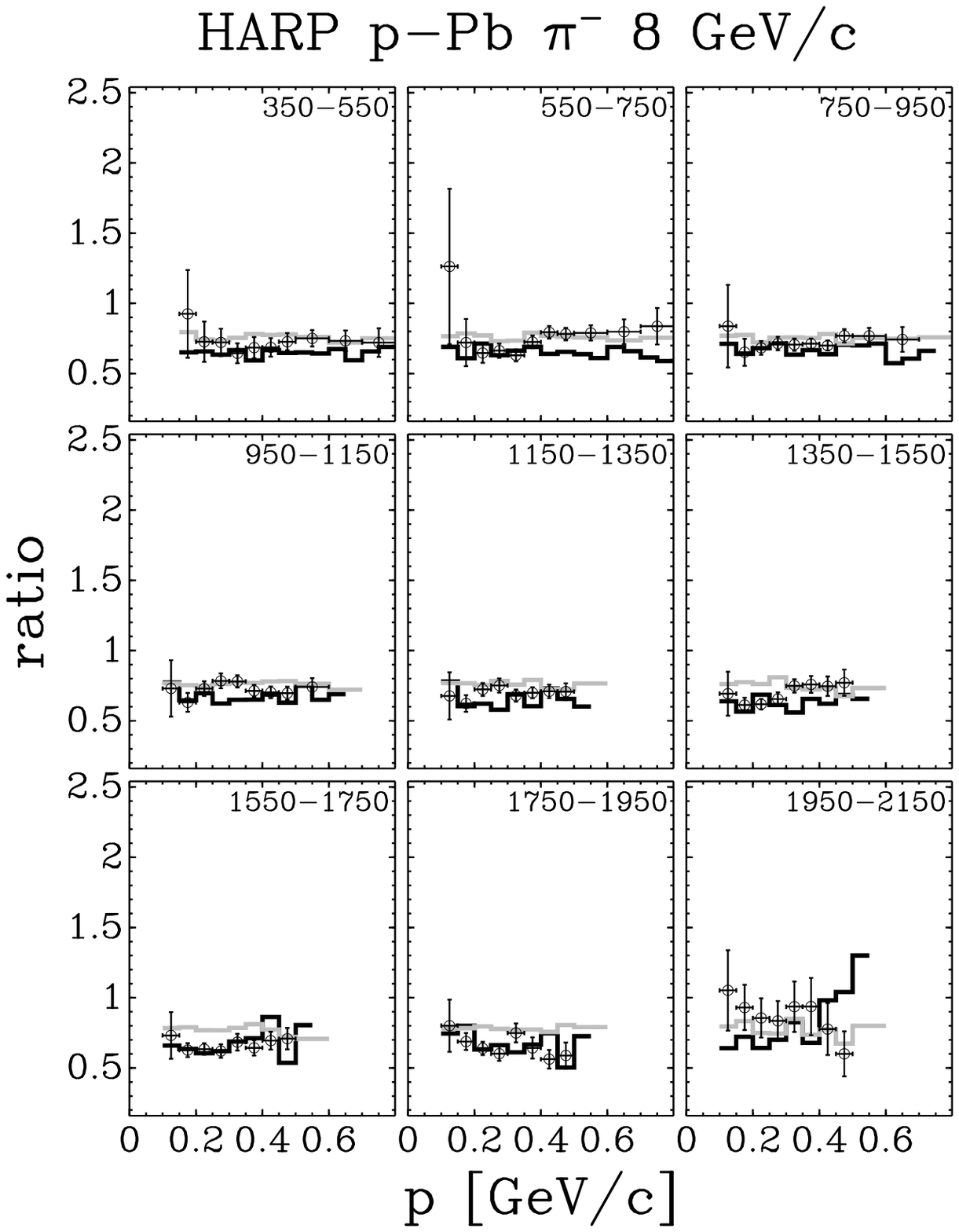}
 \end{center}
\caption{
 Comparison of the ratio of double-differential \pip(left) and \pim (right)
 production yields per target nucleon for p--Pb at 8~\GeVc taken with the
 100\%~\intlen and 5\%~\intlen target (circles) with MC predictions.
 The black curve represents the MARS prediction and the grey curve the
 GEANT4 simulation.
}
\label{fig:modelPb8}
\end{sidewaysfigure}

\begin{sidewaysfigure}[tbp!]
 \begin{center}
  \includegraphics[width=0.460\textwidth,angle=0]{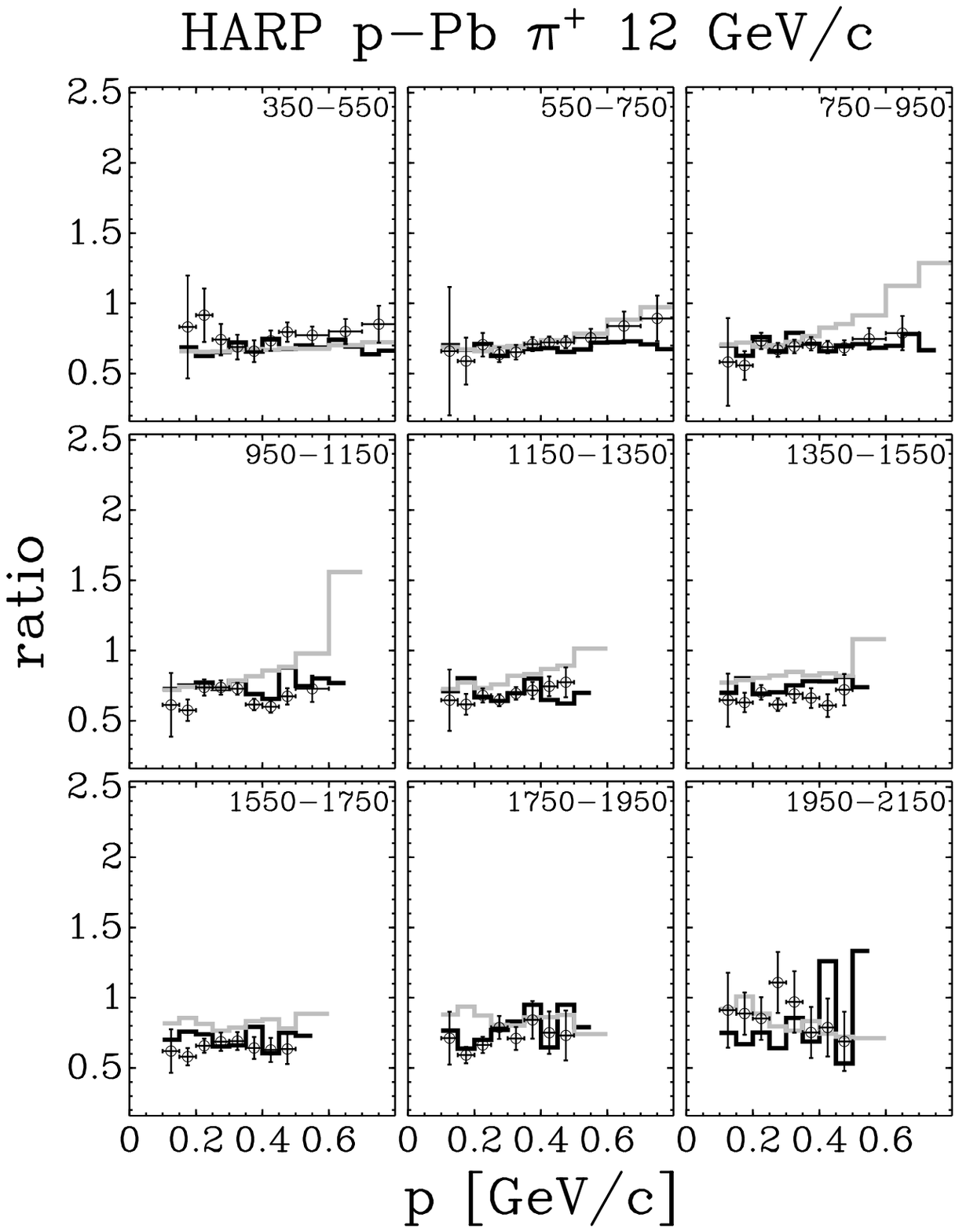}
  \includegraphics[width=0.460\textwidth,angle=0]{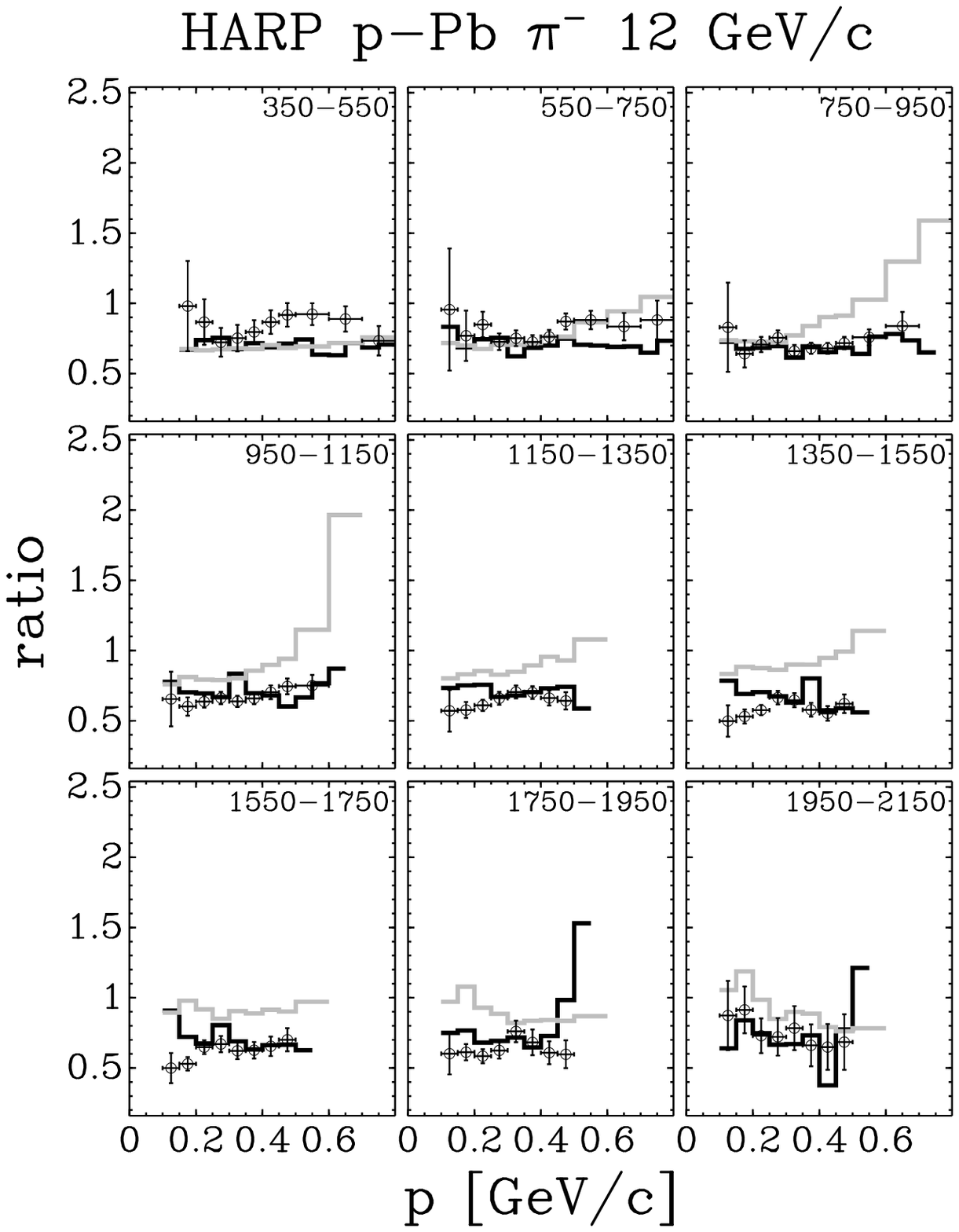}
 \end{center}
\caption{
 Comparison of the ratio of double-differential \pip(left) and \pim (right)
 production yields per target nucleon for p--Pb at 12~\GeVc taken with the
 100\%~\intlen and 5\%~\intlen target (circles) with MC predictions.
 The black curve represents the MARS prediction and the grey curve the
 GEANT4 simulation.
\label{fig:modelPb12}
}
\end{sidewaysfigure}

Our final results, obtained using the long targets can be compared 
with the previously presented data on short targets~\cite{ref:harp:la}.
We stress here that the data sets using the two sets of
targets have been analysed with methods which are as equal as
possible. 
The results are shown in  Figs.~\ref{fig:modelC5}--\ref{fig:modelPb12}.
The error bars are an estimate of the uncertainties of the ratios,
taking into account the correlation of the common errors, such as
momentum scale and assumptions on background subtraction.

The ratio of the long over short-target data can reveal the effect of
the ``degrading'' of the incoming beam particles in the target.
If all interacting protons would be absorbed in each interaction, the
ratio would be roughly 0.65~\footnote{This follows from the ratio of the
integral of the exponential degrading of the beam for 1\intlen and 0.05\intlen
targets, $\frac{0.05}{1}\frac{1-\exp(-1)}{1-\exp(-0.05)}$}.
If, on the other hand, the beam particles would continue unchanged, the
ratio would be unity.
One observes that in most cases the value of the ratio stays between the
two limits mentioned above (0.65--1.0).
It is closer to unity for the C data, especially at 12~\GeVc.
In some part of the phase-space the C data displays a ratio above one. 
The Ta and Pb data are closer to 0.65.
The behaviour as function of target material is perhaps understood by the fact that the
p--C pion production cross-section at large angles is nearly independent
of incoming momentum between 5~\GeVc and 12~\GeVc while the production
cross-sections increase with beam momentum in this momentum range for
heavier nuclei, as shown in Ref.~\cite{ref:harp:la}. 
Thus for p--C forward scattered protons with lower momentum are almost as
effective in producing pions as the original beam proton.
This is not the case for the heavier nuclei.

In Figs.~\ref{fig:modelC5}--\ref{fig:modelPb12} we also show comparisons of
the ratio of the long-target data and the corresponding
$\intlen=5\%$ targets with 
a selected list of Monte Carlo generators of GEANT4~\cite{ref:geant4} and
the MARS~\cite{ref:mars} model. 
When a range of incident energies has to be simulated, GEANT4 offers the
possibility to define a ``physics list'' of models allowing the user to
describe the different energies with optimized models.
The physics list defined for the present comparisons uses an
intra-nuclear cascade model (the ``Bertini
model''~\cite{ref:bert,ref:bert1}) for the lower energies.
The Bertini model is based on the cascade code reported in \cite{ref:bert2}
and hadron collisions are assumed to proceed according to free-space partial
cross sections corrected for nuclear field effects and final state
distributions measured for the incident particle types.
At higher energies, instead, a parton string model
(QGSP)~\cite{ref:QGSP} is used.
The MARS code system~\cite{ref:mars} uses as basic model an inclusive
approach multi-particle production originated by R. Feynman. Above 5~GeV
phenomenological particle production models are used, while below 5~GeV
a cascade-exciton model~\cite{ref:casca} combined with the Fermi
break-up model, the coalescence model, an evaporation model and a
multi-fragmentation extension are used instead.  

An extra curve is shown for the C and Ta data 
representing the fraction
of pions produced by the original (``first generation'') beam proton
compared to the overall number of pions produced by the beam.
More explicitly, we repeat here that all particles produced by the
primary beam proton within a forward cone with half-angle 0.2~\rad are
regarded as ``beam particles''.

The comparison, between data and models is good for the heavier nuclei, 
but discrepancies for carbon are visible for the comparison with MARS.
Here MARS predicts a lower ratio than observed experimentally, while
GEANT4 provides a more accurate description of the data.
From an inspection of the line representing the ratio of pions produced by
``beam particles'' and all produced pions as calculated by MARS 
it appears that in the simulation the large majority of secondary pions
are generated by the primary protons of the beam.

\begin{sidewaysfigure}[tbp!]
 \begin{center}
  \includegraphics[width=0.30\textwidth,angle=0]{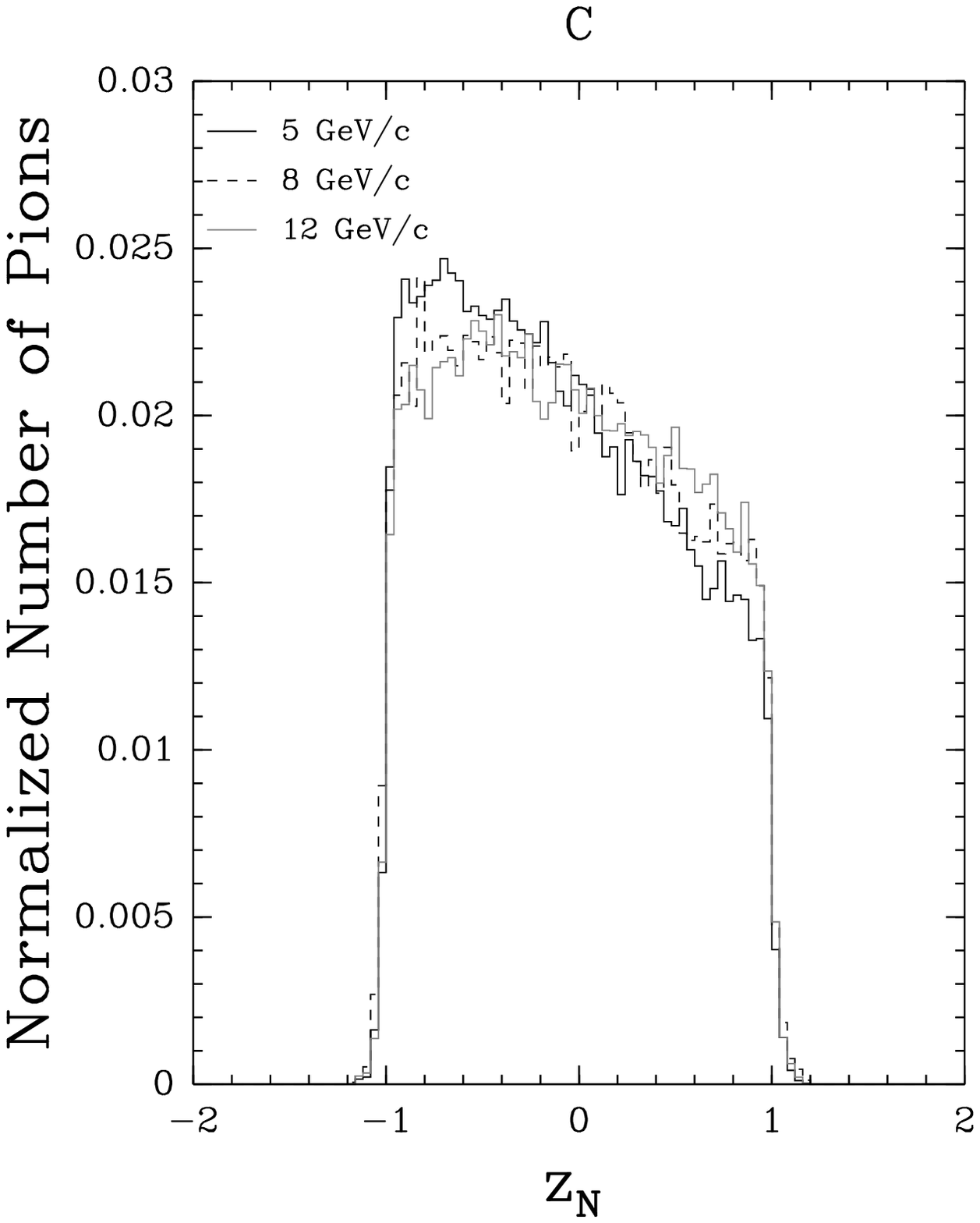}
  ~
  \includegraphics[width=0.30\textwidth,angle=0]{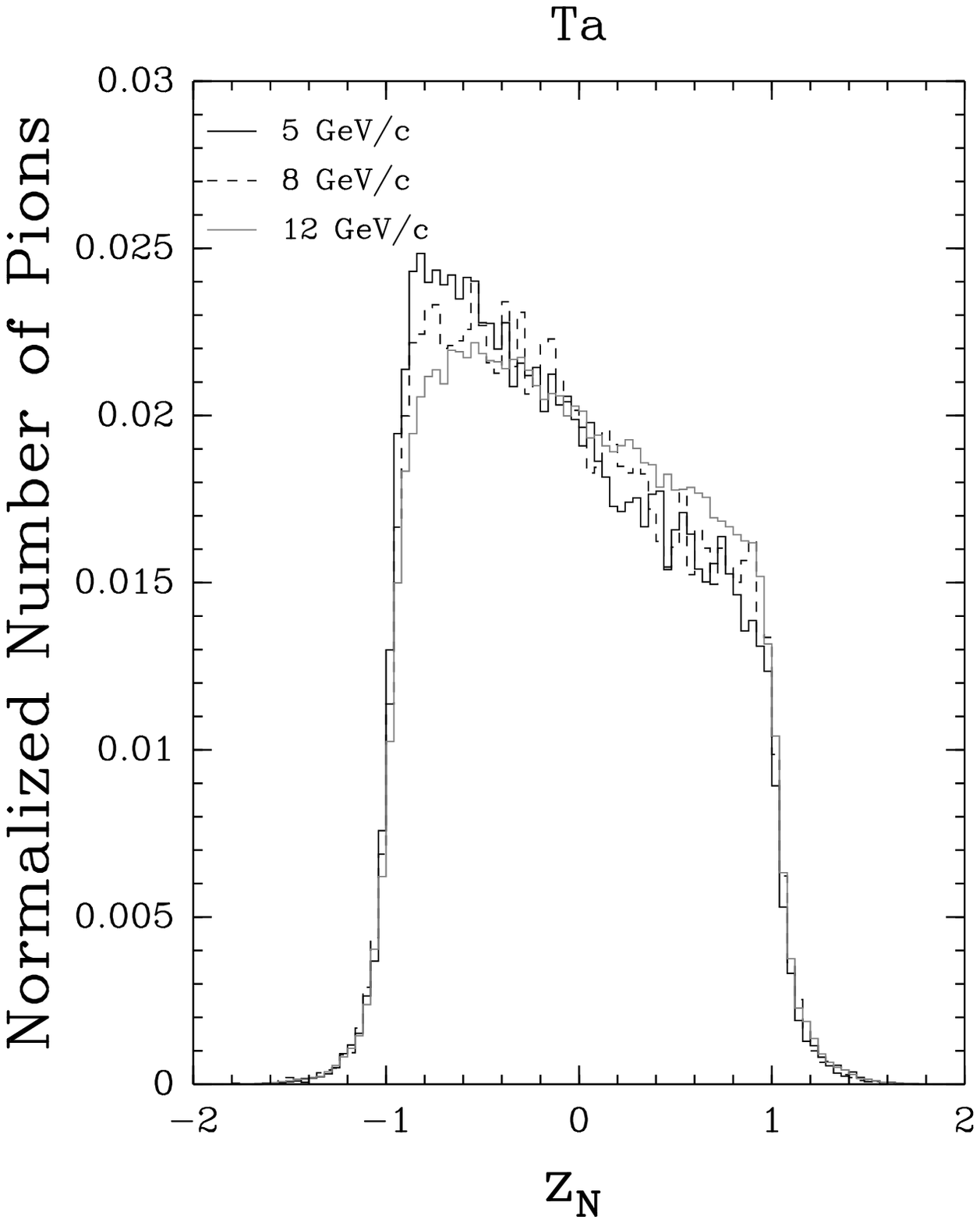}
  ~
  \includegraphics[width=0.30\textwidth,angle=0]{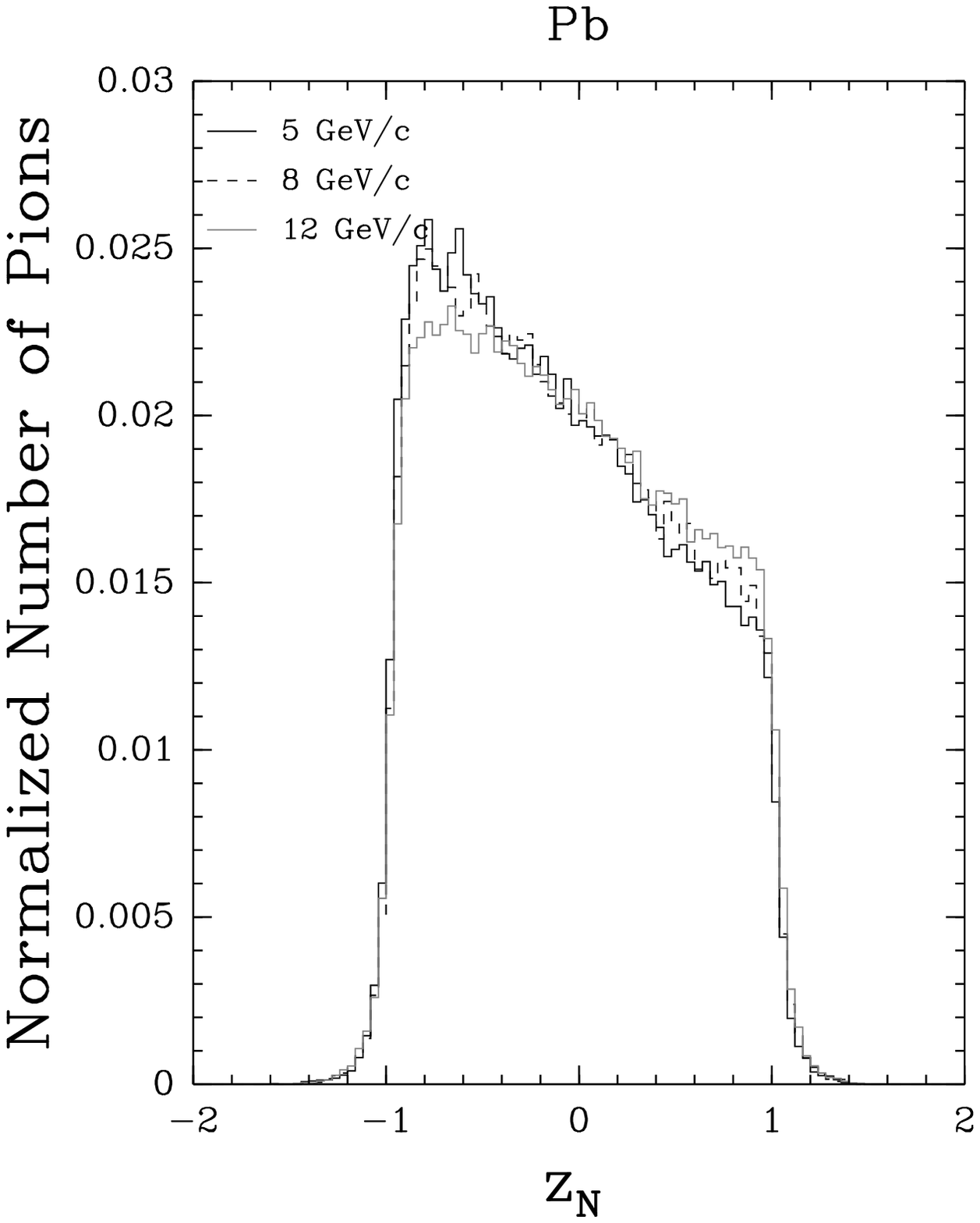}
 \end{center}
\caption{
 The raw distributions for \pipm production in
 p--C (left panel), p--Ta (middle panel) and p--Pb (right panel) 
 interactions as a function of $z_\mathrm{N}$ where all pions within the
 kinematical cuts of the analysis are counted without any correction.
 The results are given for three incident-beam momenta (solid black line:
 5~\GeVc; dashed line: 8~\GeVc; grey line: 12~\GeVc).
 The distributions are all normalized to unit area, and $z_\mathrm{N}$ is the
 normalized depth in the target which starts at -1 at the front of the
 target and ends at +1 at the back of the target.
}
\label{fig:znorm}
\end{sidewaysfigure}

The ratio increases for all target materials with increasing beam
momentum. 
This trend can be investigated by looking at the distribution of the
the origin of pion tracks as a function of the longitudinal coordinate
$z$. 
Figure~\ref{fig:znorm} shows these distributions as a function of
normalized $z$, $z_\mathrm{N}$, which runs from -1 to +1 over the length of the
target. 
We define $z$ as the value of the longitudinal coordinate of the point of closest
approach of the secondary pion track and the beam particle~\footnote{$z$
is measured using the track parameters after dynamic distortion corrections.}. 
The slopes of the $z_\mathrm{N}$ distributions are steeper for the lower beam
momenta. 
One also observes that there is a shower effect for the higher beam
momenta, i.e. the maximum is not at the very beginning of the target
indicating that some of the secondary particles are still efficient in
producing high angle pions.
Since the resolution is constant in absolute space coordinates, the
resolution tail is longer for the higher density targets in the
variable $z_\mathrm{N}$.




\afterpage{\clearpage}

\section{Summary and Conclusions}
\label{sec:summary}

An analysis of the production of pions at
large angles with respect to the beam direction for incoming protons of
 5~\GeVc, 8~\GeVc and 12~\GeVc  beam momentum impinging on long (100\%
 interaction length) carbon, tantalum and lead targets is described.   
The secondary pion yield is measured in a large angular and momentum
range and {\em effective} double-differential cross-sections are obtained.
The present measurements are important to understand the simulations
needed to design realistic pion production targets, e.g. for a neutrino
factory~\cite{ref:iss}. 
The choice has been made to correct the absorption and re-interaction of
secondary pions to make the results more universally usable for
different target geometries.
Thus the effect of the long target on the primary beam is highlighted.

The data are compared with corresponding results obtained with thin
(5\%~\intlen) targets using the same detector and analysis techniques.
It is observed that the effective attenuation of the incoming particle
beam is larger for heavier targets than for carbon.
This effect may have its origin in the stronger energy dependence of the
production cross-section for the high-$A$ targets.

The hadronic production models used for the comparison of the ratio
compare well with the data for tantalum and lead.
The GEANT4 model provides a good description for carbon, while MARS does
not describe the relatively high ratio observed in the carbon data.

\section{Acknowledgements}

We gratefully acknowledge the help and support of the PS beam staff
and of the numerous technical collaborators who contributed to the
detector design, construction, commissioning and operation.  
In particular, we would like to thank
G.~Barichello,
R.~Brocard,
K.~Burin,
V.~Carassiti,
F.~Chignoli,
D.~Conventi,
G.~Decreuse,
M.~Delattre,
C.~Detraz,  
A.~Domeniconi,
M.~Dwuznik,   
F.~Evangelisti,
B.~Friend,
A.~Iaciofano,
I.~Krasin, 
D.~Lacroix,
J.-C.~Legrand,
M.~Lobello, 
M.~Lollo,
J.~Loquet,
F.~Marinilli,
J.~Mulon,
L.~Musa,
R.~Nicholson,
A.~Pepato,
P.~Petev, 
X.~Pons,
I.~Rusinov,
M.~Scandurra,
E.~Usenko,
and
R.~van der Vlugt,
for their support in the construction of the detector.
The collaboration acknowledges the major contributions and advice of
M.~Baldo-Ceolin, 
L.~Linssen, 
M.T.~Muciaccia and A. Pullia
during the construction of the experiment.
The collaboration is indebted to 
V.~Ableev,
P.~Arce,   
F.~Bergsma,
P.~Binko,
E.~Boter,
C.~Buttar, 
M.~Calvi, 
M.~Campanelli, 
C.~Cavion, 
A.~Chukanov, 
A.~De~Min,  
M.~Doucet,
D.~D\"{u}llmann,
R.~Engel,  
V.~Ermilova, 
W.~Flegel,
P.~Gruber,  
Y.~Hayato,
P.~Hodgson, 
A.~Ichikawa,
I.~Kato, 
O.~Klimov,
T.~Kobayashi,
D.~Kustov,
M.~Laveder,  
M.~Mass,
H.~Meinhard,
T.~Nakaya,
K.~Nishikawa,
M.~Paganoni,  
F.~Paleari,  
M.~Pasquali,
J.~Pasternak, 
C.~Pattison, 
M.~Placentino,
S.~Robbins, 
G.~Santin,  
V.~Serdiouk,
S.~Simone,
A.~Tornero, 
S.~Troquereau,
S.~Ueda, 
A.~Valassi,
F.~Vannucci 
and
K.~Zuber   
for their contributions to the experiment
and to P. Dini for help in MC production. 

We acknowledge the contributions of 
V.~Ammosov,
G.~Chelkov,
D.~Dedovich,
F.~Dydak,
M.~Gostkin,
A.~Guskov, 
D.~Khartchenko, 
V.~Koreshev,
Z.~Kroumchtein,
I.~Nefedov,
A.~Semak, 
J.~Wotschack,
V.~Zaets and
A.~Zhemchugov
to the work described in this paper.

 The experiment was made possible by grants from
the Institut Interuniversitaire des Sciences Nucl\'eair\-es and the
Interuniversitair Instituut voor Kernwetenschappen (Belgium), 
Ministerio de Educacion y Ciencia, Grant FPA2003-06921-c02-02 and
Generalitat Valenciana, grant GV00-054-1,
CERN (Geneva, Switzerland), 
the German Bundesministerium f\"ur Bildung und Forschung (Germany), 
the Istituto Na\-zio\-na\-le di Fisica Nucleare (Italy), 
INR RAS (Moscow), the Russian Foundation for Basic Research (grant 08-02-00018)
and the Particle Physics and Astronomy Research Council (UK).
We gratefully acknowledge their support.
This work was supported in part by the Swiss National Science Foundation
and the Swiss Agency for Development and Cooperation in the framework of
the programme SCOPES - Scientific co-operation between Eastern Europe
and Switzerland.



\clearpage

\begin{appendix}

\section{Pion production yield data}
\label{app:data}
\begin{table}[hp!]
\begin{center}
  \caption{\label{tab:xsec-pip-c}
    HARP results for the double-differential $\pi^+$ production
    yield per target nucleon in the laboratory system,
    $\mathrm{d}^2\sigma^{\pi^+}/(\mathrm{d}p\mathrm{d}\theta)$ for p--C interactions. Each row refers to a
    different $(p_{\hbox{\small min}} \le p<p_{\hbox{\small max}},
    \theta_{\hbox{\small min}} \le \theta<\theta_{\hbox{\small max}})$ bin,
    where $p$ and $\theta$ are the pion momentum and polar angle, respectively.
    The central value as well as the square-root of the diagonal elements
    of the covariance matrix are given.}
\vspace{2mm}
\begin{tabular}{rrrr|r@{$\pm$}lr@{$\pm$}lr@{$\pm$}l}
\hline
$\theta_{\hbox{\small min}}$ &
$\theta_{\hbox{\small max}}$ &
$p_{\hbox{\small min}}$ &
$p_{\hbox{\small max}}$ &
\multicolumn{6}{c}{$\mathrm{d}^2\sigma^{\pi^+}/(\mathrm{d}p\mathrm{d}\theta)$}\\
(rad) & (rad) & (\GeVc) & (\GeVc) &
\multicolumn{6}{c}{($\barn/(\GeVc \cdot \rad)$)}\\
  &  &  &
&\multicolumn{2}{c}{\GeVc}
&\multicolumn{2}{c}{\GeVc}
&\multicolumn{2}{c}{\GeVc}
\\
\hline
 0.35 & 0.55 & 0.15 & 0.20& 0.126 &  0.028& 0.121 &  0.031& 0.164 &  0.041\\ 
      &      & 0.20 & 0.25& 0.156 &  0.019& 0.215 &  0.027& 0.239 &  0.031\\ 
      &      & 0.25 & 0.30& 0.187 &  0.019& 0.232 &  0.023& 0.294 &  0.029\\ 
      &      & 0.30 & 0.35& 0.199 &  0.020& 0.275 &  0.026& 0.295 &  0.032\\ 
      &      & 0.35 & 0.40& 0.210 &  0.018& 0.269 &  0.024& 0.333 &  0.029\\ 
      &      & 0.40 & 0.45& 0.221 &  0.017& 0.287 &  0.022& 0.342 &  0.028\\ 
      &      & 0.45 & 0.50& 0.223 &  0.013& 0.297 &  0.021& 0.368 &  0.027\\ 
      &      & 0.50 & 0.60& 0.213 &  0.015& 0.295 &  0.020& 0.371 &  0.025\\ 
      &      & 0.60 & 0.70& 0.178 &  0.021& 0.264 &  0.029& 0.343 &  0.037\\ 
      &      & 0.70 & 0.80& 0.119 &  0.023& 0.210 &  0.032& 0.279 &  0.044\\ 
\hline  
 0.55 & 0.75 & 0.10 & 0.15& 0.086 &  0.028& 0.087 &  0.029& 0.096 &  0.037\\ 
      &      & 0.15 & 0.20& 0.140 &  0.012& 0.170 &  0.019& 0.198 &  0.019\\ 
      &      & 0.20 & 0.25& 0.187 &  0.018& 0.224 &  0.019& 0.273 &  0.030\\ 
      &      & 0.25 & 0.30& 0.203 &  0.017& 0.265 &  0.021& 0.307 &  0.024\\ 
      &      & 0.30 & 0.35& 0.197 &  0.014& 0.255 &  0.019& 0.297 &  0.023\\ 
      &      & 0.35 & 0.40& 0.189 &  0.014& 0.237 &  0.013& 0.280 &  0.017\\ 
      &      & 0.40 & 0.45& 0.192 &  0.012& 0.233 &  0.012& 0.293 &  0.017\\ 
      &      & 0.45 & 0.50& 0.160 &  0.010& 0.224 &  0.014& 0.264 &  0.019\\ 
      &      & 0.50 & 0.60& 0.127 &  0.011& 0.183 &  0.015& 0.228 &  0.017\\ 
      &      & 0.60 & 0.70& 0.087 &  0.013& 0.138 &  0.018& 0.175 &  0.024\\ 
      &      & 0.70 & 0.80& 0.050 &  0.011& 0.097 &  0.019& 0.115 &  0.024\\ 
\hline  
 0.75 & 0.95 & 0.10 & 0.15& 0.100 &  0.022& 0.092 &  0.021& 0.097 &  0.024\\ 
      &      & 0.15 & 0.20& 0.153 &  0.012& 0.190 &  0.017& 0.218 &  0.023\\ 
      &      & 0.20 & 0.25& 0.188 &  0.017& 0.213 &  0.019& 0.233 &  0.019\\ 
      &      & 0.25 & 0.30& 0.178 &  0.010& 0.222 &  0.015& 0.252 &  0.024\\ 
      &      & 0.30 & 0.35& 0.154 &  0.009& 0.189 &  0.010& 0.236 &  0.013\\ 
      &      & 0.35 & 0.40& 0.141 &  0.007& 0.179 &  0.009& 0.223 &  0.011\\ 
      &      & 0.40 & 0.45& 0.109 &  0.006& 0.160 &  0.009& 0.189 &  0.011\\ 
      &      & 0.45 & 0.50& 0.092 &  0.007& 0.132 &  0.008& 0.160 &  0.010\\ 
      &      & 0.50 & 0.60& 0.067 &  0.007& 0.095 &  0.010& 0.120 &  0.012\\ 
      &      & 0.60 & 0.70& 0.038 &  0.007& 0.062 &  0.010& 0.073 &  0.013\\ 
\hline  
 0.95 & 1.15 & 0.10 & 0.15& 0.086 &  0.015& 0.110 &  0.021& 0.115 &  0.021\\ 
      &      & 0.15 & 0.20& 0.158 &  0.017& 0.199 &  0.019& 0.210 &  0.022\\ 
      &      & 0.20 & 0.25& 0.166 &  0.012& 0.184 &  0.012& 0.217 &  0.015\\ 
      &      & 0.25 & 0.30& 0.140 &  0.008& 0.167 &  0.009& 0.186 &  0.012\\ 
      &      & 0.30 & 0.35& 0.107 &  0.006& 0.136 &  0.007& 0.160 &  0.009\\ 
      &      & 0.35 & 0.40& 0.092 &  0.006& 0.105 &  0.006& 0.133 &  0.008\\ 
      &      & 0.40 & 0.45& 0.072 &  0.005& 0.088 &  0.006& 0.105 &  0.008\\ 
      &      & 0.45 & 0.50& 0.051 &  0.006& 0.075 &  0.007& 0.085 &  0.007\\ 
      &      & 0.50 & 0.60& 0.029 &  0.004& 0.046 &  0.007& 0.059 &  0.007\\ 
\hline
\end{tabular}
\end{center}
\end{table}

\begin{table}[hp!]
\begin{center}
\begin{tabular}{rrrr|r@{$\pm$}lr@{$\pm$}lr@{$\pm$}l}
\hline
$\theta_{\hbox{\small min}}$ &
$\theta_{\hbox{\small max}}$ &
$p_{\hbox{\small min}}$ &
$p_{\hbox{\small max}}$ &
\multicolumn{6}{c}{${\mathrm{d}}^2\sigma^{\pi^+}/(\mathrm{d}p\mathrm{d}\theta)$}\\
(rad) & (rad) & (\GeVc) & (\GeVc) &
\multicolumn{6}{c}{(\barn/($\GeVc \cdot \rad$))}\\
  &  &  &
&\multicolumn{2}{c}{\GeVc}
&\multicolumn{2}{c}{\GeVc}
&\multicolumn{2}{c}{\GeVc}
\\
\hline
 1.15 & 1.35 & 0.10 & 0.15& 0.101 &  0.015& 0.113 &  0.020& 0.146 &  0.024\\ 
      &      & 0.15 & 0.20& 0.164 &  0.015& 0.161 &  0.019& 0.188 &  0.023\\ 
      &      & 0.20 & 0.25& 0.136 &  0.013& 0.152 &  0.011& 0.181 &  0.014\\ 
      &      & 0.25 & 0.30& 0.097 &  0.007& 0.124 &  0.009& 0.136 &  0.010\\ 
      &      & 0.30 & 0.35& 0.071 &  0.004& 0.095 &  0.007& 0.102 &  0.008\\ 
      &      & 0.35 & 0.40& 0.054 &  0.004& 0.071 &  0.005& 0.077 &  0.005\\ 
      &      & 0.40 & 0.45& 0.040 &  0.004& 0.052 &  0.004& 0.056 &  0.005\\ 
      &      & 0.45 & 0.50& 0.024 &  0.004& 0.038 &  0.005& 0.042 &  0.005\\ 
\hline  
 1.35 & 1.55 & 0.10 & 0.15& 0.102 &  0.016& 0.140 &  0.021& 0.140 &  0.022\\ 
      &      & 0.15 & 0.20& 0.138 &  0.012& 0.152 &  0.013& 0.194 &  0.016\\ 
      &      & 0.20 & 0.25& 0.102 &  0.007& 0.139 &  0.011& 0.155 &  0.012\\ 
      &      & 0.25 & 0.30& 0.072 &  0.006& 0.091 &  0.008& 0.108 &  0.009\\ 
      &      & 0.30 & 0.35& 0.044 &  0.004& 0.064 &  0.005& 0.081 &  0.006\\ 
      &      & 0.35 & 0.40& 0.031 &  0.003& 0.045 &  0.005& 0.055 &  0.006\\ 
      &      & 0.40 & 0.45& 0.023 &  0.003& 0.030 &  0.004& 0.038 &  0.005\\ 
      &      & 0.45 & 0.50& 0.015 &  0.002& 0.018 &  0.003& 0.024 &  0.004\\ 
\hline  
 1.55 & 1.75 & 0.10 & 0.15& 0.100 &  0.015& 0.122 &  0.019& 0.125 &  0.022\\ 
      &      & 0.15 & 0.20& 0.123 &  0.012& 0.141 &  0.013& 0.155 &  0.015\\ 
      &      & 0.20 & 0.25& 0.095 &  0.008& 0.107 &  0.009& 0.114 &  0.010\\ 
      &      & 0.25 & 0.30& 0.055 &  0.006& 0.062 &  0.006& 0.080 &  0.008\\ 
      &      & 0.30 & 0.35& 0.032 &  0.003& 0.041 &  0.003& 0.046 &  0.004\\ 
      &      & 0.35 & 0.40& 0.023 &  0.002& 0.027 &  0.004& 0.034 &  0.004\\ 
      &      & 0.40 & 0.45& 0.013 &  0.002& 0.017 &  0.003& 0.025 &  0.003\\ 
      &      & 0.45 & 0.50& 0.007 &  0.001& 0.010 &  0.002& 0.016 &  0.003\\ 
\hline  
 1.75 & 1.95 & 0.10 & 0.15& 0.101 &  0.018& 0.122 &  0.018& 0.126 &  0.022\\ 
      &      & 0.15 & 0.20& 0.116 &  0.013& 0.130 &  0.017& 0.153 &  0.019\\ 
      &      & 0.20 & 0.25& 0.080 &  0.008& 0.082 &  0.010& 0.096 &  0.011\\ 
      &      & 0.25 & 0.30& 0.047 &  0.007& 0.047 &  0.006& 0.063 &  0.008\\ 
      &      & 0.30 & 0.35& 0.023 &  0.004& 0.028 &  0.004& 0.040 &  0.005\\ 
      &      & 0.35 & 0.40& 0.013 &  0.002& 0.021 &  0.003& 0.029 &  0.004\\ 
      &      & 0.40 & 0.45& 0.007 &  0.001& 0.011 &  0.002& 0.017 &  0.003\\ 
      &      & 0.45 & 0.50& 0.005 &  0.001& 0.005 &  0.002& 0.009 &  0.002\\ 
\hline  
 1.95 & 2.15 & 0.10 & 0.15& 0.110 &  0.021& 0.131 &  0.025& 0.134 &  0.027\\ 
      &      & 0.15 & 0.20& 0.129 &  0.017& 0.132 &  0.018& 0.146 &  0.020\\ 
      &      & 0.20 & 0.25& 0.067 &  0.012& 0.090 &  0.015& 0.088 &  0.015\\ 
      &      & 0.25 & 0.30& 0.037 &  0.006& 0.048 &  0.009& 0.068 &  0.011\\ 
      &      & 0.30 & 0.35& 0.023 &  0.004& 0.024 &  0.004& 0.028 &  0.006\\ 
      &      & 0.35 & 0.40& 0.012 &  0.003& 0.014 &  0.003& 0.016 &  0.003\\ 
      &      & 0.40 & 0.45& 0.006 &  0.002& 0.010 &  0.003& 0.011 &  0.003\\ 
      &      & 0.45 & 0.50& 0.003 &  0.001& 0.004 &  0.001& 0.005 &  0.002\\ 
%
\end{tabular}
\end{center}
\end{table}
\begin{table}[hp!]
\begin{center}
  \caption{\label{tab:xsec-pim-c}
    HARP results for the double-differential $\pi^-$ production
    yield per target nucleon in the laboratory system,
    $\mathrm{d}^2\sigma^{\pi^-}/(\mathrm{d}p\mathrm{d}\theta)$ for p--C interactions. Each row refers to a
    different $(p_{\hbox{\small min}} \le p<p_{\hbox{\small max}},
    \theta_{\hbox{\small min}} \le \theta<\theta_{\hbox{\small max}})$ bin,
    where $p$ and $\theta$ are the pion momentum and polar angle, respectively.
    The central value as well as the square-root of the diagonal elements
    of the covariance matrix are given.}
\vspace{2mm}
\begin{tabular}{rrrr|r@{$\pm$}lr@{$\pm$}lr@{$\pm$}l}
\hline
$\theta_{\hbox{\small min}}$ &
$\theta_{\hbox{\small max}}$ &
$p_{\hbox{\small min}}$ &
$p_{\hbox{\small max}}$ &
\multicolumn{6}{c}{$\mathrm{d}^2\sigma^{\pi^-}/(\mathrm{d}p\mathrm{d}\theta)$}
\\
(rad) & (rad) & (\GeVc) & (\GeVc) &
\multicolumn{6}{c}{($\barn/(\GeVc \cdot \rad)$)}
\\
  &  &  &
&\multicolumn{2}{c}{\GeVc}
&\multicolumn{2}{c}{\GeVc}
&\multicolumn{2}{c}{\GeVc}
\\
\hline
 0.35 & 0.55 & 0.15 & 0.20& 0.099 &  0.023& 0.160 &  0.035& 0.202 &  0.049\\ 
      &      & 0.20 & 0.25& 0.108 &  0.014& 0.175 &  0.023& 0.244 &  0.031\\ 
      &      & 0.25 & 0.30& 0.121 &  0.013& 0.189 &  0.018& 0.275 &  0.030\\ 
      &      & 0.30 & 0.35& 0.132 &  0.012& 0.192 &  0.019& 0.286 &  0.025\\ 
      &      & 0.35 & 0.40& 0.124 &  0.010& 0.180 &  0.014& 0.236 &  0.018\\ 
      &      & 0.40 & 0.45& 0.115 &  0.009& 0.183 &  0.016& 0.238 &  0.019\\ 
      &      & 0.45 & 0.50& 0.114 &  0.007& 0.180 &  0.012& 0.235 &  0.017\\ 
      &      & 0.50 & 0.60& 0.104 &  0.007& 0.172 &  0.012& 0.237 &  0.016\\ 
      &      & 0.60 & 0.70& 0.094 &  0.010& 0.157 &  0.014& 0.231 &  0.021\\ 
      &      & 0.70 & 0.80& 0.078 &  0.011& 0.147 &  0.019& 0.206 &  0.026\\ 
\hline  
 0.55 & 0.75 & 0.10 & 0.15& 0.073 &  0.024& 0.102 &  0.033& 0.114 &  0.040\\ 
      &      & 0.15 & 0.20& 0.100 &  0.010& 0.140 &  0.014& 0.200 &  0.019\\ 
      &      & 0.20 & 0.25& 0.130 &  0.012& 0.177 &  0.018& 0.218 &  0.019\\ 
      &      & 0.25 & 0.30& 0.109 &  0.008& 0.168 &  0.012& 0.230 &  0.017\\ 
      &      & 0.30 & 0.35& 0.117 &  0.010& 0.170 &  0.013& 0.204 &  0.015\\ 
      &      & 0.35 & 0.40& 0.118 &  0.007& 0.148 &  0.008& 0.183 &  0.011\\ 
      &      & 0.40 & 0.45& 0.096 &  0.006& 0.139 &  0.007& 0.178 &  0.009\\ 
      &      & 0.45 & 0.50& 0.083 &  0.005& 0.139 &  0.008& 0.180 &  0.010\\ 
      &      & 0.50 & 0.60& 0.075 &  0.005& 0.128 &  0.008& 0.176 &  0.011\\ 
      &      & 0.60 & 0.70& 0.063 &  0.007& 0.103 &  0.010& 0.144 &  0.014\\ 
      &      & 0.70 & 0.80& 0.047 &  0.007& 0.086 &  0.013& 0.118 &  0.021\\ 
\hline  
 0.75 & 0.95 & 0.10 & 0.15& 0.064 &  0.012& 0.080 &  0.014& 0.093 &  0.019\\ 
      &      & 0.15 & 0.20& 0.104 &  0.009& 0.144 &  0.015& 0.166 &  0.016\\ 
      &      & 0.20 & 0.25& 0.115 &  0.009& 0.165 &  0.015& 0.181 &  0.018\\ 
      &      & 0.25 & 0.30& 0.116 &  0.008& 0.135 &  0.009& 0.172 &  0.011\\ 
      &      & 0.30 & 0.35& 0.091 &  0.005& 0.123 &  0.009& 0.164 &  0.009\\ 
      &      & 0.35 & 0.40& 0.090 &  0.005& 0.123 &  0.007& 0.146 &  0.009\\ 
      &      & 0.40 & 0.45& 0.077 &  0.004& 0.111 &  0.006& 0.138 &  0.008\\ 
      &      & 0.45 & 0.50& 0.067 &  0.004& 0.094 &  0.006& 0.123 &  0.007\\ 
      &      & 0.50 & 0.60& 0.048 &  0.005& 0.074 &  0.006& 0.096 &  0.008\\ 
      &      & 0.60 & 0.70& 0.031 &  0.004& 0.059 &  0.007& 0.075 &  0.009\\ 
\hline  
 0.95 & 1.15 & 0.10 & 0.15& 0.066 &  0.011& 0.081 &  0.011& 0.082 &  0.015\\ 
      &      & 0.15 & 0.20& 0.122 &  0.011& 0.128 &  0.012& 0.167 &  0.016\\ 
      &      & 0.20 & 0.25& 0.101 &  0.007& 0.139 &  0.011& 0.158 &  0.013\\ 
      &      & 0.25 & 0.30& 0.083 &  0.005& 0.127 &  0.006& 0.159 &  0.010\\ 
      &      & 0.30 & 0.35& 0.069 &  0.004& 0.106 &  0.006& 0.131 &  0.008\\ 
      &      & 0.35 & 0.40& 0.062 &  0.004& 0.082 &  0.005& 0.103 &  0.007\\ 
      &      & 0.40 & 0.45& 0.048 &  0.004& 0.064 &  0.004& 0.079 &  0.005\\ 
      &      & 0.45 & 0.50& 0.038 &  0.003& 0.054 &  0.004& 0.065 &  0.005\\ 
      &      & 0.50 & 0.60& 0.026 &  0.003& 0.042 &  0.004& 0.053 &  0.005\\ 
\hline
\end{tabular}
\end{center}
\end{table}

\begin{table}[hp!]
\begin{center}
\begin{tabular}{rrrr|r@{$\pm$}lr@{$\pm$}lr@{$\pm$}l}
\hline
$\theta_{\hbox{\small min}}$ &
$\theta_{\hbox{\small max}}$ &
$p_{\hbox{\small min}}$ &
$p_{\hbox{\small max}}$ &
\multicolumn{6}{c}{$\mathrm{d}^2\sigma^{\pi^-}/(\mathrm{d}p\mathrm{d}\theta)$}
\\
(rad) & (rad) & (\GeVc) & (\GeVc) &
\multicolumn{6}{c}{(\barn/($\GeVc \cdot \rad$))}
\\
  &  &  &
&\multicolumn{2}{c}{\GeVc}
&\multicolumn{2}{c}{\GeVc}
&\multicolumn{2}{c}{\GeVc}
\\
\hline
 1.15 & 1.35 & 0.10 & 0.15& 0.063 &  0.008& 0.088 &  0.012& 0.100 &  0.015\\ 
      &      & 0.15 & 0.20& 0.090 &  0.008& 0.136 &  0.013& 0.154 &  0.018\\ 
      &      & 0.20 & 0.25& 0.080 &  0.005& 0.120 &  0.008& 0.145 &  0.010\\ 
      &      & 0.25 & 0.30& 0.077 &  0.005& 0.096 &  0.006& 0.110 &  0.008\\ 
      &      & 0.30 & 0.35& 0.052 &  0.004& 0.066 &  0.005& 0.092 &  0.007\\ 
      &      & 0.35 & 0.40& 0.040 &  0.003& 0.051 &  0.004& 0.072 &  0.005\\ 
      &      & 0.40 & 0.45& 0.029 &  0.003& 0.041 &  0.003& 0.056 &  0.004\\ 
      &      & 0.45 & 0.50& 0.023 &  0.002& 0.032 &  0.003& 0.044 &  0.004\\ 
\hline  
 1.35 & 1.55 & 0.10 & 0.15& 0.077 &  0.009& 0.085 &  0.013& 0.113 &  0.015\\ 
      &      & 0.15 & 0.20& 0.089 &  0.008& 0.118 &  0.010& 0.142 &  0.013\\ 
      &      & 0.20 & 0.25& 0.076 &  0.006& 0.092 &  0.007& 0.124 &  0.010\\ 
      &      & 0.25 & 0.30& 0.058 &  0.004& 0.075 &  0.006& 0.087 &  0.008\\ 
      &      & 0.30 & 0.35& 0.043 &  0.003& 0.052 &  0.004& 0.068 &  0.005\\ 
      &      & 0.35 & 0.40& 0.031 &  0.003& 0.037 &  0.004& 0.056 &  0.005\\ 
      &      & 0.40 & 0.45& 0.022 &  0.002& 0.026 &  0.003& 0.039 &  0.005\\ 
      &      & 0.45 & 0.50& 0.014 &  0.002& 0.021 &  0.002& 0.029 &  0.003\\ 
\hline  
 1.55 & 1.75 & 0.10 & 0.15& 0.057 &  0.007& 0.080 &  0.012& 0.096 &  0.015\\ 
      &      & 0.15 & 0.20& 0.081 &  0.008& 0.107 &  0.011& 0.123 &  0.013\\ 
      &      & 0.20 & 0.25& 0.067 &  0.006& 0.087 &  0.007& 0.093 &  0.009\\ 
      &      & 0.25 & 0.30& 0.043 &  0.005& 0.055 &  0.005& 0.063 &  0.006\\ 
      &      & 0.30 & 0.35& 0.025 &  0.002& 0.036 &  0.003& 0.047 &  0.004\\ 
      &      & 0.35 & 0.40& 0.019 &  0.002& 0.029 &  0.003& 0.036 &  0.004\\ 
      &      & 0.40 & 0.45& 0.012 &  0.002& 0.022 &  0.002& 0.023 &  0.003\\ 
      &      & 0.45 & 0.50& 0.009 &  0.001& 0.016 &  0.002& 0.014 &  0.002\\ 
\hline  
 1.75 & 1.95 & 0.10 & 0.15& 0.068 &  0.007& 0.093 &  0.013& 0.101 &  0.017\\ 
      &      & 0.15 & 0.20& 0.084 &  0.008& 0.105 &  0.011& 0.118 &  0.013\\ 
      &      & 0.20 & 0.25& 0.057 &  0.006& 0.076 &  0.008& 0.088 &  0.010\\ 
      &      & 0.25 & 0.30& 0.035 &  0.004& 0.052 &  0.006& 0.059 &  0.007\\ 
      &      & 0.30 & 0.35& 0.022 &  0.003& 0.029 &  0.005& 0.035 &  0.005\\ 
      &      & 0.35 & 0.40& 0.016 &  0.002& 0.018 &  0.002& 0.024 &  0.003\\ 
      &      & 0.40 & 0.45& 0.009 &  0.002& 0.014 &  0.002& 0.016 &  0.003\\ 
      &      & 0.45 & 0.50& 0.005 &  0.001& 0.010 &  0.002& 0.010 &  0.002\\ 
\hline  
 1.95 & 2.15 & 0.10 & 0.15& 0.082 &  0.016& 0.099 &  0.019& 0.105 &  0.020\\ 
      &      & 0.15 & 0.20& 0.084 &  0.012& 0.103 &  0.014& 0.113 &  0.016\\ 
      &      & 0.20 & 0.25& 0.059 &  0.010& 0.086 &  0.013& 0.097 &  0.015\\ 
      &      & 0.25 & 0.30& 0.032 &  0.005& 0.043 &  0.007& 0.053 &  0.009\\ 
      &      & 0.30 & 0.35& 0.019 &  0.004& 0.028 &  0.006& 0.042 &  0.008\\ 
      &      & 0.35 & 0.40& 0.011 &  0.002& 0.018 &  0.004& 0.024 &  0.005\\ 
      &      & 0.40 & 0.45& 0.006 &  0.002& 0.010 &  0.003& 0.017 &  0.004\\ 
      &      & 0.45 & 0.50& 0.003 &  0.001& 0.006 &  0.002& 0.011 &  0.003\\ 
%
\end{tabular}
\end{center}
\end{table}

\clearpage
\begin{table}[hp!]
\begin{center}
  \caption{\label{tab:xsec-pip-ta}
    HARP results for the double-differential $\pi^+$ production
    yield per target nucleon in the laboratory system,
    $\mathrm{d}^2\sigma^{\pi^+}/(\mathrm{d}p\mathrm{d}\theta)$ for p--Ta interactions. Each row refers to a
    different $(p_{\hbox{\small min}} \le p<p_{\hbox{\small max}},
    \theta_{\hbox{\small min}} \le \theta<\theta_{\hbox{\small max}})$ bin,
    where $p$ and $\theta$ are the pion momentum and polar angle, respectively.
    The central value as well as the square-root of the diagonal elements
    of the covariance matrix are given.}
\vspace{2mm}
\begin{tabular}{rrrr|r@{$\pm$}lr@{$\pm$}lr@{$\pm$}l}
\hline
$\theta_{\hbox{\small min}}$ &
$\theta_{\hbox{\small max}}$ &
$p_{\hbox{\small min}}$ &
$p_{\hbox{\small max}}$ &
\multicolumn{6}{c}{$\mathrm{d}^2\sigma^{\pi^+}/(\mathrm{d}p\mathrm{d}\theta)$}\\
(rad) & (rad) & (\GeVc) & (\GeVc) &
\multicolumn{6}{c}{($\barn/(\GeVc \cdot \rad)$)}\\
  &  &  &
&\multicolumn{2}{c}{\GeVc}
&\multicolumn{2}{c}{\GeVc}
&\multicolumn{2}{c}{\GeVc}
\\
\hline 
 0.35 & 0.55 & 0.15 & 0.20& 0.69 &  0.28& 0.98 &  0.40& 1.35 &  0.54\\ 
      &      & 0.20 & 0.25& 0.82 &  0.17& 1.36 &  0.30& 1.96 &  0.42\\ 
      &      & 0.25 & 0.30& 0.83 &  0.14& 1.42 &  0.23& 2.18 &  0.35\\ 
      &      & 0.30 & 0.35& 0.83 &  0.13& 1.35 &  0.20& 2.08 &  0.30\\ 
      &      & 0.35 & 0.40& 0.78 &  0.11& 1.37 &  0.18& 2.00 &  0.27\\ 
      &      & 0.40 & 0.45& 0.82 &  0.09& 1.35 &  0.16& 1.99 &  0.24\\ 
      &      & 0.45 & 0.50& 0.80 &  0.07& 1.33 &  0.13& 2.00 &  0.19\\ 
      &      & 0.50 & 0.60& 0.68 &  0.06& 1.24 &  0.10& 1.95 &  0.16\\ 
      &      & 0.60 & 0.70& 0.50 &  0.06& 1.04 &  0.12& 1.75 &  0.19\\ 
      &      & 0.70 & 0.80& 0.43 &  0.07& 0.87 &  0.14& 1.45 &  0.22\\ 
\hline  
 0.55 & 0.75 & 0.10 & 0.15& 0.54 &  0.36& 0.62 &  0.43& 0.82 &  0.58\\ 
      &      & 0.15 & 0.20& 0.71 &  0.22& 0.98 &  0.36& 1.35 &  0.48\\ 
      &      & 0.20 & 0.25& 0.88 &  0.13& 1.47 &  0.22& 2.09 &  0.32\\ 
      &      & 0.25 & 0.30& 0.93 &  0.10& 1.47 &  0.13& 2.22 &  0.20\\ 
      &      & 0.30 & 0.35& 0.84 &  0.07& 1.37 &  0.12& 2.00 &  0.16\\ 
      &      & 0.35 & 0.40& 0.77 &  0.06& 1.36 &  0.09& 1.90 &  0.12\\ 
      &      & 0.40 & 0.45& 0.80 &  0.05& 1.30 &  0.08& 1.87 &  0.11\\ 
      &      & 0.45 & 0.50& 0.76 &  0.04& 1.20 &  0.06& 1.75 &  0.08\\ 
      &      & 0.50 & 0.60& 0.57 &  0.05& 1.02 &  0.07& 1.51 &  0.11\\ 
      &      & 0.60 & 0.70& 0.34 &  0.04& 0.72 &  0.10& 1.08 &  0.15\\ 
      &      & 0.70 & 0.80& 0.27 &  0.05& 0.48 &  0.09& 0.76 &  0.13\\ 
\hline  
 0.75 & 0.95 & 0.10 & 0.15& 0.56 &  0.30& 0.61 &  0.32& 0.85 &  0.44\\ 
      &      & 0.15 & 0.20& 0.77 &  0.18& 1.08 &  0.27& 1.47 &  0.37\\ 
      &      & 0.20 & 0.25& 0.96 &  0.08& 1.58 &  0.16& 2.19 &  0.21\\ 
      &      & 0.25 & 0.30& 0.97 &  0.06& 1.49 &  0.09& 2.06 &  0.13\\ 
      &      & 0.30 & 0.35& 0.81 &  0.05& 1.30 &  0.08& 1.72 &  0.10\\ 
      &      & 0.35 & 0.40& 0.66 &  0.04& 1.12 &  0.06& 1.46 &  0.08\\ 
      &      & 0.40 & 0.45& 0.55 &  0.04& 0.97 &  0.04& 1.33 &  0.06\\ 
      &      & 0.45 & 0.50& 0.46 &  0.04& 0.78 &  0.05& 1.11 &  0.06\\ 
      &      & 0.50 & 0.60& 0.31 &  0.03& 0.56 &  0.05& 0.80 &  0.08\\ 
      &      & 0.60 & 0.70& 0.17 &  0.03& 0.35 &  0.05& 0.49 &  0.07\\ 
\hline  
 0.95 & 1.15 & 0.10 & 0.15& 0.65 &  0.26& 0.75 &  0.30& 1.07 &  0.43\\ 
      &      & 0.15 & 0.20& 0.80 &  0.15& 1.20 &  0.24& 1.59 &  0.32\\ 
      &      & 0.20 & 0.25& 0.95 &  0.07& 1.56 &  0.12& 2.07 &  0.15\\ 
      &      & 0.25 & 0.30& 0.77 &  0.05& 1.33 &  0.08& 1.76 &  0.10\\ 
      &      & 0.30 & 0.35& 0.60 &  0.03& 0.97 &  0.06& 1.34 &  0.07\\ 
      &      & 0.35 & 0.40& 0.46 &  0.02& 0.77 &  0.04& 1.06 &  0.05\\ 
      &      & 0.40 & 0.45& 0.35 &  0.02& 0.61 &  0.04& 0.86 &  0.04\\ 
      &      & 0.45 & 0.50& 0.26 &  0.03& 0.45 &  0.04& 0.68 &  0.05\\ 
      &      & 0.50 & 0.60& 0.15 &  0.02& 0.28 &  0.03& 0.41 &  0.05\\ 
\hline
\end{tabular}
\end{center}
\end{table}

\begin{table}[hp!]
\begin{center}
\begin{tabular}{rrrr|r@{$\pm$}lr@{$\pm$}lr@{$\pm$}l}
\hline
$\theta_{\hbox{\small min}}$ &
$\theta_{\hbox{\small max}}$ &
$p_{\hbox{\small min}}$ &
$p_{\hbox{\small max}}$ &
\multicolumn{6}{c}{$\mathrm{d}^2\sigma^{\pi^+}/(\mathrm{d}p\mathrm{d}\theta)$}\\
(rad) & (rad) & (\GeVc) & (\GeVc) &
\multicolumn{6}{c}{(\barn/($\GeVc \cdot \rad$))}\\
  &  &  &
&\multicolumn{2}{c}{\GeVc}
&\multicolumn{2}{c}{\GeVc}
&\multicolumn{2}{c}{\GeVc}
\\
\hline
 1.15 & 1.35 & 0.10 & 0.15& 0.62 &  0.19& 0.88 &  0.30& 1.19 &  0.39\\ 
      &      & 0.15 & 0.20& 0.78 &  0.11& 1.25 &  0.21& 1.66 &  0.27\\ 
      &      & 0.20 & 0.25& 0.84 &  0.06& 1.41 &  0.09& 1.87 &  0.12\\ 
      &      & 0.25 & 0.30& 0.54 &  0.04& 0.98 &  0.07& 1.33 &  0.08\\ 
      &      & 0.30 & 0.35& 0.39 &  0.02& 0.68 &  0.04& 0.94 &  0.05\\ 
      &      & 0.35 & 0.40& 0.28 &  0.02& 0.51 &  0.03& 0.67 &  0.04\\ 
      &      & 0.40 & 0.45& 0.22 &  0.02& 0.39 &  0.03& 0.51 &  0.04\\ 
      &      & 0.45 & 0.50& 0.16 &  0.02& 0.27 &  0.03& 0.36 &  0.04\\ 
\hline  
 1.35 & 1.55 & 0.10 & 0.15& 0.73 &  0.23& 0.99 &  0.33& 1.29 &  0.44\\ 
      &      & 0.15 & 0.20& 0.77 &  0.12& 1.25 &  0.19& 1.61 &  0.24\\ 
      &      & 0.20 & 0.25& 0.74 &  0.06& 1.13 &  0.08& 1.59 &  0.10\\ 
      &      & 0.25 & 0.30& 0.44 &  0.04& 0.71 &  0.04& 1.01 &  0.06\\ 
      &      & 0.30 & 0.35& 0.28 &  0.02& 0.49 &  0.03& 0.67 &  0.04\\ 
      &      & 0.35 & 0.40& 0.19 &  0.02& 0.36 &  0.03& 0.49 &  0.04\\ 
      &      & 0.40 & 0.45& 0.13 &  0.02& 0.23 &  0.02& 0.33 &  0.03\\ 
      &      & 0.45 & 0.50& 0.08 &  0.01& 0.15 &  0.02& 0.21 &  0.03\\ 
\hline  
 1.55 & 1.75 & 0.10 & 0.15& 0.76 &  0.24& 1.02 &  0.31& 1.31 &  0.39\\ 
      &      & 0.15 & 0.20& 0.73 &  0.10& 1.19 &  0.15& 1.46 &  0.19\\ 
      &      & 0.20 & 0.25& 0.64 &  0.04& 0.92 &  0.06& 1.33 &  0.08\\ 
      &      & 0.25 & 0.30& 0.39 &  0.03& 0.54 &  0.04& 0.77 &  0.05\\ 
      &      & 0.30 & 0.35& 0.24 &  0.02& 0.34 &  0.03& 0.48 &  0.03\\ 
      &      & 0.35 & 0.40& 0.14 &  0.02& 0.24 &  0.02& 0.33 &  0.03\\ 
      &      & 0.40 & 0.45& 0.09 &  0.01& 0.16 &  0.02& 0.21 &  0.03\\ 
      &      & 0.45 & 0.50& 0.05 &  0.01& 0.10 &  0.02& 0.13 &  0.02\\ 
\hline  
 1.75 & 1.95 & 0.10 & 0.15& 0.73 &  0.22& 0.98 &  0.28& 1.22 &  0.35\\ 
      &      & 0.15 & 0.20& 0.66 &  0.08& 1.07 &  0.14& 1.31 &  0.17\\ 
      &      & 0.20 & 0.25& 0.48 &  0.03& 0.73 &  0.06& 1.00 &  0.07\\ 
      &      & 0.25 & 0.30& 0.28 &  0.02& 0.36 &  0.03& 0.53 &  0.05\\ 
      &      & 0.30 & 0.35& 0.18 &  0.02& 0.23 &  0.02& 0.32 &  0.02\\ 
      &      & 0.35 & 0.40& 0.10 &  0.01& 0.15 &  0.01& 0.21 &  0.02\\ 
      &      & 0.40 & 0.45& 0.05 &  0.01& 0.09 &  0.01& 0.12 &  0.02\\ 
      &      & 0.45 & 0.50& 0.02 &  0.01& 0.06 &  0.01& 0.07 &  0.01\\ 
\hline  
 1.95 & 2.15 & 0.10 & 0.15& 0.70 &  0.22& 0.95 &  0.28& 1.15 &  0.35\\ 
      &      & 0.15 & 0.20& 0.62 &  0.09& 0.95 &  0.15& 1.17 &  0.19\\ 
      &      & 0.20 & 0.25& 0.39 &  0.05& 0.64 &  0.07& 0.85 &  0.09\\ 
      &      & 0.25 & 0.30& 0.18 &  0.02& 0.28 &  0.04& 0.39 &  0.05\\ 
      &      & 0.30 & 0.35& 0.10 &  0.01& 0.16 &  0.02& 0.20 &  0.02\\ 
      &      & 0.35 & 0.40& 0.05 &  0.01& 0.09 &  0.01& 0.12 &  0.02\\ 
      &      & 0.40 & 0.45& 0.03 &  0.01& 0.05 &  0.01& 0.07 &  0.01\\ 
      &      & 0.45 & 0.50& 0.01 &  0.01& 0.03 &  0.01& 0.04 &  0.01\\ 
%
\end{tabular}
\end{center}
\end{table}
\begin{table}[hp!]
\begin{center}
  \caption{\label{tab:xsec-piym-ta}
    HARP results for the double-differential $\pi^-$ production
    yield per target nucleon in the laboratory system,
    $\mathrm{d}^2\sigma^{\pi^-}/(\mathrm{d}p\mathrm{d}\theta)$ for p--Ta interactions. Each row refers to a
    different $(p_{\hbox{\small min}} \le p<p_{\hbox{\small max}},
    \theta_{\hbox{\small min}} \le \theta<\theta_{\hbox{\small max}})$ bin,
    where $p$ and $\theta$ are the pion momentum and polar angle, respectively.
    The central value as well as the square-root of the diagonal elements
    of the covariance matrix are given.}
\vspace{2mm}
\begin{tabular}{rrrr|r@{$\pm$}lr@{$\pm$}lr@{$\pm$}l}
\hline
$\theta_{\hbox{\small min}}$ &
$\theta_{\hbox{\small max}}$ &
$p_{\hbox{\small min}}$ &
$p_{\hbox{\small max}}$ &
\multicolumn{6}{c}{$\mathrm{d}^2\sigma^{\pi^-}/(\mathrm{d}p\mathrm{d}\theta)$}
\\
(rad) & (rad) & (\GeVc) & (\GeVc) &
\multicolumn{6}{c}{($\barn/(\GeVc \cdot \rad)$)}
\\
  &  &  &
&\multicolumn{2}{c}{\GeVc}
&\multicolumn{2}{c}{\GeVc}
&\multicolumn{2}{c}{\GeVc}
\\
\hline 
 0.35 & 0.55 & 0.15 & 0.20& 0.89 &  0.31& 1.37 &  0.52& 2.01 &  0.74\\ 
      &      & 0.20 & 0.25& 0.81 &  0.18& 1.50 &  0.33& 2.24 &  0.47\\ 
      &      & 0.25 & 0.30& 0.77 &  0.12& 1.48 &  0.23& 2.14 &  0.32\\ 
      &      & 0.30 & 0.35& 0.65 &  0.09& 1.37 &  0.18& 1.95 &  0.27\\ 
      &      & 0.35 & 0.40& 0.60 &  0.08& 1.19 &  0.14& 1.74 &  0.21\\ 
      &      & 0.40 & 0.45& 0.56 &  0.06& 1.06 &  0.12& 1.62 &  0.18\\ 
      &      & 0.45 & 0.50& 0.54 &  0.05& 0.96 &  0.10& 1.48 &  0.15\\ 
      &      & 0.50 & 0.60& 0.47 &  0.04& 0.89 &  0.07& 1.37 &  0.11\\ 
      &      & 0.60 & 0.70& 0.34 &  0.04& 0.79 &  0.08& 1.27 &  0.12\\ 
      &      & 0.70 & 0.80& 0.28 &  0.04& 0.73 &  0.10& 1.18 &  0.15\\ 
\hline  
 0.55 & 0.75 & 0.10 & 0.15& 0.90 &  0.45& 1.25 &  0.71& 1.87 &  1.06\\ 
      &      & 0.15 & 0.20& 0.80 &  0.23& 1.31 &  0.38& 1.96 &  0.54\\ 
      &      & 0.20 & 0.25& 0.79 &  0.11& 1.43 &  0.20& 2.13 &  0.27\\ 
      &      & 0.25 & 0.30& 0.72 &  0.07& 1.31 &  0.12& 1.95 &  0.17\\ 
      &      & 0.30 & 0.35& 0.63 &  0.05& 1.16 &  0.10& 1.69 &  0.13\\ 
      &      & 0.35 & 0.40& 0.57 &  0.04& 1.09 &  0.08& 1.55 &  0.11\\ 
      &      & 0.40 & 0.45& 0.49 &  0.03& 0.95 &  0.05& 1.39 &  0.07\\ 
      &      & 0.45 & 0.50& 0.44 &  0.03& 0.85 &  0.04& 1.30 &  0.06\\ 
      &      & 0.50 & 0.60& 0.38 &  0.02& 0.75 &  0.04& 1.14 &  0.06\\ 
      &      & 0.60 & 0.70& 0.26 &  0.03& 0.63 &  0.06& 0.96 &  0.09\\ 
      &      & 0.70 & 0.80& 0.21 &  0.03& 0.51 &  0.07& 0.80 &  0.10\\ 
\hline  
 0.75 & 0.95 & 0.10 & 0.15& 0.89 &  0.31& 1.30 &  0.55& 1.81 &  0.77\\ 
      &      & 0.15 & 0.20& 0.82 &  0.18& 1.31 &  0.28& 1.94 &  0.40\\ 
      &      & 0.20 & 0.25& 0.77 &  0.07& 1.37 &  0.11& 1.99 &  0.15\\ 
      &      & 0.25 & 0.30& 0.67 &  0.05& 1.17 &  0.07& 1.71 &  0.10\\ 
      &      & 0.30 & 0.35& 0.57 &  0.04& 0.99 &  0.06& 1.38 &  0.07\\ 
      &      & 0.35 & 0.40& 0.52 &  0.03& 0.85 &  0.04& 1.20 &  0.06\\ 
      &      & 0.40 & 0.45& 0.45 &  0.02& 0.73 &  0.04& 1.06 &  0.05\\ 
      &      & 0.45 & 0.50& 0.37 &  0.02& 0.65 &  0.03& 0.93 &  0.04\\ 
      &      & 0.50 & 0.60& 0.27 &  0.02& 0.50 &  0.04& 0.74 &  0.05\\ 
      &      & 0.60 & 0.70& 0.18 &  0.02& 0.35 &  0.04& 0.54 &  0.06\\ 
\hline  
 0.95 & 1.15 & 0.10 & 0.15& 0.97 &  0.33& 1.38 &  0.45& 1.92 &  0.65\\ 
      &      & 0.15 & 0.20& 0.86 &  0.14& 1.29 &  0.20& 1.90 &  0.30\\ 
      &      & 0.20 & 0.25& 0.75 &  0.05& 1.24 &  0.08& 1.80 &  0.10\\ 
      &      & 0.25 & 0.30& 0.60 &  0.03& 0.96 &  0.05& 1.44 &  0.08\\ 
      &      & 0.30 & 0.35& 0.48 &  0.03& 0.77 &  0.04& 1.14 &  0.05\\ 
      &      & 0.35 & 0.40& 0.39 &  0.02& 0.64 &  0.03& 0.91 &  0.04\\ 
      &      & 0.40 & 0.45& 0.32 &  0.02& 0.50 &  0.02& 0.74 &  0.03\\ 
      &      & 0.45 & 0.50& 0.24 &  0.02& 0.43 &  0.02& 0.61 &  0.03\\ 
      &      & 0.50 & 0.60& 0.17 &  0.02& 0.31 &  0.03& 0.44 &  0.03\\ 
\hline
\end{tabular}
\end{center}
\end{table}

\begin{table}[hp!]
\begin{center}
\begin{tabular}{rrrr|r@{$\pm$}lr@{$\pm$}lr@{$\pm$}l}
\hline
$\theta_{\hbox{\small min}}$ &
$\theta_{\hbox{\small max}}$ &
$p_{\hbox{\small min}}$ &
$p_{\hbox{\small max}}$ &
\multicolumn{6}{c}{$\mathrm{d}^2\sigma^{\pi^-}/(\mathrm{d}p\mathrm{d}\theta)$}
\\
(rad) & (rad) & (\GeVc) & (\GeVc) &
\multicolumn{6}{c}{(\barn/($\GeVc \cdot \rad$))}
\\
  &  &  &
&\multicolumn{2}{c}{\GeVc}
&\multicolumn{2}{c}{\GeVc}
&\multicolumn{2}{c}{\GeVc}
\\
\hline
 1.15 & 1.35 & 0.10 & 0.15& 1.08 &  0.34& 1.68 &  0.49& 2.12 &  0.66\\ 
      &      & 0.15 & 0.20& 0.82 &  0.12& 1.32 &  0.17& 1.81 &  0.24\\ 
      &      & 0.20 & 0.25& 0.65 &  0.04& 1.09 &  0.06& 1.55 &  0.08\\ 
      &      & 0.25 & 0.30& 0.44 &  0.03& 0.77 &  0.05& 1.16 &  0.05\\ 
      &      & 0.30 & 0.35& 0.35 &  0.02& 0.58 &  0.03& 0.91 &  0.04\\ 
      &      & 0.35 & 0.40& 0.29 &  0.02& 0.47 &  0.02& 0.67 &  0.03\\ 
      &      & 0.40 & 0.45& 0.22 &  0.02& 0.38 &  0.02& 0.50 &  0.03\\ 
      &      & 0.45 & 0.50& 0.15 &  0.02& 0.30 &  0.02& 0.39 &  0.03\\ 
\hline  
 1.35 & 1.55 & 0.10 & 0.15& 1.02 &  0.25& 1.69 &  0.46& 2.00 &  0.53\\ 
      &      & 0.15 & 0.20& 0.73 &  0.09& 1.29 &  0.16& 1.60 &  0.20\\ 
      &      & 0.20 & 0.25& 0.50 &  0.03& 1.00 &  0.06& 1.26 &  0.07\\ 
      &      & 0.25 & 0.30& 0.36 &  0.02& 0.65 &  0.04& 0.84 &  0.05\\ 
      &      & 0.30 & 0.35& 0.29 &  0.02& 0.47 &  0.03& 0.63 &  0.03\\ 
      &      & 0.35 & 0.40& 0.22 &  0.02& 0.34 &  0.02& 0.47 &  0.03\\ 
      &      & 0.40 & 0.45& 0.16 &  0.02& 0.25 &  0.02& 0.35 &  0.03\\ 
      &      & 0.45 & 0.50& 0.10 &  0.01& 0.16 &  0.02& 0.25 &  0.02\\ 
\hline  
 1.55 & 1.75 & 0.10 & 0.15& 1.02 &  0.28& 1.50 &  0.40& 1.87 &  0.49\\ 
      &      & 0.15 & 0.20& 0.64 &  0.08& 1.08 &  0.12& 1.37 &  0.15\\ 
      &      & 0.20 & 0.25& 0.42 &  0.03& 0.78 &  0.05& 0.99 &  0.06\\ 
      &      & 0.25 & 0.30& 0.30 &  0.02& 0.50 &  0.03& 0.62 &  0.04\\ 
      &      & 0.30 & 0.35& 0.20 &  0.02& 0.35 &  0.03& 0.45 &  0.03\\ 
      &      & 0.35 & 0.40& 0.14 &  0.01& 0.25 &  0.02& 0.35 &  0.02\\ 
      &      & 0.40 & 0.45& 0.09 &  0.01& 0.17 &  0.02& 0.25 &  0.02\\ 
      &      & 0.45 & 0.50& 0.06 &  0.01& 0.12 &  0.01& 0.18 &  0.02\\ 
\hline  
 1.75 & 1.95 & 0.10 & 0.15& 0.92 &  0.22& 1.30 &  0.35& 1.70 &  0.44\\ 
      &      & 0.15 & 0.20& 0.57 &  0.06& 0.87 &  0.10& 1.15 &  0.13\\ 
      &      & 0.20 & 0.25& 0.33 &  0.03& 0.55 &  0.04& 0.71 &  0.05\\ 
      &      & 0.25 & 0.30& 0.22 &  0.02& 0.34 &  0.02& 0.42 &  0.03\\ 
      &      & 0.30 & 0.35& 0.14 &  0.01& 0.22 &  0.02& 0.30 &  0.02\\ 
      &      & 0.35 & 0.40& 0.09 &  0.01& 0.14 &  0.01& 0.23 &  0.02\\ 
      &      & 0.40 & 0.45& 0.06 &  0.01& 0.10 &  0.01& 0.17 &  0.01\\ 
      &      & 0.45 & 0.50& 0.04 &  0.01& 0.07 &  0.01& 0.12 &  0.01\\ 
\hline  
 1.95 & 2.15 & 0.10 & 0.15& 0.86 &  0.26& 1.34 &  0.38& 1.67 &  0.45\\ 
      &      & 0.15 & 0.20& 0.55 &  0.08& 0.78 &  0.11& 1.07 &  0.15\\ 
      &      & 0.20 & 0.25& 0.26 &  0.03& 0.41 &  0.05& 0.58 &  0.06\\ 
      &      & 0.25 & 0.30& 0.14 &  0.02& 0.23 &  0.02& 0.33 &  0.03\\ 
      &      & 0.30 & 0.35& 0.09 &  0.01& 0.14 &  0.02& 0.21 &  0.02\\ 
      &      & 0.35 & 0.40& 0.06 &  0.01& 0.09 &  0.01& 0.15 &  0.01\\ 
      &      & 0.40 & 0.45& 0.03 &  0.01& 0.06 &  0.01& 0.09 &  0.01\\ 
      &      & 0.45 & 0.50& 0.02 &  0.01& 0.05 &  0.01& 0.06 &  0.01\\ 
%
\end{tabular}
\end{center}
\end{table}

\clearpage
\begin{table}[hp!]
\begin{center}
  \caption{\label{tab:xsec-pip-pb}
    HARP results for the double-differential $\pi^+$ production
    yield per target nucleon in the laboratory system,
    $\mathrm{d}^2\sigma^{\pi^+}/(\mathrm{d}p\mathrm{d}\theta)$ for p--Pb interactions. Each row refers to a
    different $(p_{\hbox{\small min}} \le p<p_{\hbox{\small max}},
    \theta_{\hbox{\small min}} \le \theta<\theta_{\hbox{\small max}})$ bin,
    where $p$ and $\theta$ are the pion momentum and polar angle, respectively.
    The central value as well as the square-root of the diagonal elements
    of the covariance matrix are given.}
\vspace{2mm}
\begin{tabular}{rrrr|r@{$\pm$}lr@{$\pm$}lr@{$\pm$}l}
\hline
$\theta_{\hbox{\small min}}$ &
$\theta_{\hbox{\small max}}$ &
$p_{\hbox{\small min}}$ &
$p_{\hbox{\small max}}$ &
\multicolumn{6}{c}{$\mathrm{d}^2\sigma^{\pi^+}/(\mathrm{d}p\mathrm{d}\theta)$}\\
(rad) & (rad) & (\GeVc) & (\GeVc) &
\multicolumn{6}{c}{($\barn/(\GeVc \cdot \rad)$)}\\
  &  &  &
&\multicolumn{2}{c}{\GeVc}
&\multicolumn{2}{c}{\GeVc}
&\multicolumn{2}{c}{\GeVc}
\\
\hline
0.35 & 0.55 & 0.15 & 0.20& 0.64 &  0.24& 1.01 &  0.43& 1.17 &  0.51\\ 
      &      & 0.20 & 0.25& 0.79 &  0.16& 1.51 &  0.30& 1.89 &  0.38\\ 
      &      & 0.25 & 0.30& 0.84 &  0.13& 1.50 &  0.22& 2.01 &  0.29\\ 
      &      & 0.30 & 0.35& 0.83 &  0.10& 1.33 &  0.16& 1.83 &  0.22\\ 
      &      & 0.35 & 0.40& 0.84 &  0.09& 1.35 &  0.15& 1.90 &  0.21\\ 
      &      & 0.40 & 0.45& 0.79 &  0.07& 1.39 &  0.13& 1.89 &  0.17\\ 
      &      & 0.45 & 0.50& 0.77 &  0.05& 1.40 &  0.11& 1.94 &  0.15\\ 
      &      & 0.50 & 0.60& 0.74 &  0.06& 1.44 &  0.11& 1.98 &  0.15\\ 
      &      & 0.60 & 0.70& 0.65 &  0.08& 1.27 &  0.15& 1.78 &  0.19\\ 
      &      & 0.70 & 0.80& 0.50 &  0.09& 0.97 &  0.17& 1.52 &  0.23\\ 
\hline 
 0.55 & 0.75 & 0.10 & 0.15& 0.58 &  0.30& 0.65 &  0.42& 0.68 &  0.47\\ 
      &      & 0.15 & 0.20& 0.70 &  0.18& 1.06 &  0.30& 1.31 &  0.37\\ 
      &      & 0.20 & 0.25& 0.91 &  0.10& 1.58 &  0.17& 2.06 &  0.23\\ 
      &      & 0.25 & 0.30& 0.89 &  0.07& 1.53 &  0.12& 2.02 &  0.15\\ 
      &      & 0.30 & 0.35& 0.88 &  0.06& 1.42 &  0.11& 1.85 &  0.13\\ 
      &      & 0.35 & 0.40& 0.86 &  0.05& 1.48 &  0.09& 1.91 &  0.12\\ 
      &      & 0.40 & 0.45& 0.76 &  0.04& 1.48 &  0.08& 1.89 &  0.09\\ 
      &      & 0.45 & 0.50& 0.70 &  0.05& 1.29 &  0.08& 1.77 &  0.09\\ 
      &      & 0.50 & 0.60& 0.57 &  0.05& 1.11 &  0.08& 1.50 &  0.12\\ 
      &      & 0.60 & 0.70& 0.40 &  0.06& 0.86 &  0.11& 1.18 &  0.14\\ 
      &      & 0.70 & 0.80& 0.29 &  0.05& 0.59 &  0.12& 0.89 &  0.16\\ 
\hline 
 0.75 & 0.95 & 0.10 & 0.15& 0.62 &  0.26& 0.73 &  0.35& 0.76 &  0.40\\ 
      &      & 0.15 & 0.20& 0.80 &  0.14& 1.23 &  0.22& 1.58 &  0.28\\ 
      &      & 0.20 & 0.25& 1.04 &  0.08& 1.65 &  0.12& 2.17 &  0.15\\ 
      &      & 0.25 & 0.30& 0.89 &  0.06& 1.53 &  0.10& 1.82 &  0.11\\ 
      &      & 0.30 & 0.35& 0.88 &  0.05& 1.34 &  0.07& 1.68 &  0.10\\ 
      &      & 0.35 & 0.40& 0.72 &  0.05& 1.15 &  0.07& 1.49 &  0.08\\ 
      &      & 0.40 & 0.45& 0.56 &  0.04& 1.02 &  0.06& 1.34 &  0.07\\ 
      &      & 0.45 & 0.50& 0.44 &  0.03& 0.85 &  0.06& 1.12 &  0.07\\ 
      &      & 0.50 & 0.60& 0.31 &  0.03& 0.64 &  0.06& 0.83 &  0.08\\ 
      &      & 0.60 & 0.70& 0.18 &  0.03& 0.42 &  0.07& 0.56 &  0.08\\ 
\hline
 0.95 & 1.15 & 0.10 & 0.15& 0.68 &  0.21& 0.94 &  0.32& 0.96 &  0.35\\ 
      &      & 0.15 & 0.20& 0.89 &  0.12& 1.42 &  0.18& 1.75 &  0.22\\ 
      &      & 0.20 & 0.25& 1.03 &  0.08& 1.54 &  0.11& 2.00 &  0.14\\ 
      &      & 0.25 & 0.30& 0.81 &  0.05& 1.23 &  0.08& 1.55 &  0.09\\ 
      &      & 0.30 & 0.35& 0.63 &  0.04& 0.96 &  0.05& 1.29 &  0.06\\ 
      &      & 0.35 & 0.40& 0.51 &  0.03& 0.74 &  0.04& 1.02 &  0.05\\ 
      &      & 0.40 & 0.45& 0.36 &  0.03& 0.60 &  0.04& 0.81 &  0.04\\ 
      &      & 0.45 & 0.50& 0.25 &  0.03& 0.50 &  0.04& 0.66 &  0.05\\ 
      &      & 0.50 & 0.60& 0.15 &  0.02& 0.34 &  0.04& 0.43 &  0.05\\ 
\hline
\end{tabular}
\end{center}
\end{table}

\begin{table}[hp!]
\begin{center}
\begin{tabular}{rrrr|r@{$\pm$}lr@{$\pm$}lr@{$\pm$}l}
\hline
$\theta_{\hbox{\small min}}$ &
$\theta_{\hbox{\small max}}$ &
$p_{\hbox{\small min}}$ &
$p_{\hbox{\small max}}$ &
\multicolumn{6}{c}{$\mathrm{d}^2\sigma^{\pi^+}/(\mathrm{d}p\mathrm{d}\theta)$}\\
(rad) & (rad) & (\GeVc) & (\GeVc) &
\multicolumn{6}{c}{(\barn/($\GeVc \cdot \rad$))}\\
  &  &  &
&\multicolumn{2}{c}{\GeVc}
&\multicolumn{2}{c}{\GeVc}
&\multicolumn{2}{c}{\GeVc}
\\
\hline
 1.15 & 1.35 & 0.10 & 0.15& 0.73 &  0.20& 1.11 &  0.33& 1.15 &  0.38\\ 
      &      & 0.15 & 0.20& 0.89 &  0.11& 1.54 &  0.17& 1.80 &  0.21\\ 
      &      & 0.20 & 0.25& 0.89 &  0.06& 1.37 &  0.09& 1.75 &  0.11\\ 
      &      & 0.25 & 0.30& 0.64 &  0.04& 0.88 &  0.06& 1.22 &  0.07\\ 
      &      & 0.30 & 0.35& 0.41 &  0.03& 0.63 &  0.04& 0.90 &  0.04\\ 
      &      & 0.35 & 0.40& 0.33 &  0.03& 0.49 &  0.03& 0.65 &  0.04\\ 
      &      & 0.40 & 0.45& 0.24 &  0.03& 0.40 &  0.03& 0.49 &  0.04\\ 
      &      & 0.45 & 0.50& 0.16 &  0.02& 0.28 &  0.03& 0.34 &  0.04\\ 
\hline 
 1.35 & 1.55 & 0.10 & 0.15& 0.81 &  0.20& 1.14 &  0.29& 1.20 &  0.34\\ 
      &      & 0.15 & 0.20& 0.90 &  0.10& 1.37 &  0.13& 1.61 &  0.17\\ 
      &      & 0.20 & 0.25& 0.76 &  0.06& 1.13 &  0.07& 1.43 &  0.09\\ 
      &      & 0.25 & 0.30& 0.45 &  0.03& 0.72 &  0.05& 0.96 &  0.06\\ 
      &      & 0.30 & 0.35& 0.30 &  0.02& 0.49 &  0.04& 0.62 &  0.05\\ 
      &      & 0.35 & 0.40& 0.24 &  0.02& 0.36 &  0.03& 0.42 &  0.04\\ 
      &      & 0.40 & 0.45& 0.16 &  0.02& 0.23 &  0.03& 0.26 &  0.03\\ 
      &      & 0.45 & 0.50& 0.11 &  0.01& 0.17 &  0.03& 0.19 &  0.03\\ 
\hline 
 1.55 & 1.75 & 0.10 & 0.15& 0.88 &  0.20& 1.16 &  0.27& 1.22 &  0.30\\ 
      &      & 0.15 & 0.20& 0.82 &  0.07& 1.22 &  0.12& 1.34 &  0.13\\ 
      &      & 0.20 & 0.25& 0.66 &  0.05& 0.97 &  0.07& 1.14 &  0.07\\ 
      &      & 0.25 & 0.30& 0.37 &  0.03& 0.57 &  0.05& 0.70 &  0.05\\ 
      &      & 0.30 & 0.35& 0.23 &  0.02& 0.36 &  0.03& 0.45 &  0.03\\ 
      &      & 0.35 & 0.40& 0.17 &  0.01& 0.26 &  0.02& 0.30 &  0.03\\ 
      &      & 0.40 & 0.45& 0.12 &  0.01& 0.19 &  0.02& 0.20 &  0.02\\ 
      &      & 0.45 & 0.50& 0.07 &  0.01& 0.11 &  0.02& 0.11 &  0.02\\ 
\hline 
 1.75 & 1.95 & 0.10 & 0.15& 0.88 &  0.21& 1.13 &  0.28& 1.19 &  0.31\\ 
      &      & 0.15 & 0.20& 0.76 &  0.07& 1.10 &  0.10& 1.19 &  0.10\\ 
      &      & 0.20 & 0.25& 0.53 &  0.04& 0.79 &  0.06& 0.87 &  0.06\\ 
      &      & 0.25 & 0.30& 0.25 &  0.02& 0.39 &  0.04& 0.50 &  0.04\\ 
      &      & 0.30 & 0.35& 0.14 &  0.02& 0.22 &  0.02& 0.29 &  0.02\\ 
      &      & 0.35 & 0.40& 0.08 &  0.01& 0.14 &  0.02& 0.18 &  0.02\\ 
      &      & 0.40 & 0.45& 0.05 &  0.01& 0.09 &  0.01& 0.09 &  0.02\\ 
      &      & 0.45 & 0.50& 0.03 &  0.01& 0.05 &  0.01& 0.06 &  0.01\\ 
\hline 
 1.95 & 2.15 & 0.10 & 0.15& 0.96 &  0.27& 1.13 &  0.31& 1.30 &  0.37\\ 
      &      & 0.15 & 0.20& 0.83 &  0.14& 1.10 &  0.18& 1.26 &  0.20\\ 
      &      & 0.20 & 0.25& 0.52 &  0.09& 0.76 &  0.13& 0.90 &  0.15\\ 
      &      & 0.25 & 0.30& 0.22 &  0.04& 0.40 &  0.07& 0.49 &  0.09\\ 
      &      & 0.30 & 0.35& 0.11 &  0.02& 0.20 &  0.04& 0.23 &  0.05\\ 
      &      & 0.35 & 0.40& 0.07 &  0.01& 0.11 &  0.02& 0.12 &  0.02\\ 
      &      & 0.40 & 0.45& 0.03 &  0.01& 0.07 &  0.02& 0.08 &  0.02\\ 
      &      & 0.45 & 0.50& 0.02 &  0.01& 0.04 &  0.01& 0.04 &  0.01\\ 
%
\end{tabular}
\end{center}
\end{table}
\begin{table}[hp!]
\begin{center}
  \caption{\label{tab:xsec-pim-pb}
    HARP results for the double-differential $\pi^-$ production
    yield per target nucleon in the laboratory system,
    $\mathrm{d}^2\sigma^{\pi^-}/(\mathrm{d}p\mathrm{d}\theta)$ for p--Pb interactions. Each row refers to a
    different $(p_{\hbox{\small min}} \le p<p_{\hbox{\small max}},
    \theta_{\hbox{\small min}} \le \theta<\theta_{\hbox{\small max}})$ bin,
    where $p$ and $\theta$ are the pion momentum and polar angle, respectively.
    The central value as well as the square-root of the diagonal elements
    of the covariance matrix are given.}
\vspace{2mm}
\begin{tabular}{rrrr|r@{$\pm$}lr@{$\pm$}lr@{$\pm$}l}
\hline
$\theta_{\hbox{\small min}}$ &
$\theta_{\hbox{\small max}}$ &
$p_{\hbox{\small min}}$ &
$p_{\hbox{\small max}}$ &
\multicolumn{6}{c}{$\mathrm{d}^2\sigma^{\pi^-}/(\mathrm{d}p\mathrm{d}\theta)$}
\\
(rad) & (rad) & (\GeVc) & (\GeVc) &
\multicolumn{6}{c}{($\barn/(\GeVc \cdot \rad)$)}
\\
  &  &  &
&\multicolumn{2}{c}{\GeVc}
&\multicolumn{2}{c}{\GeVc}
&\multicolumn{2}{c}{\GeVc}
\\
\hline 
 0.35 & 0.55 & 0.15 & 0.20& 0.83 &  0.26& 1.38 &  0.46& 1.89 &  0.60\\ 
      &      & 0.20 & 0.25& 0.82 &  0.15& 1.37 &  0.27& 2.08 &  0.37\\ 
      &      & 0.25 & 0.30& 0.73 &  0.10& 1.48 &  0.19& 1.96 &  0.26\\ 
      &      & 0.30 & 0.35& 0.62 &  0.08& 1.23 &  0.13& 1.83 &  0.21\\ 
      &      & 0.35 & 0.40& 0.60 &  0.07& 1.16 &  0.12& 1.65 &  0.16\\ 
      &      & 0.40 & 0.45& 0.63 &  0.06& 1.10 &  0.10& 1.54 &  0.13\\ 
      &      & 0.45 & 0.50& 0.55 &  0.04& 1.05 &  0.09& 1.55 &  0.12\\ 
      &      & 0.50 & 0.60& 0.50 &  0.03& 1.02 &  0.08& 1.48 &  0.11\\ 
      &      & 0.60 & 0.70& 0.41 &  0.04& 0.90 &  0.09& 1.31 &  0.12\\ 
      &      & 0.70 & 0.80& 0.36 &  0.05& 0.76 &  0.11& 1.13 &  0.15\\ 
\hline 
 0.55 & 0.75 & 0.10 & 0.15& 0.95 &  0.40& 1.61 &  0.70& 1.67 &  0.75\\ 
      &      & 0.15 & 0.20& 0.85 &  0.20& 1.40 &  0.32& 1.86 &  0.42\\ 
      &      & 0.20 & 0.25& 0.84 &  0.08& 1.41 &  0.15& 2.03 &  0.20\\ 
      &      & 0.25 & 0.30& 0.78 &  0.06& 1.39 &  0.10& 1.88 &  0.13\\ 
      &      & 0.30 & 0.35& 0.70 &  0.05& 1.20 &  0.08& 1.72 &  0.11\\ 
      &      & 0.35 & 0.40& 0.62 &  0.04& 1.12 &  0.07& 1.48 &  0.08\\ 
      &      & 0.40 & 0.45& 0.57 &  0.03& 1.07 &  0.05& 1.40 &  0.07\\ 
      &      & 0.45 & 0.50& 0.51 &  0.03& 1.00 &  0.05& 1.39 &  0.07\\ 
      &      & 0.50 & 0.60& 0.43 &  0.03& 0.88 &  0.06& 1.27 &  0.08\\ 
      &      & 0.60 & 0.70& 0.33 &  0.03& 0.71 &  0.08& 1.00 &  0.11\\ 
      &      & 0.70 & 0.80& 0.24 &  0.04& 0.57 &  0.09& 0.83 &  0.12\\ 
\hline 
 0.75 & 0.95 & 0.10 & 0.15& 0.88 &  0.31& 1.51 &  0.53& 1.50 &  0.57\\ 
      &      & 0.15 & 0.20& 0.92 &  0.14& 1.48 &  0.21& 1.84 &  0.27\\ 
      &      & 0.20 & 0.25& 0.85 &  0.06& 1.44 &  0.10& 1.93 &  0.12\\ 
      &      & 0.25 & 0.30& 0.78 &  0.05& 1.30 &  0.08& 1.69 &  0.10\\ 
      &      & 0.30 & 0.35& 0.65 &  0.04& 1.09 &  0.06& 1.43 &  0.07\\ 
      &      & 0.35 & 0.40& 0.55 &  0.03& 0.88 &  0.04& 1.20 &  0.05\\ 
      &      & 0.40 & 0.45& 0.46 &  0.02& 0.77 &  0.04& 1.02 &  0.04\\ 
      &      & 0.45 & 0.50& 0.40 &  0.02& 0.72 &  0.04& 0.93 &  0.04\\ 
      &      & 0.50 & 0.60& 0.28 &  0.02& 0.59 &  0.04& 0.76 &  0.05\\ 
      &      & 0.60 & 0.70& 0.19 &  0.02& 0.44 &  0.05& 0.60 &  0.07\\ 
\hline 
 0.95 & 1.15 & 0.10 & 0.15& 1.00 &  0.26& 1.58 &  0.43& 1.61 &  0.47\\ 
      &      & 0.15 & 0.20& 0.98 &  0.11& 1.47 &  0.15& 1.85 &  0.19\\ 
      &      & 0.20 & 0.25& 0.77 &  0.05& 1.37 &  0.09& 1.72 &  0.10\\ 
      &      & 0.25 & 0.30& 0.61 &  0.04& 1.19 &  0.08& 1.41 &  0.08\\ 
      &      & 0.30 & 0.35& 0.48 &  0.03& 0.91 &  0.05& 1.12 &  0.05\\ 
      &      & 0.35 & 0.40& 0.41 &  0.02& 0.69 &  0.04& 0.92 &  0.04\\ 
      &      & 0.40 & 0.45& 0.35 &  0.02& 0.56 &  0.03& 0.76 &  0.04\\ 
      &      & 0.45 & 0.50& 0.26 &  0.02& 0.47 &  0.03& 0.61 &  0.04\\ 
      &      & 0.50 & 0.60& 0.18 &  0.02& 0.37 &  0.03& 0.46 &  0.04\\ 
\hline
\end{tabular}
\end{center}
\end{table}

\begin{table}[hp!]
\begin{center}
\begin{tabular}{rrrr|r@{$\pm$}lr@{$\pm$}lr@{$\pm$}l}
\hline
$\theta_{\hbox{\small min}}$ &
$\theta_{\hbox{\small max}}$ &
$p_{\hbox{\small min}}$ &
$p_{\hbox{\small max}}$ &
\multicolumn{6}{c}{$\mathrm{d}^2\sigma^{\pi^-}/(\mathrm{d}p\mathrm{d}\theta)$}
\\
(rad) & (rad) & (\GeVc) & (\GeVc) &
\multicolumn{6}{c}{(\barn/($\GeVc \cdot \rad$))}
\\
  &  &  &
&\multicolumn{2}{c}{\GeVc}
&\multicolumn{2}{c}{\GeVc}
&\multicolumn{2}{c}{\GeVc}
\\
\hline
 1.15 & 1.35 & 0.10 & 0.15& 1.07 &  0.26& 1.64 &  0.40& 1.84 &  0.47\\ 
      &      & 0.15 & 0.20& 0.93 &  0.08& 1.39 &  0.13& 1.74 &  0.16\\ 
      &      & 0.20 & 0.25& 0.68 &  0.04& 1.25 &  0.08& 1.41 &  0.08\\ 
      &      & 0.25 & 0.30& 0.56 &  0.03& 0.95 &  0.06& 1.08 &  0.06\\ 
      &      & 0.30 & 0.35& 0.41 &  0.03& 0.65 &  0.04& 0.80 &  0.04\\ 
      &      & 0.35 & 0.40& 0.30 &  0.02& 0.50 &  0.02& 0.63 &  0.03\\ 
      &      & 0.40 & 0.45& 0.22 &  0.02& 0.40 &  0.02& 0.50 &  0.03\\ 
      &      & 0.45 & 0.50& 0.17 &  0.02& 0.33 &  0.03& 0.39 &  0.03\\ 
\hline 
 1.35 & 1.55 & 0.10 & 0.15& 1.02 &  0.20& 1.66 &  0.38& 1.91 &  0.42\\ 
      &      & 0.15 & 0.20& 0.81 &  0.06& 1.30 &  0.11& 1.53 &  0.13\\ 
      &      & 0.20 & 0.25& 0.60 &  0.04& 0.94 &  0.05& 1.11 &  0.05\\ 
      &      & 0.25 & 0.30& 0.46 &  0.03& 0.69 &  0.05& 0.86 &  0.05\\ 
      &      & 0.30 & 0.35& 0.34 &  0.02& 0.52 &  0.03& 0.59 &  0.03\\ 
      &      & 0.35 & 0.40& 0.24 &  0.02& 0.39 &  0.03& 0.42 &  0.03\\ 
      &      & 0.40 & 0.45& 0.15 &  0.02& 0.28 &  0.02& 0.32 &  0.02\\ 
      &      & 0.45 & 0.50& 0.09 &  0.01& 0.21 &  0.02& 0.25 &  0.02\\ 
\hline 
 1.55 & 1.75 & 0.10 & 0.15& 1.00 &  0.20& 1.59 &  0.36& 1.74 &  0.37\\ 
      &      & 0.15 & 0.20& 0.69 &  0.06& 1.13 &  0.09& 1.33 &  0.11\\ 
      &      & 0.20 & 0.25& 0.51 &  0.03& 0.73 &  0.04& 0.91 &  0.06\\ 
      &      & 0.25 & 0.30& 0.32 &  0.02& 0.51 &  0.04& 0.64 &  0.04\\ 
      &      & 0.30 & 0.35& 0.23 &  0.01& 0.36 &  0.03& 0.42 &  0.03\\ 
      &      & 0.35 & 0.40& 0.17 &  0.01& 0.23 &  0.02& 0.30 &  0.02\\ 
      &      & 0.40 & 0.45& 0.11 &  0.01& 0.17 &  0.01& 0.22 &  0.02\\ 
      &      & 0.45 & 0.50& 0.07 &  0.01& 0.13 &  0.01& 0.17 &  0.01\\ 
\hline 
 1.75 & 1.95 & 0.10 & 0.15& 1.02 &  0.22& 1.45 &  0.34& 1.60 &  0.38\\ 
      &      & 0.15 & 0.20& 0.60 &  0.05& 0.99 &  0.08& 1.17 &  0.10\\ 
      &      & 0.20 & 0.25& 0.36 &  0.03& 0.54 &  0.04& 0.68 &  0.05\\ 
      &      & 0.25 & 0.30& 0.21 &  0.02& 0.34 &  0.03& 0.43 &  0.03\\ 
      &      & 0.30 & 0.35& 0.13 &  0.01& 0.25 &  0.02& 0.30 &  0.02\\ 
      &      & 0.35 & 0.40& 0.10 &  0.01& 0.16 &  0.02& 0.20 &  0.02\\ 
      &      & 0.40 & 0.45& 0.06 &  0.01& 0.11 &  0.01& 0.14 &  0.01\\ 
      &      & 0.45 & 0.50& 0.04 &  0.01& 0.08 &  0.01& 0.10 &  0.01\\ 
\hline 
 1.95 & 2.15 & 0.10 & 0.15& 1.19 &  0.31& 1.67 &  0.45& 1.79 &  0.50\\ 
      &      & 0.15 & 0.20& 0.68 &  0.10& 1.10 &  0.19& 1.27 &  0.22\\ 
      &      & 0.20 & 0.25& 0.35 &  0.06& 0.56 &  0.09& 0.69 &  0.11\\ 
      &      & 0.25 & 0.30& 0.18 &  0.03& 0.33 &  0.05& 0.39 &  0.06\\ 
      &      & 0.30 & 0.35& 0.11 &  0.02& 0.21 &  0.04& 0.23 &  0.04\\ 
      &      & 0.35 & 0.40& 0.06 &  0.01& 0.13 &  0.03& 0.15 &  0.03\\ 
      &      & 0.40 & 0.45& 0.04 &  0.01& 0.08 &  0.02& 0.10 &  0.02\\ 
      &      & 0.45 & 0.50& 0.03 &  0.01& 0.05 &  0.01& 0.07 &  0.02\\ 
%
\end{tabular}
\end{center}
\end{table}

\clearpage


\end{appendix}


\begin{thebibliography}{999}

 \bibitem{harp-prop}
	 M.G.~Catanesi {\it et al.}, HARP Collaboration,
	 ``Proposal to study hadron production
	 for the neutrino factory and for the atmospheric
	 neutrino flux'',
	 CERN-SPSC/99-35 (1999).
	 
 \bibitem{ref:nufact}
	 A.~Blondel {\it et al.},
	 CERN-2004-002, ECFA/04/230; \\
	 M.~M.~Alsharoa {\it et al.}, Phys. Rev. St. Accel. Beams 6, 081001 (2003).
	 
 \bibitem{Battistoni}  
	 G. Battistoni,
	 Nucl. Phys. Proc. Suppl. {\bf B100} (2001) 101.
	 
 \bibitem{Stanev}  T.~Stanev, Rapporteur's talk at the 26th Int.
	 Cosmic Ray Conference (Salt Lake City, Utah, USA; eds. B.L.~Dingus
	 {\it et al.}, AIP Conf. Proceedings 516, (2000) 247).
	 
 \bibitem{Gaisser} T.K.~Gaisser, Nucl. Phys.
	 Proc. Suppl. {\bf B87} (2000) 145.
	 
 \bibitem{Engel} 
	 R.~Engel, T.K.~Gaisser and T.~Stanev,
	 Phys. Lett. {\bf B472} (2000) 113.
	 
 \bibitem{ref:physrep} 
	 M.~Bonesini and A.~Guglielmi, Phys. Rep. {\bf 433} (2006) 66.
	 
 \bibitem{ref:k2k}  
	 M.H.~Ahn {\it et al.}, K2K Collaboration, 
	 Phys. Rev. Lett. {\bf 90} (2003) 041801.

 \bibitem{ref:k2kfinal}
	 M.~H.~Ahn {\it et al.},  K2K Collaboration,
	 Phys.\ Rev.\ {\bf D74} (2006) 072003,
	 arXiv:hep-ex/0606032.
	 
 \bibitem{ref:miniboone}
	 A. A. Aguilar-Arevalo,
	 The MiniBooNE Collaboration, ``A Search for Electron Neutrino
	 Appearance at the $\Delta m^2 \ \sim \ 1 \ \mathrm{eV}^2$
	 Scale``, arXiv:0704.1500. \\
	 E.~Church {\it et al.}, BooNe Collaboration,
	 ``A proposal for an experiment to measure muon-neutrino $\to$
	 electron-neutrino oscillations and muon-neutrino disappearance at the
	 Fermilab Booster: BooNE'',
	 FERMILAB-PROPOSAL-0898, (1997).
	 
 \bibitem{ref:sciboone}
	 A.~A.~Aguilar-Arevalo {\it et al.},  SciBooNE Collaboration,
	 ``Bringing the SciBar detector to the Booster neutrino beam,''
	 FERMILAB-PROPOSAL-0954, (2006),
	 arXiv:hep-ex/0601022.
	 
 \bibitem{ref:harp:alPaper}
	 M.~G.~Catanesi {\it et al.}, HARP Collaboration,
	 Nucl.\ Phys.\ {\bf B732} (2006) 1,
	 arXiv:hep-ex/0510039.
	 
 \bibitem{ref:harp:bePaper}
	 M.~G.~Catanesi {\it et al.},  HARP Collaboration,
	 Eur. Phys. J. {\bf C52} (2007) 29,
	 arXiv:hep-ex/0702024.
	 
 \bibitem{ref:harp:carbonfw} 
	 M.~G.~Catanesi {\it et al.}, HARP Collaboration,
	 Astropart.\ Phys.\  {\bf 29} (2008) 257,
	 arXiv:0802.0657 [astro-ph].
	 
 \bibitem{ref:harp:o2n2} 
	 M.~G.~Catanesi {\it et al.}, HARP Collaboration,
	 Astropart.\ Phys.\  {\bf 30} (2008) 124,
	 arXiv:0807.1025 [hep-ex].
	 
 \bibitem{ref:harp:forward_pi}
	 M.~Apollonio {\it et al.}, HARP Collaboration,
	 Nucl.\ Phys.\  A {\bf 821} (2009) 118,
	 arXiv:0902.2105 [hep-ex].
	 
 \bibitem{ref:harp:la}
	 M.G. Catanesi {\it et al.}, HARP Collaboration,
	 Phys. Rev. {\bf C77}, (2008)  055207,
	 arXiv:0805.2871 [hep-ex].
	 
 \bibitem{ref:harp:la:pions}
	 M.~Apollonio {\it et al.}, HARP Collaboration,
	 arXiv:0907.1428 [hep-ex] (2009), 
	 to be published in Phys. Rev. C.

 \bibitem{ref:harp:tantalum}
	 M.G. Catanesi {\it et al.}, HARP Collaboration,
	 Eur. Phys. J. {\bf C51}(2007) 787,
	 arXiv:0706.1600 [hep-ex].
	 
 \bibitem{ref:harp:cacotin}
	 M.G. Catanesi {\it et al.}, HARP Collaboration,
	 Eur. Phys. J. {\bf C53 } (2008) 177,
	 arXiv:0709.3464 [hep-ex].
	 
 \bibitem{ref:harp:bealpb}
	 M.G. Catanesi {\it et al.}, HARP Collaboration,
	 Eur. Phys. J. {\bf C54 } (2008) 37,
	 arXiv:0709.3458 [hep-ex]. 

 \bibitem{ref:cdp:la}
	 A.~Bolshakova  {\it et al.}, 
	 Eur. Phys. J. C 62, 293-317 (2009); 
	 Eur. Phys. J. C 62, 697-754 (2009); 
	 arXiv:0906.0471 [hep-ex];
	 arXiv:0906.3653 [hep-ex].
	 
 \bibitem{ref:tpcmom} 
	 M.~G.~Catanesi {\it et al.}, HARP Collaboration,
	 JINST 3 (2008) P04007,
	 arXiv:0709.2806 [hep-ex].
	 
 \bibitem{ref:dyndist} 
	 A.~Bagulya {\it et al.},
	 JINST 4 (2009) P11014,
	 arXiv:0903.4762 [hep-ex].

 \bibitem{ref:harpTech}
	 M.G.~Catanesi {\it et al.}, HARP Collaboration, 
	 Nucl.\ Instrum.\ Meth.\ {\bf A571} (2007) 527; {\bf A571} (2007) 564.
	 
 \bibitem{NOMAD_NIM_DC}
	 M. Anfreville {\it et al.}, Nucl.\ Instrum.\ Meth.\ {\bf A481}
	 (2002) 339.
	 
 \bibitem{ref:tofPaper}
	 M.~Baldo-Ceolin {\it et al.},
	 Nucl.\ Instrum.\ Meth.\ {\bf A532} (2004) 548; \\
	 M.~Bonesini {\it et al.},
	 IEEE Trans. Nucl. Sci. NS-50 (2003) 1053.
	 
 \bibitem{ref:tpc:ieee}E.~Radicioni, 
	 presented at NSS2004, 
	 IEEE Transaction on Nuclear Science, Vol 52, N 6 (2005) 2986. 
	 
 \bibitem{ref:rpc} M.~Bogomilov {\it et al.}, Nucl. Instrum. Methods 
	 {\bf A508} (2003) 152; \\
	 G.~Barr {\it et al.}, Nucl. Instrum. Methods {\bf A533}
	 (2004) 214; \\
	 M.~Bogomilov {\it et al.},
	 IEEE Transaction on Nuclear Science {\bf 54} (2007) 342; \\
	 A.~Artamonov  {\it et al.}, 
	 JINST 2 (2007) P10004,
	 arXiv:0709.3756 [hep-ex].

 \bibitem{ref:t9}
	 L.~Durieu, A.~Mueller and M.~Martini,
	 PAC-2001-TPAH142
	 {\it Presented at IEEE Particle Accelerator Conference (PAC2001),
	 Chicago, Illinois, 18-22 Jun 2001};\\
	 L.~Durieu {\it et al.}, 
	 Proceedings of PAC'97, Vancouver, (1997);  \\
	 L.~Durieu, O.~Fernando, 
	 CERN PS/PA Note 96-38.
	 
\bibitem{ref:tpc:dydak}
	V.~Ammosov {\it et al.}, 
	Nucl. Instrum. Methods {\bf A588} (2008) 294.
	 
 \bibitem{dagostini}
	 G.~D'Agostini,
	 Nucl.\ Instrum.\ Meth.\ {\bf A362} (1995) 487.
	 
 \bibitem{ref:geant4}
	 S.~Agostinelli {\it et al.},  GEANT4 Collaboration,
	 Nucl.\ Instrum.\ Meth.\ {\bf A506} (2003) 250.

 \bibitem{ref:E910} 
	 I. Chemakin {\it et al.}, E910 Collaboration, 
	 Phys. Rev. {\bf C65} (2002) 024904.
	 
 \bibitem{ref:gallmeister}
	 K.~Gallmeister and U.~Mosel,
	 Nucl.\ Phys.\  A {\bf 826} (2009) 151,
	 arXiv:0901.1770 [hep-ex] (2009).
	 
 \bibitem{ref:mars} 
	 N.V. Mokhov, S.I. Striganov,
	 ``MARS overview'', FERMILAB-CONF-07-008-AD, 2007. 
	 
 \bibitem{ref:bert} 
	 D.H. Wright {\it et al.}, AIP Conf. Proc. 896 (2007) 11.
 \bibitem{ref:bert1} 
	 A. Heikkinen {\it et al.}, arXiv:nucl-th/0306008.
 \bibitem{ref:bert2} 
	 H.W. Bertini, P. Guthrie, Nucl. Phys. {\bf A169} (1971) 670.
 \bibitem{ref:QGSP} 
	 G. Folger and H.P. Wellisch, arXiv:nucl-th/0306007.
 
 \bibitem{ref:casca} 
	 S.G. Mashnik {\it et al.}, LANL report LA-UR-05-7321, 2005.

 \bibitem{ref:iss}
	 K.~Long,
	 Nucl. Phys. {\bf B} (Proc. Suppl.), {\bf 154} (2006) 111;
	 ISS/2005/01,
	 ``An international scoping study of a Neutrino Factory and
	 super-beam facility'',
	 {\tt http://www.hep.ph.ic.ac.uk/iss/iss-notes/ISS\_Doc1\_v02\_13-7-2005.pdf}.
	 
	 %
\end{thebibliography}
\end{document}